\documentclass[prx,reprint,columns,notitlepage,superscriptaddress]{revtex4-1}
\usepackage{blindtext}
\usepackage{hyperref}
\usepackage{amsmath,ulem, bm}
\usepackage{lipsum}
\usepackage{eucal}
\usepackage[usenames,dvipsnames]{color}
\usepackage{graphicx}
\usepackage{titletoc}
\usepackage{amssymb,amsfonts,amsmath}
\usepackage{bm}
\usepackage{enumerate}
\usepackage{graphicx}
\usepackage[caption=false]{subfig}
\usepackage{comment}
\usepackage{natbib}
\usepackage{lineno}
\usepackage{multirow,array} 
\usepackage{makecell}

\usepackage[toc,page,header]{appendix}
\usepackage{minitoc}

\begin{document}


\title{Emergence of irregular activity in networks of strongly coupled conductance-based neurons
} 
\author{A.~Sanzeni}
\affiliation{Department of Neurobiology, Duke University, Durham, NC, USA}
\affiliation{National institute of Mental Health Intramural Program, NIH, Bethesda, MD, USA}
\author{M.H.~Histed}
\affiliation{National institute of Mental Health Intramural Program, NIH, Bethesda, MD, USA}
\author{N.~Brunel}
\affiliation{Department of Neurobiology, Duke University, Durham, NC, USA}
\affiliation{Department of Physics, Duke University, Durham, NC, USA}

\begin{abstract}
Cortical neurons are characterized by irregular firing and a broad distribution of rates. 
The balanced state model explains these observations with a cancellation of mean excitatory and inhibitory currents, which makes fluctuations drive firing. 
In networks of neurons with current-based synapses, the balanced state emerges dynamically 
if coupling is strong, i.e. if the mean number of synapses per neuron $K$ is large and synaptic efficacy is of order $1/\sqrt{K}$.
When synapses are conductance-based, current fluctuations are suppressed when coupling is strong, questioning the applicability of the balanced state idea to biological neural networks.
We analyze networks of strongly coupled conductance-based neurons and show that asynchronous irregular activity and broad distributions of rates emerge if synaptic efficacy is of order $1/\log(K)$. 
In such networks, unlike in the standard balanced state model, current fluctuations are small and firing is maintained by a drift-diffusion balance.
This balance emerges dynamically, without fine tuning, if inputs are smaller than a critical value, which depends on synaptic time constants and coupling strength, and is  significantly more robust to connection heterogeneities than the classical balanced state model.
Our analysis makes experimentally testable predictions of how the network response properties should evolve as input increases. 

\end{abstract}
\maketitle

\section{Introduction}
Each neuron in  cortex receives inputs from hundreds to thousands of pre-synaptic neurons. 
If these inputs were to sum to produce a large net current, the central limit theorem argues that fluctuations should be small compared to the mean, leading to regular firing, as observed during \textit{in vitro} experiments under constant current injection~\cite{Douglas1991,Mainen1992}. 
Cortical activity, however, is highly irregular, with a coefficient of variation of  interspike intervals (CV of ISI) close to one~\cite{softky93b,compte03}. 
To explain the observed irregularity, it has been proposed that neural networks operate in a balanced state, where strong feedforward and recurrent excitatory inputs are canceled by recurrent inhibition and firing is driven by fluctuations \cite{shadlen94,shadlen98}. 
At the single neuron level, in order for this state to emerge, input currents must satisfy two constraints.
First, excitatory and inhibitory currents must be fine tuned so to produce an average input below threshold.
Specifically, if $K$ and $J$ represent the average number of input connections per neuron and synaptic efficacy, respectively, the difference between excitatory and inhibitory presynaptic inputs must be of order $1/KJ$. 
Second, input fluctuations should be large enough to drive firing.

It has been shown that the balanced state emerges dynamically (without fine tuning) in randomly connected networks of binary units~\cite{vanvreeswijk96b,vanvreeswijk98} and networks of current-based spiking neurons~\cite{Amit1997,brunel00}, provided that coupling is strong, and recurrent inhibition is powerful enough to counterbalance instabilities due to recurrent excitation.
However, these results have all been derived assuming that the firing of a presynaptic neuron produces a fixed amount of synaptic current, hence neglecting the dependence of synaptic current on the membrane potential, a key aspect of neuronal biophysics.  In real synapses, synaptic inputs are mediated by changes in conductance, due to opening of synaptic receptor-channels on the membrane, and synaptic currents are proportional to the product of synaptic conductance and a driving force which depends on the membrane potential. Models that incorporate this description are referred to as `conductance-based synapses'.

Large synaptic conductances has been shown to have major effects on the stationary~\cite{Richardson2004} and dynamical~\cite{Capaday2006} response of single cells, and form the basis of the `high-conductance state' \cite{Destexhe2001,Shelley2002,Rudolph2003,Lerchner04,Meffin2004,Rudolph2005,Kumar2008} that has been argued to describe well {\it in vivo} data~\cite{Pare1998,Destexhe1999,Destexhe2003} (but see \cite{Li2020} and Discussion).
At the network level, conductance modulation plays a role in controlling signal propagation~\cite{Vogels05},  input summation~\cite{Histed17}, and firing statistics~\cite{Cavallari2014}. 
However, most of the previously mentioned studies rely exclusively on numerical simulations, and in spite of a few attempts at analytical descriptions of networks of conductance-based neurons~\cite{Brunel2001,Meffin2004,Zerlaut2018,Ebsch2018,Capone2019,diVolo2019}, an understanding of the behavior of such networks when coupling is strong is still lacking. 

Here, we investigate  networks of strongly coupled conductance-based neurons. 
We find that, for  synapses of order $1/\sqrt{K}$, fluctuations are too weak to sustain firing, questioning the relevance of the balanced state idea to cortical dynamics.
Our analysis, on the other hand,  shows that stronger synapses (of order $1/\log(K)$) generate irregular firing {when coupling is strong}.
We characterize the properties of networks with such a scaling, showing that they match properties observed in cortex, and discuss constraints induced by  synaptic time constant.
The model generates qualitatively different predictions compared to the current-based model, which could be tested experimentally. 

\section{Models of single neuron and network dynamics} 
\textbf{Membrane potential dynamics.} 
We study the dynamics of networks of leaky integrate-and-fire (LIF) neurons with conductance-based synaptic inputs.
The membrane potential $V_j$ of the $j$-th neuron in the network follows the equation
\begin{equation}\label{eq:membrane}
\begin{split}
\mathcal{C}_j \frac{dV_j}{dt}=- \sum_{A=L,E,I}g^j_A \left( V_j -E_{A}\right) \, , 
\end{split}
\end{equation}
where $\mathcal{C}_j$ is the neuronal capacitance; $E_{L}$, $E_{E}$ and $E_{I}$ are the reversal potentials of the leak, excitatory and inhibitory currents; while $g^j_{L}$, $g^j_{E}$ and $g^j_{I}$ are the leak, excitatory and inhibitory conductances.
Assuming instantaneous synapses (the case of finite synaptic time constants is discussed at the end of the results section), excitatory and inhibitory conductances are given by
\begin{equation}\label{eq:g_instant}
\begin{split}
\frac{g^j_{E,I}}{g^j_L}= \tau_j \sum_m a_{jm} \sum_{n} \delta(t-t_m^n) \, .
\end{split}
\end{equation}
In Eq.~\eqref{eq:g_instant}, $\tau_j=\mathcal{C}_j/{g^j_L}$ is the single neuron membrane time constant,  $a_{jm}$  are dimensionless measures of synaptic strength between neuron $j$ and neuron $m$, $\sum_{n} \delta(t-t_m^n)$ represents the sum of all the spikes generated at times $t_m^n$ by neuron $m$. 
Every time the membrane potential $V_j$ reaches the firing threshold $\theta$, the $j$th neuron emits a spike, its membrane potential is set to a reset $V_r$, and stays at that value for a refractory period $\tau_{rp}$; after this time the dynamics resumes, following Eq.~(\ref{eq:membrane}).

We use $a_{jm}=a$ ($a\, g$) for all excitatory (inhibitory)  synapses.
In the homogeneous case, each neuron receives synaptic inputs from $K_E=K$  ($K_I=\gamma K$) excitatory (inhibitory) cells.
In the network case, each neuron receives additional $K_X=K$ excitatory  inputs from an external population firing with Poisson statistics with rate $\nu_{X}$. 
We use excitatory and inhibitory neurons with the same biophysical properties,   hence the above assumptions imply that the firing rates of excitatory and inhibitory neurons are equal, $\nu=\nu_E=\nu_I$.
Models taking into account the biophysical diversity between the excitatory and inhibitory populations are discussed in Appendix~\ref{SI:response_strong_coupling}.
When heterogeneity is taken into account, the above defined values of $K_{E,I,X}$ represent the means of Gaussian distributions.
We use the following single neuron parameters:
$\tau_{rp}=2$ms, $\theta=-55$mV, $V_r=-65$mV, $E_E=0$mV, $E_I=-75$mV, $E_L=-80$mV,  $\tau_j =\tau_L=20$ms.
We explore various scalings of $a$ with $K$ and, in all cases, we assume that $a\ll 1$. 
When $a\ll 1$, an incoming spike produced by an excitatory presynaptic neuron produces a jump in the membrane potential of amplitude $a(E_E-V)$, where $V$ is the voltage just before spike arrival. 
In cortex, $V\sim{-60}$mV and average amplitudes of post-synaptic potentials are in the order $0.5-1.0$mV~\cite{Bruno1622,Markram1997,Sjostrom2001,Holmgren2003,Lefort2009a,Perin2011,Jiang2015}. Thus, we expect realistic values of $a$ to be in the order of $0.01$.

\textbf{Diffusion and effective time constant approximations.}  
We assume that each cell receives projections from a large number of cells ($K\gg 1$), neurons are sparsely connected and fire approximately as Poisson processes, each incoming spike provides a small change in conductance ($a\ll 1$), and that temporal correlations in synaptic inputs can be neglected.
Under these assumptions,  we can use the diffusion approximation, and approximate the conductances as
\begin{equation}\label{eq:conductance}
\begin{split}
\frac{g_{E}}{g_L}=  a \tau_L \left[K r_E +\sqrt{ K r_E } \zeta_E \right] \, , \\
\frac{g_I}{g_L}= a  g \tau_L \left[  \gamma K  r_I +\sqrt{\gamma K  r_I} \zeta_I \right]\, .
\end{split}
\end{equation}
where $r_E$ and $r_I$ are the firing rates of pre-synaptic E and I neurons, respectively, and $\zeta_{E}$ and $\zeta_{I}$  are independent Gaussian white noise terms  with zero mean and unit variance density. 
In the single neuron case, we take $r_E=\nu_X$, $r_I=\eta \nu_X$ where $\eta$ represents the ratio of I/E input rate. 
In the network case, $r_E=\nu_X+\nu$, $r_I=\nu$ where $\nu_X$ is the external rate, while $\nu$ is the firing rate of excitatory and inhibitory neurons in the network, determined self-consistently (see below).
We point out that, for some activity levels, the assumption of Poisson pre-synaptic firing made in the derivation of Eq.~\eqref{eq:conductance}  breaks down, as neurons in the network show interspike intervals with CV significantly different from one  (e.g. see Fig.~\ref{fig:compare_r_solutions}C).
However, comparisons between mean field results and numerical simulations  (see  Appendix~\ref{SI:MFT_vs_SIM}) show that neglecting non-Poissonianity (as well as other contributions discussed above Eq.~\eqref{eq:conductance}) generates quantitative  but not qualitative discrepancies, with magnitude that decreases with coupling strength.
Moreover,  in Appendix~\ref{SI:single_neuron_response_at_strong_coupling}, we show that if $a\ll1$ the firing of neurons in the network matches that of a Poisson process with refractory period and hence,  when $\nu\ll 1/\tau_{rp}$, deviations from Poissonianity become negligible.

Using the diffusion approximation, Eq.~\eqref{eq:membrane} reduces to 
\begin{equation}\label{eq:intro_V}
    \tau \frac{d V}{d t}=-V+\mu + \sigma(V) \sqrt{\tau}\zeta,
\end{equation}
where $\zeta$ is a white noise term, with zero mean and unit variance density, while
\begin{equation}\label{eq:param_MM}
\begin{split}\tau^{-1}={\tau_L}^{-1}+ a K \, (r_E  +r_I g \gamma)  \, ,  \\
\mu=\tau\{E_L/\tau_L+a K [r_E E_E +r_I g \gamma E_I]\}\, , \\
\sigma^2(V)= a^2K\tau  \left[r_E\left(V-{E_E}\right)^2 +g^2 \gamma r_I \left(V-{E_I}\right)^2 \right]\, .
\end{split}
\end{equation}
In Eq.~\eqref{eq:intro_V}, $\tau$ is an effective membrane time constant,  while $\mu$ and $\sigma^2(V)$ represent the average and the variance of the synaptic current generated by incoming spikes, respectively. 

The noise term in Eq.~\eqref{eq:intro_V} can be decomposed into an additive and a multiplicative component. 
The latter has an effect on membrane voltage statistics that is of the same order of the contribution coming from synaptic shot noise~\cite{Richardson2005}, a factor which has been neglected in deriving Eq.~\eqref{eq:conductance}. 
Therefore, for a consistent analysis, we  neglect the multiplicative component of the noise in the above derivation; this leads to an equation of the form of Eq.~\eqref{eq:intro_V} with the substitution
\begin{equation}\label{eq:effective_tau_approx}
   \sigma(V)\rightarrow \sigma(\mu) \, .
\end{equation}
This approach has been termed the effective time constant approximation~\cite{Richardson2005}. 
Note that the substitution of Eq.~\eqref{eq:effective_tau_approx} greatly simplifies mathematical expressions but it is not a necessary ingredient for the results presented in this paper.
In fact, all our results can be obtained without having to resort to this approximation (see Appendix~\ref{SI:mean_field_response}, \ref{SI:single_neuron_response_at_strong_coupling} and \ref{SI:response_strong_coupling}).

\textbf{Current-based model.}  
The previous definitions and results translate directly to current-based models, with the only exception that the dependency of excitatory and inhibitory synaptic currents on the membrane potential are neglected (see~\cite{brunel00} for more details).
Therefore, Eq.~\eqref{eq:membrane} becomes
\begin{equation}\label{eq:membrane_curr}
{\tau}_j \frac{dV_j}{dt}=- V_j+ I^j_{E}-I^j_{I}\, ,
\end{equation}
where
\begin{equation*}
I^j_A =\tau_j \sum_m J_{jm} \sum_{n} \delta(t-t_m^n) 
\end{equation*}
represent the excitatory and inhibitory input currents. 
Starting from Eq.~\eqref{eq:membrane_curr}, making  assumptions analogous to those discussed above and using the diffusion approximation~\cite{brunel00}, the dynamics of current-based neurons is given by an equation of the form of Eq.~\eqref{eq:intro_V}
with 
\begin{equation}\label{eq:param_curr}
\begin{split}
\tau=\tau_L\, , \quad
\mu=\tau J K \left[r_E-g \gamma  r_I\right] \, , \\
\sigma^2 =\tau J^2 K  \left[r_E+g^2\gamma r_I\right] \, ;
\end{split}
\end{equation}
Note that, unlike what happens in conductance-based models, $\tau$ is a fixed parameter and does not depend on network firing rate or external drive.
Another difference between the current-based and conductance-based models is that in the latter, but not the former, model $\sigma$  depends on $V$; as we discussed above, this difference is neglected in the main text, where we use the effective time constant approximation.

\section{Behavior of single neuron response for large $K$}
\begin{figure*}[!ht]
\includegraphics[width=18cm]{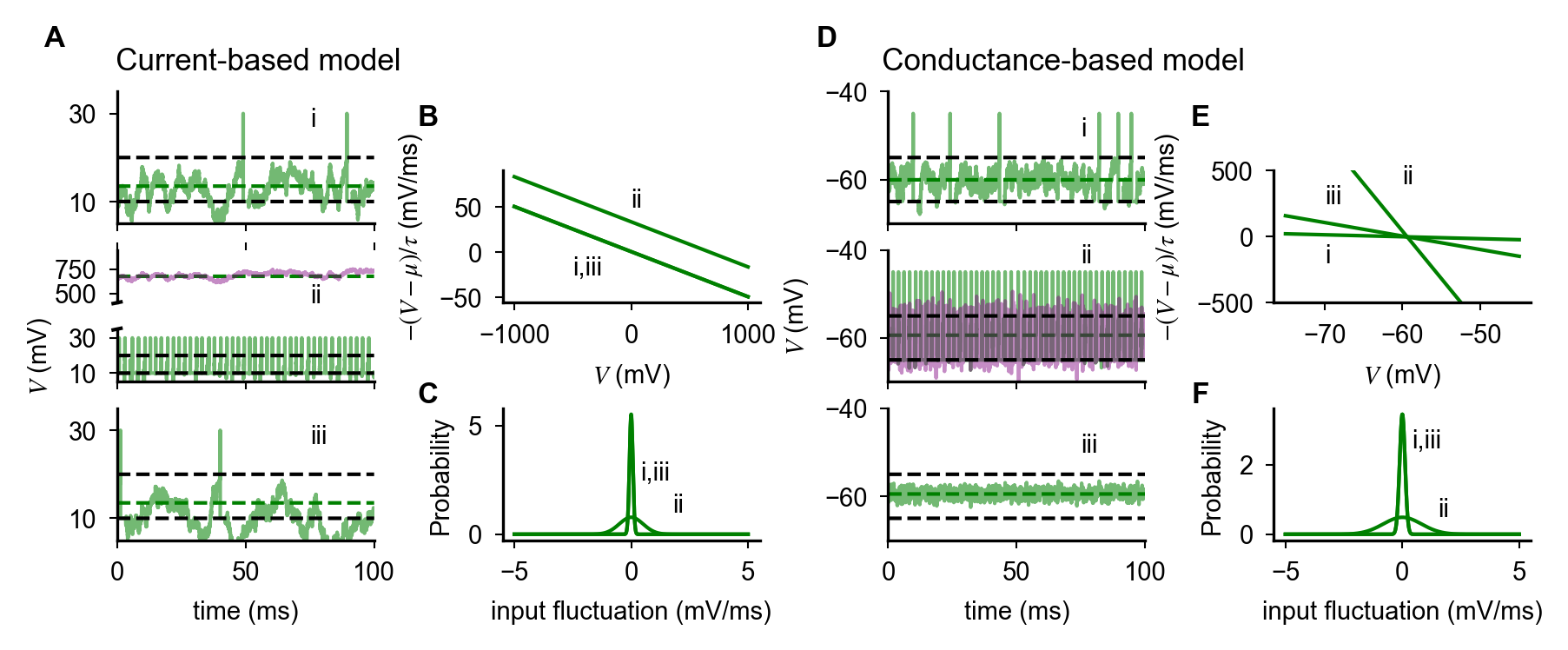}
\caption{{\bf Effects of coupling strength on the firing behavior of current-based and conductance-based neurons.}
({\bf A}) Membrane potential of a single current-based neuron for 
(i) $J=0.3$mV, $K=10^3$, $g=\gamma=1$, $\eta$ such that $1-g \gamma   \eta=0.075$;  (ii) with $K=5\,10^4$; (iii) with $K=5\,10^4$ and scaled synaptic efficacy ($J\sim1/\sqrt{K}$, which gives $J=0.04$mV) and input difference $1-g \gamma   \eta=0.01$;
({\bf B},{\bf C}) Effect of coupling strength on drift force and input noise in a current-based neuron.
({\bf D}) Membrane potential of a single conductance-based neuron for fixed input difference ($g 1-\gamma   \eta=-2.8$) and
(i) $a=0.01$, $K=10^3$;  (ii)  $K=5\,10^4$; (iii)  $K=5\,10^4$ and scaled synaptic efficacy ($a\sim1/\sqrt{K}$,  $a=0.001$).
({\bf E},{\bf F}) Effect of coupling strength on drift force and input noise in a conductance-based neuron.
In panels {\bf A} and {\bf D}, dashed lines represent threshold and reset (black) and equilibrium value of membrane potential (green).
In panels {\bf A}ii and {\bf D}ii, light purple traces represent dynamics in the absence of spiking mechanism. 
Input fluctuations in {\bf C} and {\bf F} represent input noise per unit time, i.e. the integral of $\sigma \sqrt{\tau} \zeta$ of Eq.~\eqref{eq:intro_V} computed over an interval $\Delta t$ and normalized over  $\Delta t$.
}
\label{fig:single_neuron_saturation} 
\end{figure*}

We start our analysis investigating the effects of synaptic conductance on single neuron response.
We  consider a neuron receiving $K$ ($\gamma K$) excitatory (inhibitory) inputs, each with synaptic efficacy $J$ ($gJ$), 
from cells firing with Poisson statistics with a rate
\begin{equation}\label{eq:single_neuron_inputs}
r_E=  \nu_X\, , \quad r_I=\eta  \nu_X \, ,
\end{equation}
and analyze its membrane potential dynamics in the frameworks of current-based and conductance-based models. 
In both models, the membrane potential $V$ follows a stochastic differential equation of the form of Eq.~\eqref{eq:intro_V}; differences emerge in the dependency of $\tau$, $\mu$ and $\sigma$ on the parameters characterizing the connectivity, $K$ and $J$. 
In particular,  in the current-based model,  
the different terms in Eq.~\eqref{eq:param_curr} can be writen as
\begin{equation*}
    \tau\sim \tau_0^{curr}\, , \quad \mu\sim K J \mu_0^{curr}\, , \quad \sigma\sim \sqrt{K} J \sigma_0^{curr}\, ;
\end{equation*}
where $\tau_0^{curr}$, $\mu_0^{curr}$, and $\sigma_0^{curr}$ are independent of $J$ and $K$. 
In the conductance-based model, the efficacy of excitatory and inhibitory synapses depend on the membrane potential as $J=a (E_{E,I}-V)$; the different terms in Eq.~\eqref{eq:intro_V}, under the assumption that $Ka\gg1$, become of order
\begin{equation*}
    \tau\sim  \frac{\tau_0^{cond}}{K a }\, , \quad \mu\sim  \mu_0^{cond}\, , \quad \sigma\sim \sqrt{a} \sigma_0^{cond} \, .
\end{equation*}
Here, all these terms depend on parameters in a completely different way than in the current-based case. 
As we will show below, these differences drastically modify how the neural response changes as $K$ and $J$ are varied and hence the size of $J$ ensuring finite response for a given value of $K$.


The dynamics of a current-based neuron is shown in Fig.~\ref{fig:single_neuron_saturation}Ai, with parameters leading to irregular firing. 
Because of the chosen parameter values, the mean excitatory and inhibitory inputs approximately cancel each other, generating subthreshold average input and fluctuation-driven spikes, which leads to irregularity of firing.
If all parameters are fixed while $K$ is increased ($J\sim K^0$), the response changes drastically (Fig.~\ref{fig:single_neuron_saturation}Aii), since the mean input becomes much larger than threshold and firing becomes regular. 
To understand this effect, we analyze how terms in Eq.~\eqref{eq:intro_V} are modified as $K$ increases. 
The evolution of the membrane potential in time is determined by two terms: 
a  drift term $-(V-\mu)/\tau$, which drives the membrane potential toward its mean value $\mu$, and a noise term $\sigma/\sqrt{\tau}$, which leads to fluctuations around this mean value. 
Increasing $K$ modifies the equilibrium value $\mu$ of the drift force and the input noise, which increase proportionally to $K J (1-\gamma g \eta) $ and $ K J^2 (\gamma g^2 \eta +1) $, respectively  (Fig.~\ref{fig:single_neuron_saturation}B,C). 

This observation suggests that, to preserve irregular firing as $K$ is increased, two ingredients are needed.
First, the rates of excitatory and inhibitory inputs must be fine tuned to maintain a mean input below threshold; this can be achieved choosing  $\gamma g \eta-1\sim 1/K J $.
Second, the amplitude of input fluctuations should be preserved; this can be achieved scaling synaptic efficacy as $J\sim1/\sqrt{K}$. 
Once these two conditions are met, irregular firing is restored  (Fig.~\ref{fig:single_neuron_saturation}Aiii).
Importantly, in a network with $J\sim1/\sqrt{K}$, irregular firing emerges without fine tuning, since rates dynamically adjust to balance excitatory and inhibitory inputs and maintain  mean inputs below threshold~\cite{vanvreeswijk96b,vanvreeswijk98}.

We now show that this solution does not work once synaptic conductance is taken into account.
The  dynamics of a conductance-based neuron in response to the inputs described above is shown in Fig.~\ref{fig:single_neuron_saturation}Di.
As in the current-based neuron, it features irregular firing, with mean input below threshold and spiking driven by fluctuations, and firing becomes regular  for larger  $K$, leaving all other parameters unchanged  (Fig.~\ref{fig:single_neuron_saturation}Dii).
However, unlike the current-based neuron, input remains below threshold at large $K$; regular firing is produced by large fluctuations, which saturate response and  produce spikes that are regularly spaced  because of the refractory period.
These observations can be understood looking at the equation for the membrane potential dynamics: increasing $K$ leaves invariant the equilibrium value of the membrane potential $\mu$ but increases the drift force and the input noise amplitude as $K a $ and $\sqrt{K}a$,  respectively (Fig.~\ref{fig:single_neuron_saturation}E,F). 
Since the equilibrium membrane potential is fixed below threshold, response properties are determined by the interplay between drift force and input noise, which have opposite effects on the probability of spike generation.
The response saturation observed in  Fig.~\ref{fig:single_neuron_saturation}Dii shows that, as $K$ increases at fixed $a$, fluctuations dominate over drift force. 
On the other hand, using the scaling $a\sim1/\sqrt{K}$ leaves the amplitude of fluctuations unchanged, but generates a restoring force of order $\sqrt{K}$  (Fig.~\ref{fig:single_neuron_saturation}E) which dominates and  completely abolishes firing  at strong coupling (Fig.~\ref{fig:single_neuron_saturation}Diii). 

Results in Fig.~\ref{fig:single_neuron_saturation} show that the response of a conductance-based neuron when $K$ is large depends on the balance between drift force and input noise. The scalings $a\sim O(1)$ and $a\sim 1/\sqrt{K}$ leave one of the two contributions dominate;  suggesting that an intermediate scaling could keep a balance between them. 
Below we derive such a scaling, showing that it preserves firing rate and CV of ISI when $K$ becomes large.

\begin{figure*}[!htb]
\centering
\includegraphics[width=17.8cm]{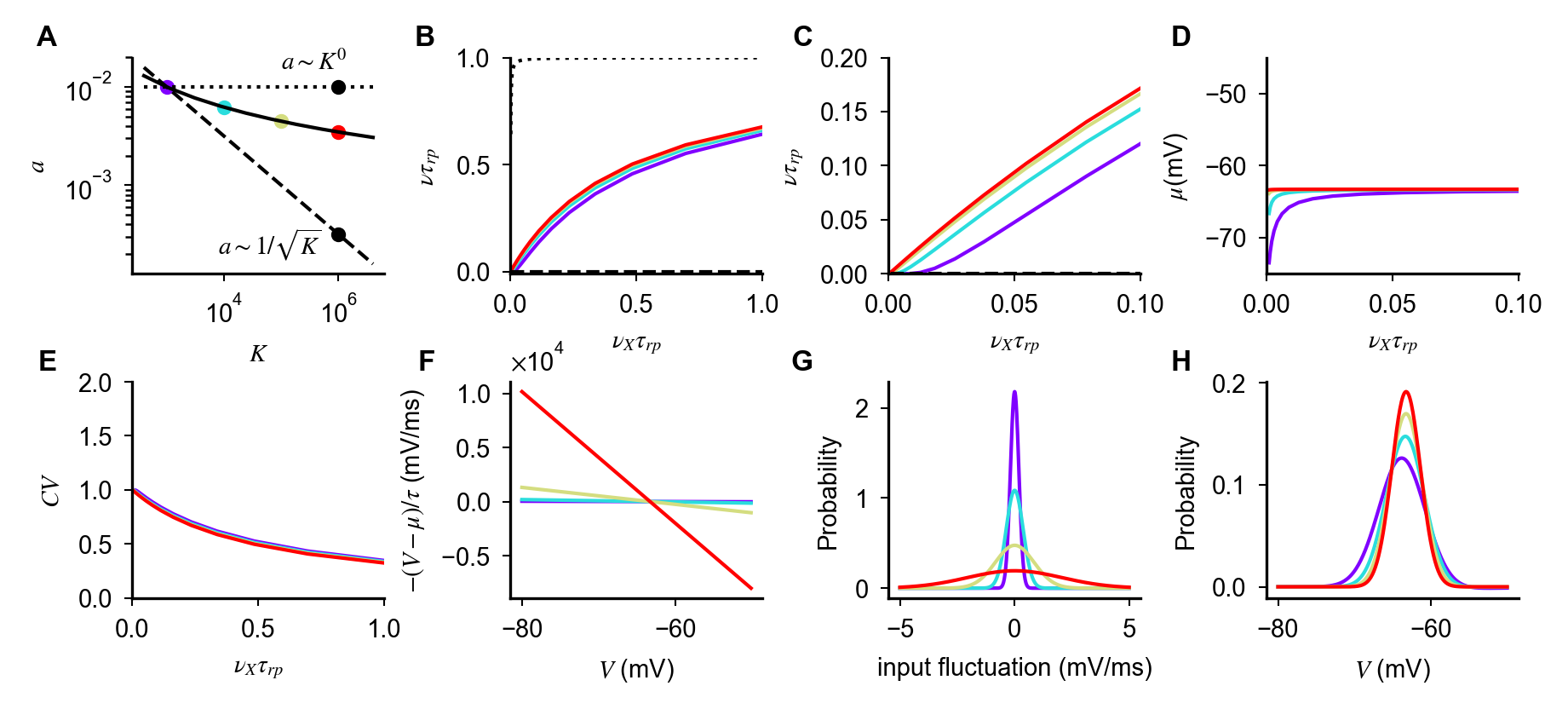}
\caption{{\textbf{The scaling of Eq.~\eqref{eq:scaling_sub} {preserves the} response of a single conductance-based neuron {for large $K$.}}} 
({\bf A}) Scaling relation preserving firing in conductance-based neurons (Eq.~\eqref{eq:scaling_sub}, solid line); constant scaling ($a\sim K^0$, dotted line) and scaling of the balanced state model  ($a\sim 1/\sqrt{K}$, dashed line) are shown as a comparison.
Colored dots indicate values of $a,K$  used in the subsequent panels. 
({\bf B}-{\bf H}) Response of conductance-based neurons, for different values of coupling strength and synaptic efficacy  (colored lines).
The scaling of Eq.~\eqref{eq:scaling_sub}  preserves how firing rate ({\bf B},{\bf C}); equilibrium value of the membrane potential ({\bf D});  and CV of the inter-spike interval distribution  ({\bf E}) depend on external input rate $\nu_X$.
This invariance is achieved increasing the drift force  ({\bf F}) and input fluctuation  ({\bf G}) in a way that weakly decreases (logarithmically in $K$)  membrane potential fluctuations ({\bf H}). 
Different scalings either saturate or suppress response ({\bf B}, black lines correspond to $K=10^5$ and $a$ values as in panel {\bf A}).
Parameters: $a=0.01$ for $K=10^3$, $g=12$, $\eta=1.8$, $\gamma=1/4$.
}
\label{fig:single_neuron_scaling}
\end{figure*}

\section{A scaling relation that preserves single neuron response for large $K$}
We analyze under what conditions the response of a single conductance-based neuron is preserved when  $K$ is large. 
For a LIF neuron driven 
described by Eqs.~(\ref{eq:intro_V}, \ref{eq:param_MM}, \ref{eq:effective_tau_approx}),  the single cell transfer function, i.e.~the dependency of the firing rate $\nu$ on the external drive $\nu_X$, is given by \cite{Siegert1951,Ricciardi1977}
\begin{equation}\label{eq:nu}
\nu=\left[\tau_{rp}+\tau \sqrt{\pi} \int_{v_{min}}^{v_{max}}dx\exp(x^2)\left(1+\text{erf}(x) \right)\right]^{-1}\, ,
\end{equation}
with
\begin{equation}\label{eq:terms_model_A_MFT}
v(x)=\frac{x-\mu}{\sigma} \, , \quad
v_{min}=v({V_r})\, , \quad
v_{max}=v(\theta) \, .
\end{equation}
In the biologically relevant case of  $a\ll1$, Eq.~\eqref{eq:nu} simplifies significantly.
In fact, the distance between the equilibrium membrane potential measured in units of noise $u_{max}$
is of order $1/\sqrt{a}$ (except for inputs $\nu_X\ll1/aK\tau_L$, where it is of order $1/a \sqrt{K\nu_X\tau_L}\gg 1/\sqrt{a}$.)
Therefore $u_{max}$ is large when $a$ is small; in this limit, the firing rate is given by Kramers escape rate \cite{GardinerBook09}, and Eq.~\eqref{eq:nu} becomes: 
\begin{equation}\label{eq:single_neuron}
\begin{split}
\nu=\frac1{\tau_{rp}+\frac{\mathcal{Q}}{\nu_X}}  \, , \quad \mathcal{Q}=\frac{\bar{\tau} \sqrt{\pi}}{\sqrt{a}K \bar{v}} \exp\left(\frac{\bar{v}^2}a\right) \, ,
\end{split}
\end{equation}
{where we have defined $\bar{v}^2=a v_{max}^2$ and $\bar \tau = aK\nu_{X}\tau$. 
The motivation to introduce $\bar v$ and $\bar \tau$  is that they remain of order 1 in the small $a$ limit, provided the external inputs $\nu_X$ are at least of order $1/(aK\tau_L)$. When the external inputs are such that $\nu_X\gg 1/(aK \tau_L)$, these quantities become independent of $\nu_X$, $a$ and $K$ and are given by}
 \begin{equation}\label{eq:param}
\begin{split}
\bar{\tau}=\left(1+g \gamma \eta 
\right)^{-1} \, , \quad \bar{v}=\frac{\theta-\bar{\mu}}{\bar{\sigma}}\, , \\
\bar{\mu}=\bar{\tau}\left( E_E + g \gamma \eta E_I
\right)\, ,\\
\bar{\sigma}^2=\bar{\tau}\left[\left(\bar{\mu}-{E_E}\right)^2 +g^2 \gamma \eta \left(\bar{\mu}-{E_I}\right)^2\right]\, .
\end{split}
\end{equation}
The firing rate given by Eq.~\eqref{eq:single_neuron} remains finite when $a$ is small and/or $K$ is large if $\mathcal{Q}$ remains of order one;  this condition
leads to the following scaling relationship
\begin{equation}\label{eq:scaling_sub}
K\sim \frac{\bar{\tau}}{\sqrt{a} \,\bar{v}}\exp \left(\frac{\bar{v}^2}{a }  \right) \, ;
\end{equation}
i.e.~$a$ should be of order $1/\log(K)$.

In Appendix~\ref{SI:general_single_neuron}, we show that expressions analogous to Eq.~\eqref{eq:single_neuron} can be derived in integrate-and-fire neuron models which feature additional intrinsic voltage-dependent currents, as long as synapses are conductance-based and input noise is small ($a\ll1$). Examples of such models include the exponential integrate-and-fire neurons with its spike-generating exponential current \cite{Fourcaud2003}, and models with voltage-gated subthreshold currents~\cite{Li2020}.
Moreover, we show that, in these models, firing remains finite if  $a\sim1/\log(K)$, and  voltage-dependent currents generate corrections to the logarithmic scaling which are negligible when coupling is strong.

Since $\bar{v}$ and $\bar{\tau}$ vary with $\nu_X$, Eq.~\eqref{eq:scaling_sub} can be satisfied, and hence firing can be supported, only if the inputs span a small range of values, such that $\bar{\tau}$ and $\bar{v}$ are approximately constant,  or if $\nu_X \gg 1/aK\tau_L$. 
Note that, while in the strong coupling limit (i.e. when $K$ goes infinity), only the second of these two possibilities can be implemented with input rates spanning physiologically relevant values, both are are admissible when coupling is moderate (i.e. when $K$ is large but finite, a condition consistent with experimental data on cortical networks~\cite{SanzeniISN,Ahmadian2019}).   
In what follows, with the exception of the section on finite synaptic time constant, we focus on the case $\nu_X\gg 1/{aK \tau_L }$, and  investigate how different properties evolve with $K$ using the scaling defined by Eq.~\eqref{eq:scaling_sub} with  $\bar{v}$ and $\bar{\tau}$ given by Eq.~\eqref{eq:param}. 
Importantly, all the results discussed below hold for inputs outside the region $\nu_X\gg 1/{aK \tau_L }$, as long as 
 $\nu_X$ is at least of order $1/aK\tau_L$ (a necessary condition for the derivation of Eq.~\eqref{eq:single_neuron} to be valid), and that inputs span a region  small enough for the variations of $\bar{v}$ and $\bar{\tau}$ to be negligible.

In Fig.~\ref{fig:single_neuron_scaling}A, we compare the scaling defined by Eq.~\eqref{eq:scaling_sub} with the $a\sim 1/\sqrt{K}$ scaling of current-based neurons.
At low values of $K$, the values of $a$ obtained with the two scalings are similar;
at larger values of $K$, synaptic strength defined by Eq.~\eqref{eq:scaling_sub} decays as $ a\sim  1/{\log(K)}$, i.e. synapses are stronger in the conductance-based model than in the current-based model. 
Examples of single neuron transfer function computed from Eq.~\eqref{eq:nu} for different coupling strength are shown in Fig.~\ref{fig:single_neuron_scaling}B,C.
Responses are nonlinear at onset and  close to saturation.
As predicted by the theory, scaling $a$ with $K$ according to Eq.~\eqref{eq:scaling_sub} {preserves the firing} rate over a region of inputs that increases with coupling strength (Fig.~\ref{fig:single_neuron_scaling}C,D), while the {average} membrane potential remains below threshold (Fig.~\ref{fig:single_neuron_scaling}D).
The quantity $\bar{v}/\sqrt{a}$ represents the distance from threshold of the equilibrium membrane potential in units of input fluctuations; Eq.~\eqref{eq:scaling_sub} implies that this distance increases with coupling strength. 
When $K$ is very large, the effective membrane time constant, which is of order $\tau\sim 1/aK\nu_X$, becomes small and firing is driven by fluctuations that, on the time scale of this effective membrane time constant, are rare.

We next investigated if the above scaling preserves irregular firing by analyzing the CV of interspike intervals. 
This quantity is given by~\cite{brunel00} 
\begin{equation}\label{eq:CV}
CV^2=2\pi\nu^2\tau^2
\int_{v_{min}}^{v_{max}} dx \,  e^{x^2} \int_{-\infty}^x dy \,  e^{y^2} \left( 1+erf(y)\right)^2 \,   
\end{equation}
and, for the biologically relevant case of $a\ll1$ and $\mu<\theta$,  reduces to (see  Appendix~\ref{SI:single_neuron_response_at_strong_coupling} for details)  
\begin{equation}\label{eq:CV_sub}
    CV=1-\tau_{rp}\nu \, ;
\end{equation}
i.e. the CV is close to one at low rates and it decays monotonically as the neuron approaches saturation.
Critically, Eq.~\eqref{eq:CV_sub} depends on coupling strength only through $\nu$, hence any scaling relation preserving firing rate will also produce CV of order one at low rate. 
We validated numerically this result in Fig.~\ref{fig:single_neuron_scaling}E.

We now investigate  how  Eq.~\eqref{eq:scaling_sub} preserves irregular firing in conductance-based neurons. 
We have shown that increasing $K$ at fixed $a$ produces large input and membrane fluctuations, which saturate firing; the scaling $a\sim 1/\sqrt{K}$ preserves input fluctuations but, because of the strong drift force, suppresses membrane potential fluctuations, and hence firing. 
The scaling of Eq.~\eqref{eq:scaling_sub}, at every value of $K$, yields the value of $a$ that  balances the  contribution of drift and input fluctuations, so that membrane fluctuations are of the right size to preserve the rate of threshold crossing.
Note that, unlike what happens in the current-based model,  both input fluctuations and drift force increase with $K$  (Fig.~\ref{fig:single_neuron_scaling}F,G) while  the  membrane potential distribution, which is given by~\cite{Brunel1999}
\begin{equation}\label{eq:P_of_V}
P(V)=\frac{2  \nu \tau }{\sigma}
\int_{v(V)}^{v_{max}}dx \theta(x-v(V_r)) \exp\left[x^2-v(V)^2\right] \, , 
\end{equation}
slowly becomes narrower  (Fig.~\ref{fig:single_neuron_scaling}H).
This result can be understood by noticing that, when $a\ll1$ and neglecting the contribution due to the refractory period,  Eq.~\eqref{eq:P_of_V} reduces to 
\begin{equation}\label{eq:P_sub}
P(V)=\frac{1}{\sigma\sqrt{\pi}} \exp \left(-\frac{\left( V-\mu\right)^2}{\sigma^2}\right)\, .
\end{equation}
Hence, the probability distribution becomes Gaussian  when coupling is strong, with a variance proportional to $\sigma^2\sim a$. 
We note that, since $a$ is of order $1/\log{K}$, the width of the distribution becomes small only for unrealistically large values of $K$.

\section{Asynchronous irregular activity in network response at strong coupling}
We have so far considered the case of a single neuron subjected to stochastic inputs.
We now show how the above results generalize to the network case, where inputs to a neuron are produced by a combination of external and recurrent inputs.
\begin{figure}[!htb]
\centering
\includegraphics[width=8.7cm]{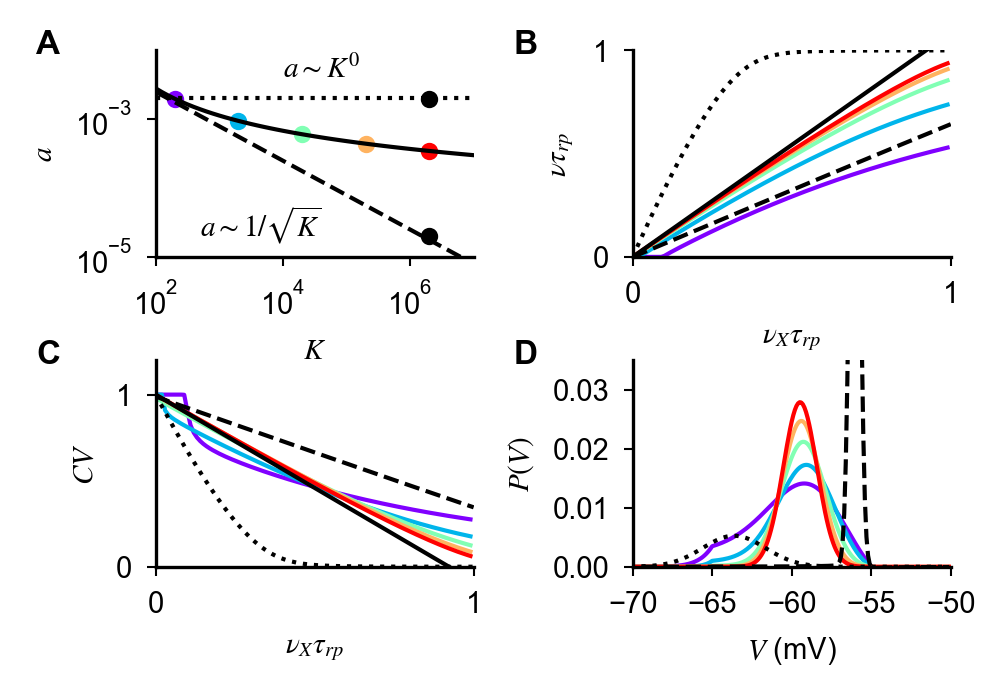}
\caption{
\textbf{
Response of networks of conductance-based neurons for large $K$.
} 
(\textbf{A}) Scaling relation defined by self-consistency condition given by Eqs
.~\eqref{eq:scaling_sub} and \eqref{eq:net_rates} (black line), values of parameters used in panels \textbf{B}-\textbf{D} (colored-dots). Constant scaling ($a\sim K^0$, dotted line) and scaling of the balanced state model  ($a\sim 1/\sqrt{K}$, dashed line) are shown for comparison.
(\textbf{B},\textbf{C}) Firing rate and CV of ISI as a function of external input, obtained from Eqs.~\eqref{eq:nu} and \eqref{eq:CV} (colored lines) with strong coupling limit solution of Eqs.~\eqref{eq:network_solution} and \eqref{eq:CV_sub} (black line).
(\textbf{D}) Probability distribution of the membrane potential obtained from \eqref{eq:P_of_V}.
In panels \textbf{B}-\textbf{D}, dotted and dashed lines represent quantities obtained with the scalings $J\sim K^0$ and $J\sim1/\sqrt{K}$, respectively, for  values of $K$ and $J$ indicated in panel \textbf{A} (black dots).
Parameters: $\gamma=1/4$ and $g=30$.
}
\label{fig:compare_r_solutions}
\end{figure}

We consider networks of recurrently connected excitatory and inhibitory neurons, firing at rate $\nu$, stimulated by an external population firing with Poisson statistics with firing rate $\nu_X$. 
Using again the diffusion approximation, the response of a single neuron in the networks is given by Eq.~\eqref{eq:nu} (and hence Eq.~\eqref{eq:single_neuron}) with 
\begin{equation}\label{eq:net_rates}
    r_E= \nu_{X}+\nu\, ,  \quad r_I= \nu \, .
\end{equation}
Eq~\eqref{eq:nu}, if all neurons in a given population are described by the same single cell parameters  and the network is in an asynchronous state in which cells fire at a constant firing rate,  provides an implicit equation whose solution is the network transfer function.
Example solutions  are shown in Fig.~\ref{fig:compare_r_solutions}B   (numerical validation of the mean field results is provided in  Appendix~\ref{SI:MFT_vs_SIM}). 
In Appendix~\ref{SI:response_strong_coupling}, we prove that firing in the network is preserved  when coupling is strong if parameters are rescaled according to Eq.~\eqref{eq:scaling_sub}.
Moreover, we show that  response nonlinearities are suppressed and the network response in the strong coupling limit (i.e. when $K$ goes infinity) is given, up to saturation, by 
\begin{equation}\label{eq:network_solution}
    \nu=\rho \nu_{X} \, .
\end{equation}
The parameter $\rho$, which is obtained solving Eq.~\eqref{eq:single_neuron} self-consistently (see Appendix~\ref{SI:response_strong_coupling} for details),  is the response gain in the strong coupling limit. 
Finally,  our derivation implies that Eq.~\eqref{eq:scaling_sub}  preserves irregular firing and creates a probability distribution of membrane potential
whose width decreases only logarithmically as $K$ increases  (Fig.~\ref{fig:compare_r_solutions}C,D and  numerical validation in  Appendix~\ref{SI:MFT_vs_SIM}), as in the single neuron case. While this logarithmic decrease is a qualitative difference with the current-based balanced state in which the width stays finite in the large $K$ limit, in practice for realistic values of $K$, realistic fluctuations of membrane potential (a few mV) can be observed in both cases.

We now turn to the question of what happens in networks with different scalings between $a$ and $K$.
Our analysis of single neuron response described above shows that scalings different from that of  Eq.~\eqref{eq:scaling_sub} fail to preserve firing for large $K$, as they let either input noise or drift dominate. However, the situation in networks might be different, since recurrent interactions could in principle adjust the statistics of input currents such that irregular firing at low rates is preserved when coupling becomes strong. Thus, we turn to the analysis of the network behavior when a scaling $a\sim K^{-\alpha}$ is assumed.
For $\alpha\leq0$,  the dominant contribution of input noise at the single neuron level (Figs.~\ref{fig:single_neuron_saturation} and \ref{fig:single_neuron_scaling}) generates saturation of response and regular firing in the network (Fig.~\ref{fig:compare_r_solutions}). 
This can be understood by noticing that, for large $K$, the factor $\mathcal{Q}$ in  Eq.~\eqref{eq:single_neuron} becomes negligible and the self-consistency condition defining the network rate is solved by $\nu=1/\tau_{rp}$.
For $\alpha>0$, the network response for large $K$ is determined by two competing elements. On the one hand, input drift dominates and tends to suppress firing (Figs.~\ref{fig:single_neuron_saturation} and \ref{fig:single_neuron_scaling}).
On the other hand, for the network to be stable, inhibition must dominate recurrent interactions~\cite{Amit1997}. Hence, any suppression in network activity reduces recurrent inhibition and tends to increase neural activity.
When these two elements conspire to generate a finite network response, the factor $\mathcal{Q}$ in Eq.~\eqref{eq:single_neuron} must be of order one and $\bar{v}\sim a\sim K^{-\alpha}$. 
In this scenario, the network activity exhibits the following features (Fig.~\ref{fig:compare_r_solutions}): (i) the mean inputs drive neurons very close to threshold ($\theta-\bar{\mu}\sim a \bar{\sigma}\sim K^{-\alpha}$); (ii) the response of the network to external inputs is linear and, up to corrections of order ${K^{-\alpha}}$, given by   
\begin{equation}\label{eq:response_large_tau_S}
   \nu=\frac{\left(E_E-\theta\right) \nu_X}{\theta(1+g\gamma)-E_E-g\gamma E_I} \, ;
\end{equation}
 (iii) firing is irregular (because of Eq.~\eqref{eq:CV_sub}); (iv) the width of the membrane potential distribution  is of order  $a\sim K^{-\alpha}$ (because of Eq.~\eqref{eq:P_sub}).
 Therefore, scalings different from that of  Eq.~\eqref{eq:scaling_sub} can produce asynchronous irregular activity in networks of conductance-based neurons, but this leads to networks with membrane potentials narrowly distributed close to 
 threshold, a property which seems at odds with what is observed in cortex~\cite{Anderson2000,Crochet2006,Poulet2008,Tan2014,Okun2015,Yu2016}.

\section{Robust lognormal distribution of firing rates in networks with heterogeneous connectivity}
Up to this point, we have assumed a number of connections equal for all  neurons.
In real  networks, however, this number fluctuates from cell to cell.
The goal of this section is to  analyze the effects of heterogeneous connectivity in networks of conductance-based neurons. 
\begin{figure}[!ht]
\includegraphics[width=8.7cm]{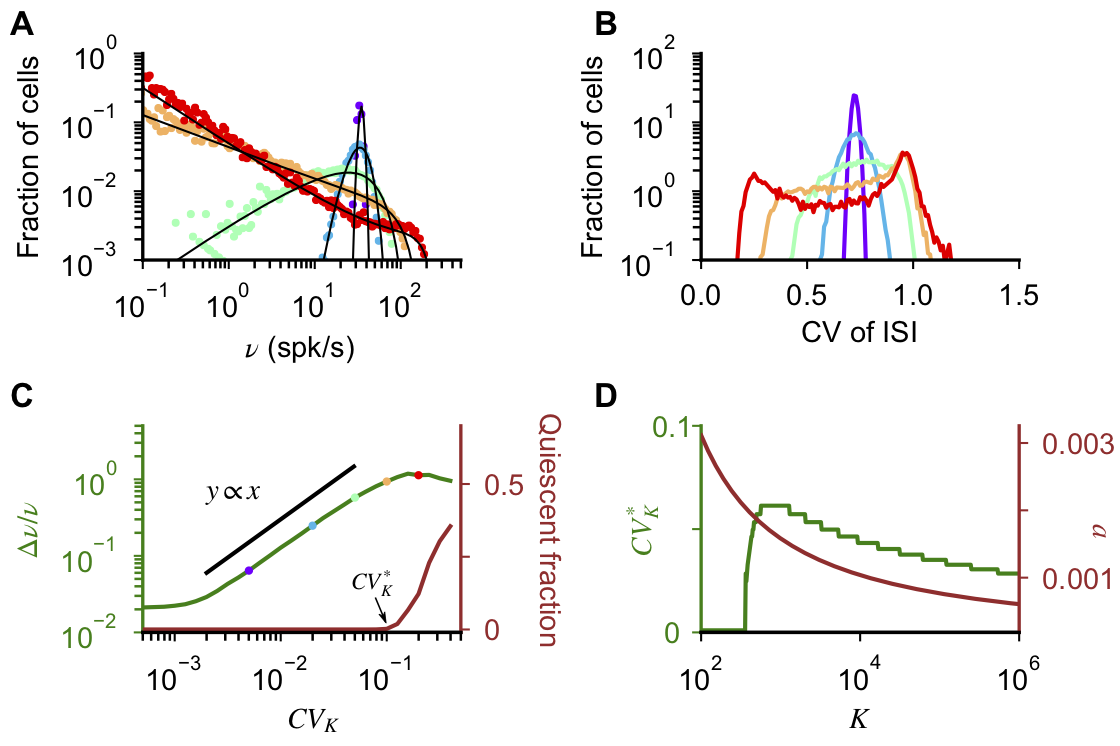}
\caption{{\bf Effects of heterogeneous connectivity on the network response.} 
({\bf A}-{\bf B}) 
Distribution of $\nu$ and  $CV$ of ISI computed from network simulations (dots) and from the mean field analysis ({\bf A}, black lines) for different values of $CV_K$ (values are indicated by dots in panel {\bf C}).
({\bf C}) $\Delta \nu/\nu$ (green, left axis) and 
fraction of quiescent cells (brown, right axis) computed from network simulations as a function of $CV_K$.
For $CV_K\lesssim CV_K^*$, $\Delta \nu/\nu$ increases linearly, as predicted by the mean field analysis; deviations from linear scaling emerge for $CV_K\gtrsim CV_K^*$, when a significant fraction of cells become quiescent.
The deviation from linear scaling at low $CV_K$ is due to sampling error in estimating the firing rate from simulations.
({\bf D}) $CV_K^*$ as a function of $K$ computed from the mean field theory (green, left axis), with $a$ rescaled according to Eq.~\eqref{eq:scaling_sub}. 
For large $K$, $CV_K^*$ decays proportionally to $a$ (brown, right axis). 
When $K$ is too low, the network is silent and $CV_K^*=0$.
In panels {\bf A}-{\bf C} $K=10^3$, $g=20$, $a=1.6 10^{-3}$, $N_E=N_X=N_I/\gamma =10 K$, $\nu_X=0.05/\tau_{rp}$.
In network simulations, the dynamics was run for 20 seconds using a time step of 50$\mu$s
Parameters in panel {\bf D} are as in Fig.~\ref{fig:compare_r_solutions}. }
\label{fig:simulations_variability}
\end{figure}

We investigated numerically the effects of connection heterogeneity as follows. 
We chose a Gaussian distribution of the number of connections per neuron, with mean $K$ and variance  $\Delta K^2$ for excitatory connections, and mean $\gamma K$ and variance $\gamma^2\Delta K^2$ for inhibitory connections. The connectivity matrix was constructed by drawing first randomly E and I in-degrees $K_{E,X,I}^i$ from these Gaussian distributions for each neuron, and then selecting at random $K_{E,X,I}^i$ E/I pre-synaptic neurons. We then simulated network dynamics and measured the distribution of rates and CV of the ISI in the population.
Results for different values of $CV_K=\Delta K/K$ are shown in  Fig.~\ref{fig:simulations_variability}A-C.
For small and moderate values of connection heterogeneity, increasing $CV_K$ broadens the distribution of rates and CV of the ISI, but both distributions remain peaked around a mean rate that is close to that  of homogeneous networks (Fig.~\ref{fig:simulations_variability}A,B).
For larger $CV_K$, on the other hand, the distribution of rates changes its shape, with a large fraction of neurons moving to very low rates, while others increase their rates  (Fig.~\ref{fig:simulations_variability}A) and the distribution of the CV of ISI becomes bimodal, with a peak at low CV corresponding to the high rate neurons, while the peak at a CV close to 1 corresponds to neurons with very low firing rates (Fig.~\ref{fig:simulations_variability}B).

To characterize more systematically the change in the distribution of rates with $CV_K$,  we measured, for each value of $CV_K$, the fraction of quiescent cells, defined as the number of cells that did not spike during 20 seconds of the simulated dynamics (Fig.~\ref{fig:simulations_variability}C). 
This analysis shows that the number of quiescent cells, and hence the distribution of rates, changes abruptly as the $CV_K$ is above a critical value $CV_K^*$. 
Importantly, unlike our definition of the fraction of quiescent cells, this abrupt change is a property of the network that is independent of the duration of the simulation.

To understand  these numerical results, we performed  a mean field analysis of the effects of connection heterogeneity on the distribution of rates (Appendix~\ref{SI:conn_variability}).
This analysis captures quantitatively numerical simulations  (Fig.~\ref{fig:simulations_variability}A) and shows that, in the limit of small $CV_K$  and $a$,  
rates in the network are given by 
\begin{equation}\label{eq:nu_variab}
\nu_i=\nu_0 \exp\left[   \Omega \, \frac{ CV_K }a  \,  z_i\right]    
\end{equation}
where $\nu_0$ is the population average in the absence of heterogeneity, $z_i$ is a Gaussian random variable, and the prefactor $\Omega$ is independent of $a$, $K$ and $\nu_X$.
The exponent in Eq.~\eqref{eq:nu_variab} represents a quenched disorder in the value of $v^i$, i.e. in the distance from threshold of the single cell $\mu^i$ in units of input noise.
As shown in Appendix~\ref{SI:conn_variability}, Eq.~\eqref{eq:nu_variab} implies that the distribution of rates is lognormal, a feature consistent with experimental observations~\cite{Hromadka2008,OConnor2010,Buszaki2014} and distributions of rates in networks of current-based LIF neurons \cite{Roxin16217}. It also implies that the variance of the  distribution ${\Delta \nu}/{\nu}$ should increase linearly with $CV_K$, a prediction which is confirmed by numerical simulations (Fig.~\ref{fig:simulations_variability}C).  
The derivation in  Appendix~\ref{SI:conn_variability} also provides an explanation for the change in the shape of the distribution for larger $CV_K$.
In fact, for larger heterogeneity, 
the small $CV_K$ approximation is not valid and fluctuations in input connectivity
produce cells for which  $\mu^i$ far from $\theta$, 
that are either firing at extremely low rate ($\mu^i<\theta$) or regularly ($\mu^i>\theta$).
The latter generates the peak at low values in the CV of the ISI seen for large values $CV_K$.

The quantity $CV_K^*$ represents the level of connection heterogeneity above which significant deviations from the asynchronous irregular state emerges, 
i.e. large  fractions of neurons show extremely low or regular firing.
Eq.~\eqref{eq:nu_variab} suggests that $CV_K^*$ should increase linearly with $a$.
We validated this prediction with our mean field model, by computing the minimal value of $CV_K$ at which 1\% of the cells fire at rate of $10^{-3}$ spk/s.
(Fig.~\ref{fig:simulations_variability}D).
Note that the derivation of  Eq.~\eqref{eq:nu_variab}  only assumes $a$ to be small and does not depend on the scaling relation between $a$ and $K$. 
On the other hand, the fact that $CV_K^*$ increases linearly with $a$ makes the state emerging in networks of conductance-based neurons with  $a\sim 1/\log(K)$  significantly more robust to connection fluctuations than that emerging with $a\sim K^{-\alpha}$, for which $CV_K^*\sim K^{-\alpha}$, and with current-based neurons, where $CV_K^*\sim 1/\sqrt{K}$~\cite{Landau}.
Note that, while in randomly connected networks $CV_K\sim 1/\sqrt{K}$, a larger degree of heterogeneity has been observed in cortical networks~\cite{daCosta2011,Furuta2011,Xue2014,Schoonover2014,Okun2015,Landau}. 
Our results show that networks of conductance-based neurons could potentially be much more robust to such heterogeneities than networks of current-based neurons.

\section{Comparison with experimental data}
\begin{figure*}[htb!]
\centering
\includegraphics[width=12.75cm]{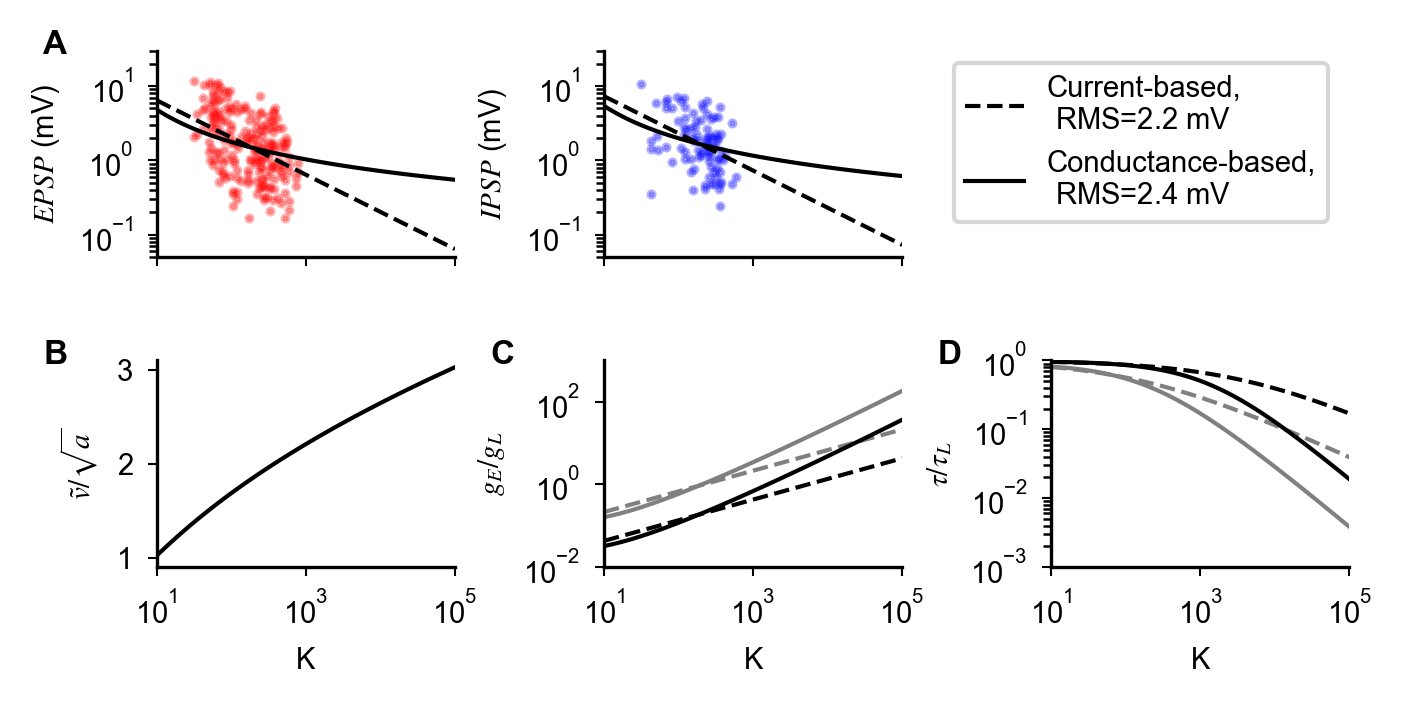}
\caption{\small{\textbf{Comparison of predictions given by  current-based and the conductance-based models in describing experimental data from cultures}.
{\bf A} Strength of excitatory  (EPSP)  and inhibitory  (IPSP) post synaptic potentials recorded in~\cite{Barral2016} are compared with best fits using scaling relationships derived from networks with current-based synapses (dashed line) and conductance-based synapses (continuous line). Root mean square (RMS) and best fit parameters are: RMS=$2.2$mV, $g=1.1$,  $J_0=20$mV, for the current-based model; and RMS=$2.4$mV, $g=3.4$, $\bar{v}=0.08$, for the conductance-based model. 
{\bf B} Value of $\bar{v}/\sqrt{a}$ predicted by the conductance-based model as a function of $K$. 
{\bf C} Ratio between excitatory and leak conductance as a function of $K$, for $\nu_E=\nu_I=\nu_X=1.$ spk/s (black) and $\nu_E=\nu_I=\nu_X=5$spk/s (gray) obtained with $a$ rescaled as Eq.~\eqref{eq:scaling_sub} (continuous line) and as $1/\sqrt{K}$ (dashed line).
{\bf D} Ratio between $\tau$ and $\tau_L$ as a function of $K$, parameters and scaling as in panel {\bf C}.
} 
}
\label{fig:fit_data}
\end{figure*}

The relation between synaptic efficacy and number of connections per neuron has been recently studied experimentally using a culture preparation~\cite{Barral2016}. 
In this paper, it was found that cultures in which $K$ was larger had weaker synapses than cultures with smaller $K$ (Fig.~\ref{fig:fit_data}).
In what follows we compare this data with the scalings expected in networks of current-based and conductance-based neurons, and discuss implications for \textit{in vivo} networks.

 In the current-based model, the strength of excitatory  and inhibitory  post synaptic potentials as a function of $K$ can be written as
$J_E=J_0/\sqrt{K}$ and $J_I=g\, J_E$. 
In the conductance-based model, these quantities become
$J_E=(V-E_E) a$ and $J_I=g(V-E_I) a$; where $a=a(K, \bar{v})$ is given by Eq.~\eqref{eq:scaling_sub} while, for the dataset of~\cite{Barral2016}, $V\sim -60$mV,  $J_E\sim J_I$,  $E_E\sim0$mV and $E_I\sim-80$mV. 
For each model, we inferred  free parameters from the data with a least-squares optimization in logarithmic scale (best fit: $g=1.1$ and $J_0=20$mV in the current-based model; $g=3.4$ and $\bar{v}=0.08$ in the conductance-based model) and computed the expected synaptic strength as a function of $K$ (lines in Fig.~\ref{fig:fit_data}A). 
Our analysis shows that the performance of the current-based and the conductance-based model in describing the data, over the range of $K$ explored in the experiment, are similar, with the former being slightly better than the latter (root mean square 2.2mV vs 2.4mV).
This result is consistent with the observation made in~\cite{Barral2016} that, when fitted with a power-law $J\sim K^{-\beta}$, data are best described by  $\beta=0.59$ but are compatible with a broad range of values (95\% confidence interval: [0.47:0.70]).
Note that even though both models give similar results for PSP amplitudes in the range of values of $K$ present in cultures ($\sim$50-1000), they give significantly different predictions for larger values of $K$. For instance, for $K=10,000$, $J_E$ is expected to be $\sim0.2$~mV in the current-based model and $\sim0.7$~mV in the conductance-based model.

In  Fig.~\ref{fig:fit_data}B, we plot the distance between the equilibrium membrane potential $\mu$ and threshold $\theta$ in units of input fluctuations, $\bar v/\sqrt{a}$ as a function of $K$ using  the value of $\bar{v}$  obtained above, and  find that the expected value in vivo, where $K\sim 10^3-10^4$, is in the range 2-3. 
In Fig.~\ref{fig:fit_data}C,D, we plot how total synaptic excitatory conductance, and effective membrane time constant, change as a function of $K$.
Both quantities change significantly faster using the conductance-based scaling ($g_E/g_L\sim K/\log(K)$; $\tau/\tau_L\sim \log(K)/K$) than what expected by the scaling of the current-based model  ($g_E/g_L\sim \sqrt{K}$; $\tau/\tau_L\sim 1/\sqrt{K}$).
For $K$ in the range $10^3-10^4$ and mean firing rates in the range 1-5 spk/s, the total synaptic conductance is found to be  in a range from about 2 to 50 times the leak conductance, while the effective membrane time constant is found to be smaller than the membrane time constant by a factor 2 to 50. We compare these values with available experimental data in the Discussion.

\section{Effect of finite synaptic time constants}
Results shown in Fig.~\ref{fig:fit_data} beg the question whether the assumption of negligible synaptic time constants we have made in our analysis is reasonable. In fact, synaptic decay time constants of experimentally recorded post-synaptic currents range from a few ms (for AMPA and GABA$_A$ receptor-mediated currents) to tens of ms (for GABA$_B$ and NMDA receptor-mediated currents, see e.g.~\cite{Destexhe1998}), i.e. they are comparable to the membrane time constant already at weak coupling, where $\tau\sim\tau_L$ is typically in the range 
10-30ms~\cite{McCormick1985,Inagaki2019}. 
In the strong coupling limit, the effective membrane time constant goes to zero, and so
this assumption clearly breaks down.
In this section, we investigate the range of validity of this assumption, and what happens once the assumption of negligible time constants is no longer valid. 

\begin{figure*}[htb!]
\centering
\includegraphics[width=13.4cm]{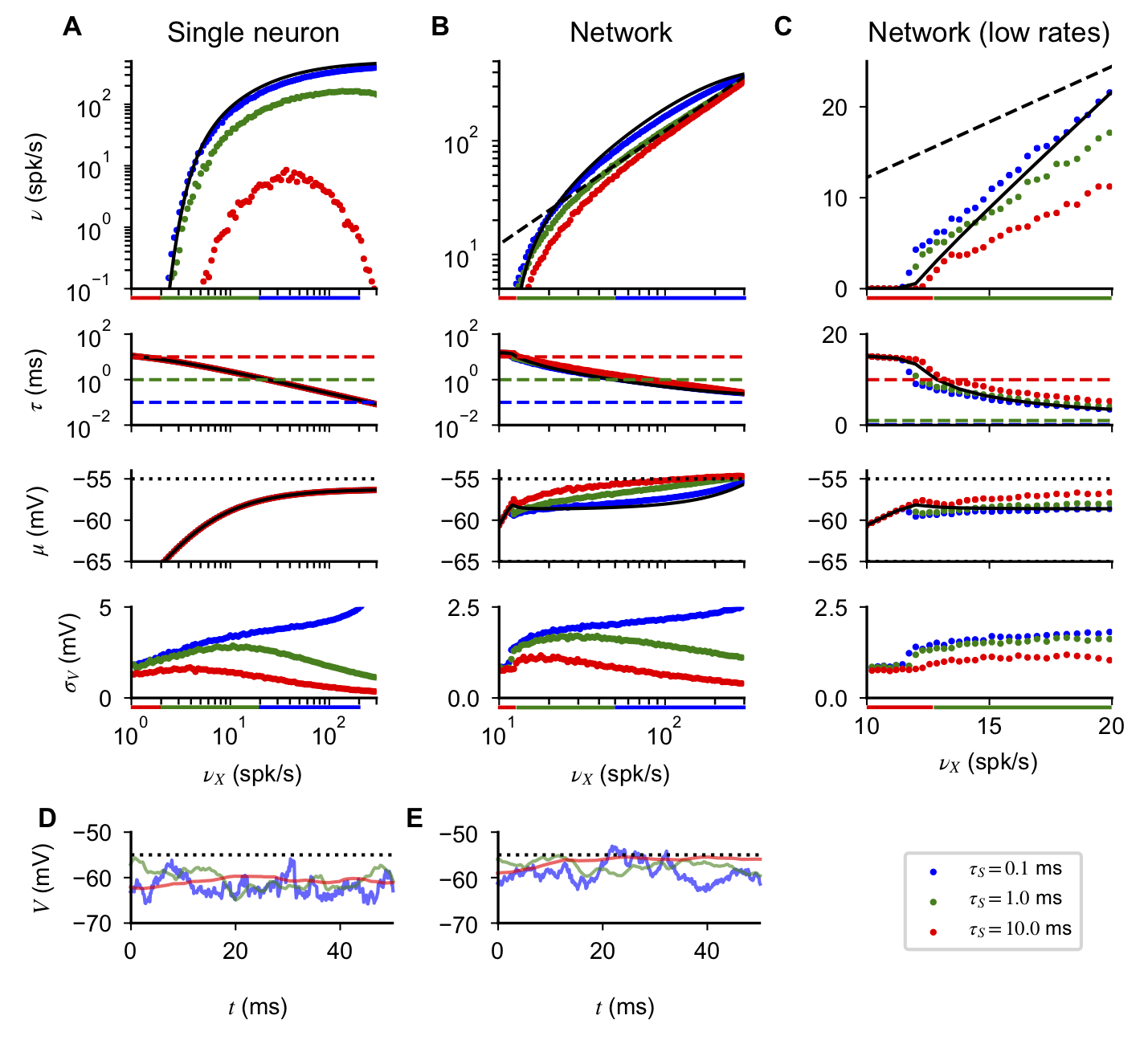}
\caption{\textbf{Effects of synaptic time constant on single neuron and network response.}
({\bf A}) Single neuron response as a function of input rate $\nu_X$, computed numerically from Eqs.~\eqref{eq:membrane},
\eqref{eq:g_dynamics_tau_S} for different synaptic time constants $\tau_S$ (indicated in the bottom right of the figure).
In all panels, black lines correspond to the prediction obtained with instantaneous synapses.
Colored bars below the first and the fourth row indicate  inputs that gives $0.1<\tau_S/\tau<1$, i.e. the region where deviations in the neural response from the case of instantaneous synapses emerge.
Firing rates (first row) match predictions obtained for instantaneous synapses for small $\tau_S/\tau$; significant deviations and response suppression emerge for larger $\tau_S/\tau$.
The effective membrane time constant ($\tau$, second row) decreases with input rate, is independent of $\tau_S$, and reaches the value $\tau_S/\tau\sim 1$ for different levels of external drive (dashed lines represent the different values of $\tau_S$).
The equilibrium value of the membrane potential ($\mu$, third row) increases with input rate and is  independent of  $\tau_S$ (dotted line represents spiking threshold).
The fluctuation of the membrane potential ($\sigma_{V}$, fourth row) has a non-monotonic relationship with input rate, and peaks at a value of $\nu_X$  for which $\tau$ is of the same order as $\tau_S$.
({\bf B}) Analogous to panel {\bf A} but in the network case.
Firing rates are no longer suppressed as $\tau_S/\tau$ increases, but approach the response scaling predicted by Eq.~\eqref{eq:response_large_tau_S} (dashed line). 
As discussed in the text, high firing rates are obtained by increasing the value of $\mu$ towards threshold. 
({\bf C}) Zoom of panel {\bf B} in the neurobiologically relevant region of low rates.
({\bf D}, {\bf E}) Examples of membrane potential dynamics for single neuron ({\bf D}) and network ({\bf E}) in the absence of spiking mechanisms ($\nu_X=5$spk/s in {\bf D} and $20$spk/s in {\bf E}). 
High frequency fluctuations are suppressed at large $\tau_S$. 
In the network case, increasing $\tau_S$ reduces recurrent inhibition and produces membrane potential trajectories which are increasingly closer to firing threshold.
Single neuron parameters: $a=0.01$, $K=10^3$, $g=8$, $\eta=1.5$, $\gamma=1/4$.
Network parameters: $a=0.0016$, $K=10^3$, $g=20$, $\gamma=1/4$.
Simulations were performed with the simulator BRIAN2~\cite{Stimberg2019}, with neurons receiving inputs from $N_{EX,IX}$= 10 $K$ independent Poisson units firing at rates $\nu_{X}, \eta \nu_{X}$, in the single neuron case, or $\nu_X$, in the network case . 
Network simulations used $N_{E,I}=10 K$ excitatory and inhibitory neurons. 
}
\label{fig:synaptic_time_constant}
\end{figure*}

With finite synaptic time constants, the temporal evolution of conductances of   Eq.~\eqref{eq:g_instant} is replaced by
\begin{equation}\label{eq:g_dynamics_tau_S}
\tau_{E,I} \frac{d {g}^j_{E,I}}{dt}=-{g}^j_{E,I}+{g^j_L} \tau_{E,I} \sum_m a_{jm} \sum_{n} \delta(t-t_m^n) \,  .
\end{equation}
It follows that the single-neuron membrane potential dynamics is described by Eqs.~(\ref{eq:membrane},\ref{eq:g_dynamics_tau_S}). Here, for simplicity, we take excitatory and inhibitory synaptic currents to have the same decay time constant $\tau_S$.
Fig.~\ref{fig:synaptic_time_constant}A shows how increasing the synaptic time constant modifies the mean firing rate of single integrate-and-fire neurons in response to  $K$ ($\gamma K$) excitatory (inhibitory) inputs with synaptic strength $a$ ($g a$) and frequency $\nu_X$ ($\eta \nu_X$). 
The figure shows that, though the mean firing rate is close to predictions obtained with instantaneous synapses for small $\tau_{S}/\tau$, deviations emerge for $\tau_S/\tau\sim1$, and firing is strongly suppressed as $\tau_S/\tau$ becomes larger. 
To understand these numerical results, we resort again to the diffusion approximation \cite{Brunel1998}, together with the effective time constant approximation \cite{Johannesma1968,Richardson2004} to derive a simplified expression of the single neuron membrane potential dynamics with finite synaptic time constant (details in Appendix~\ref{SI:finite_tau_S}); this is given by  
\begin{equation}\label{eq:syna_time_constant_main}
{\tau} \frac{dV}{dt}=-\left( V-\mu\right) +\sigma \sqrt{\frac{\tau}{\tau_S}}z \,   ,
\end{equation}
where $\tau$, $\mu$ and $\sigma$ are as in the case of negligible synaptic time constant (Eq.~\eqref{eq:param_MM}), whilst
$z$ is an Ornstein-Uhlenbeck process with correlation time $\tau_S$.
It follows that, with respect to Eq.~\eqref{eq:intro_V}, input fluctuations with frequency larger than $1/\tau_S$ are suppressed and, for large $\tau_S/\tau$, the membrane potential dynamics is given by 
\begin{equation}\label{eq:V_large_tau_S}
    V(t)=\mu+\sigma \sqrt{\frac{\tau}{\tau_S}}z(t) \, ,
\end{equation}
i.e.~the membrane potential is essentially slaved to a time dependent effective reversal potential corresponding to the r.h.s.~of Eq.~(\ref{eq:V_large_tau_S}) \cite{Shelley2002}. Note that Eq.~(\ref{eq:V_large_tau_S}) is valid only in the subthreshold regime. When the r.h.s.~of Eq.~(\ref{eq:V_large_tau_S}) exceeds the threshold, the neuron fires a burst of action potentials whose frequency, in the strong coupling limit, is close to the inverse of the refractory period~\cite{MorenoBote2004}.
As $\nu_X$ increases, the equilibrium value $\mu$ remains constant while $\tau$ decreases, leading to a suppression of membrane fluctuations (Fig.~\ref{fig:synaptic_time_constant}D), and in turn to the suppression of response observed in Fig.~\ref{fig:synaptic_time_constant}A.

In Appendix~\ref{SI:finite_tau_S}, we use existing analytical expansions~\cite{Brunel1998,MorenoBote2004,Badel2011}, as well as numerical simulations, to shows that neural responses obtained with finite $\tau_S$ are in good agreement with predictions obtained with instantaneous synapses as long as $\tau_S/\tau \lesssim0.1$.
It follows that the single neuron properties we discussed in the case of instantaneous synapses hold in the region of inputs for which $\tau_S/\tau \lesssim0.1$ (i.e. $\nu_X\lesssim0.1/a K\tau_S$) and the derivation of Eq.~\eqref{eq:scaling_sub} is valid (i.e. $\nu_X$ is at least of order $1/{aK \tau_L }$).
Thus, there is at best a narrow range of inputs for which these properties carry over to the finite synaptic constant case.
Interestingly, when biologically relevant parameters are considered (e.g. Fig.~\ref{fig:synaptic_time_constant}), inputs within this region generate firing rates that are in the experimentally observed range in cortical networks~\cite{Anderson2000,Crochet2006,Poulet2008,Tan2014,Yu2016,Li2020,Okun2015,Hromadka2008,OConnor2010,Buszaki2014}.
The analysis of Appendix~\ref{SI:finite_tau_S} also shows that, when $\tau_S/\tau\sim 1$, i.e. once the input rate $\nu_X$ is of order $1/aK\tau_S$, firing is suppressed exponentially.
The scaling relation of Eq.~\eqref{eq:scaling_sub} does not prevent this suppression,  which emerges for external rates of order $1/aK\tau_S\sim \log(K)/K \tau_S$.
Scalings of the form $a\sim K^{-\alpha}$, with $\alpha>0$, on the other hand, create a larger region of inputs for which $\tau_S/\tau \ll 1$ but, as we showed when studying the neural dynamics with instantaneous synapses, fail in generating response for large $K$.
We next asked if another scaling relation between $a$ and $K$ could prevent suppression of neural response  when $\tau_S$ is finite.
The single neuron response computed in  Appendix~\ref{SI:finite_tau_S} is a nonlinear function of the input $\nu_X$, which depends parametrically on $a$ and $K$.
It follows that, in order to preserve the single neuron response, $a$ should scale differently with $K$ for different values of $\nu_X$.
Since in cortical networks input rates, i.e. $\nu_X$, change dynamically on a time scale much shorter than that over which plasticity can modify synaptic connections, we conclude that a biologically realistic scaling between $a$ and $K$, which prevents suppression of neural response when $\tau_S$ is finite in a broad range of external inputs, does not exist.
Moreover, the membrane potential dynamics for large $K$ and $\tau_S/\tau$ (Eq.~\eqref{eq:V_large_tau_S}) becomes independent of $a$. This shows that rescaling synaptic efficacy with $K$ cannot prevent suppression of response. 

We next examined the effect of synaptic time constant on network response. 
Numerically computed responses in networks of neurons with finite synaptic time constant are shown in Fig.~\ref{fig:synaptic_time_constant}B,C.
Network response is close to the prediction obtained with instantaneous synapses for small $\tau_{S}/\tau$, and deviations emerge for $\tau_S/\tau\sim1$. 
Hence, analogously to the single neuron case, the network properties discussed in the case of instantaneous synapses remain valid for low inputs.
However, unlike the single neuron case, no suppression appears for larger $\tau_S/\tau$.
This lack of suppression in the network response, analogously to the one we discussed in networks with instantaneous synapses and $a\sim K^{-\alpha}$, is a consequence of the fact that, to have stable dynamics when $K$ is large, inhibition must dominate recurrent interactions~\cite{Amit1997}.
In this regime, any change which would produce suppression of single neuron response (e.g. increase of $\nu_X$ or $\tau_S$) lowers recurrent inhibition and increases the equilibrium value of the membrane potential $\mu$ (Fig.~\ref{fig:synaptic_time_constant}B,C,E).
The balance between these two effects determines the network firing rate and, when $\tau_S/\tau\gg 1$, generates a response which (see derivation in Appendix~\ref{SI:finite_tau_S}), up to corrections of order $1/\sqrt{K}\tau_S$, is given by Eq.~\eqref{eq:response_large_tau_S} (dashed line in Fig.~\ref{fig:synaptic_time_constant}B).
Similarly to what happens in networks with instantaneous synapses and $a\sim K^{-\alpha}$, this finite response emerges because recurrent interactions set $\mu$ very close to threshold, at a distance $\theta-\mu\sim 1/\sqrt{K}$ that matches the size of the membrane potential fluctuations (Eq.~\eqref{eq:V_large_tau_S}, $\sigma\sqrt{\tau/\tau_S}\sim 1/\sqrt{K}$). 
In addition, the network becomes much more sensitive to connection heterogeneity, with $CV_K^*\sim 1/\sqrt{K}$.
However, here the dynamics of the single neuron membrane potential is correlated over a timescale $\tau_S$ (Fig.~\ref{fig:synaptic_time_constant}E) and firing is bursty, with periods of regular spiking randomly interspersed in time. 
Moreover, the properties discussed here are independent of the scaling of $a$ with $K$, since they always emerge once $\tau_S/\tau\gg1$, a condition that is met for any scaling once  $\nu_X\gg 1/aK\tau_S$.
The specific scaling relation, on the other hand, is important to determine the input strength at which $\tau_S/\tau\sim1$. 
\begin{table*}[htb!]
\centering
\begin{tabular}{|*{6}{c|}} 
\hline 
\multicolumn{1}{|c|}{Synaptic model}&  \makecell{Ratio of synaptic\\and membrane\\ time constant (${\tau_S}/{\tau}$)} &  \makecell{Synaptic \\ strength} & \makecell{Membrane potential\\ statistics} & \makecell{Activity\\structure}  & \makecell{Heterogeneity\\of in-degree\\  supported ($CV_K^*$)}\\ 
\hline 
\hline 
\multicolumn{1}{|c|}{\makecell{Current-based\\ (balanced state model)}}&  \makecell{constant,\\independent of\\$\nu_X$, $a$, and $K$} &$J\sim\frac1{\sqrt{K}}$ &
\makecell{$\theta-\mu\sim\sigma_V\sim 1$;\\
$\tau_V\sim\tau_L$} & \makecell{Irregular firing,\\ CV of ISI$\sim1$}& $\sim\frac1{\sqrt{K}}$\\ 
\hline 
\multirow{2}{*}{\makecell{Conductance-based}} & \multirow{2}{*}{\makecell{$\ll1$ for $\nu_X\ll\frac{1}{aK \tau_S}$;\\
always satisfied\\for instantaneous\\synapses ($\tau_S=0$)}} 
                        & $a\sim\frac1{\log{K}}$ & \makecell{$\theta-\mu\sim1$;\\ $\sigma_V\sim \frac1{\sqrt{\log{K}}}$;\\
$\tau_V\sim \log(K)/K$} & \makecell{Irregular firing,\\ CV of ISI$\sim1$}& $\sim\frac1{\log{K}}$ \\
                       \cline{3-6}
&                      & \makecell{$a\sim{K}^{-\alpha}$,\\$\alpha>0$} &
\makecell{$\theta-\mu\sim\sigma_V\sim K^{\frac{-\alpha}{2}}$;\\
$\tau_V\sim K^{\alpha-1}$}
& \makecell{Irregular firing,\\ CV of ISI$\sim1$}& $\sim{K^{-\alpha}}$ \\ 
\cline{2-6}
& \multirow{1}{*}{
\makecell{$\gg1$ for $\nu_X\gg\frac{1}{aK \tau_S}$}
}  & \makecell{any scaling} &
\makecell{$\theta-\mu\sim\sigma_V\sim \frac1{\sqrt{K}}$;\\
$\tau_V\sim \tau_S$} & \makecell{Irregular bursting}
 & $\sim \frac1{\sqrt{K}}$ \\ 
\hline 
\end{tabular}
  \caption{\textbf{Overview of of networks of current-based and conductance-based neurons.} 
  Synaptic strength and time constant strongly affect response properties in networks of conductance based neurons. Properties similar to what is observed in cortex emerge in these networks if $a\sim1/\log{K}$ and input rates are lower than a critical value, which is fixed by synaptic time constant and coupling strength. The model predicts that these properties should gradually mutate as the input to the network increases and, for large inputs, should coincide with those indicated in the last line of the table. In the table, the different quantities related to the membrane potential represent:  the mean distance from threshold ($\theta-\mu$); the size of temporal fluctuations ($\sigma_V$); the membrane potential correlation time constant ($\tau_V$).
  } 
\label{tab:summary}
\end{table*}

In the previous sections, we have shown that networks of conductance-based neurons with instantaneous synapses present features similar to those observed in cortex if synaptic efficacy is of order $a\sim 1/\log(K)$, while other scalings generate network properties that are at odds with experimental data (see Tab.~\ref{tab:summary} for a summary). 
In this section, we have found that, when the synaptic time constant is considered, these properties are preserved in the model for low inputs. As the input increases, the structure of the network response evolves gradually and, for large inputs ($\nu_X\gg1/aK\tau_S$), significant deviations from the case of instantaneous synapses emerge (see Tab.~\ref{tab:summary} for a summary). 
In particular, as the input to the network increases, our analysis shows that: the membrane potential approaches threshold while  its fluctuations become smaller and temporally correlated;  firing  becomes more bursty; the network becomes more sensitive to heterogeneity in the in-degree and, if the heterogeneity is larger than that of random networks, significant fractions of neurons become quiescent or fire regularly. 
These features of the model provide a list of predictions which could be tested experimentally by measuring the evolution of the membrane potential dynamics of cells in cortex with the intensity of inputs to the network.

\section{Discussion}
In this work, we analyzed networks of strongly coupled conductance-based neurons. 
The study of this regime  is motivated by the experimental observation that in cortex $K$ is large, with single neurons receiving inputs from hundreds or thousands of pre-synaptic cells. 
We showed that the classical balanced state idea~\cite{shadlen94,shadlen98}, which was developed in the context of current-based models and features synaptic strength of order $1/\sqrt{K}$~\cite{vanvreeswijk96b,vanvreeswijk98}, results in current fluctuations of very small amplitude, which can generate firing in networks only if the mean membrane potential is extremely close to threshold. This seems problematic since intracellular recordings in cortex show large membrane potential fluctuations (see e.g.~\cite{Anderson2000,Crochet2006,Poulet2008,Tan2014,Okun2015,Yu2016}).
To overcome this problem, we introduced a new scaling relation which, in the case of instantaneous synaptic currents, maintains firing by preserving the balance of input drift and diffusion at the single neuron level.
Assuming this scaling, the network response automatically shows multiple features that are observed in cortex in vivo: irregular firing, wide distribution of rates, membrane potential with non-negligible distance from threshold and fluctuation size. 
When  finite synaptic time constants are included in the model, we showed that these properties are preserved for low inputs, but are gradually modified as inputs increase: the membrane mean  approaches threshold while its fluctuations decrease in size and develop non-negligible temporal correlations. 
These properties, which are summarized in Tab.~\ref{tab:summary}, provide a list of predictions that could be tested experimentally by analyzing the membrane potential dynamics as a function of input strength in cortical neurons.

\medskip
When synaptic time constants are negligible with respect to the membrane time constant,  our theory shows properties that are analogous to those of the classical balanced state model:  linear transfer function, CV of order one, and distribution of membrane potentials with finite width.
However, these properties emerge from a different underlying dynamics than in the current based model.
In current-based models, the mean input current is at distance of order one from threshold in units of input fluctuations.
In conductance-based models, this distance increases with coupling strength and firing is generated by large fluctuations at strong coupling.
The different operating mechanism manifests itself in two ways: the strength of synapses needed to sustain firing and the robustness to connection heterogeneity, as we discuss in the next paragraphs.

The scaling relation determines how strong synapses should be to allow firing at a given firing rate, for a given a value of $K$. 
In current-based neurons, irregular firing is produced as long as synaptic strengths are of order $1/\sqrt{K}$.
In conductance-based neurons,  stronger synapses are needed, with a scaling which approaches  $1/\log(K)$ for large $K$. 
We showed that both scaling relations are in agreement with data obtained from culture preparations~\cite{Barral2016}, which are limited to relatively small networks, and argued that differences might be important \textit{in vivo}, where $K$ should be larger.

In current-based models, the mean input current must be set at an appropriate level  to produce irregular firing; this constraint is realized by recurrent dynamics in networks with random connectivity and strong enough inhibition~\cite{Amit1997,vanvreeswijk96b,vanvreeswijk98}. 
However, in networks with structural heterogeneity, with connection heterogeneity larger than $1/\sqrt{K}$, the variability in mean input currents produces significant departures from the asynchronous irregular state, with large fractions of neurons that become silent or fire regularly~\cite{Landau}. 
This problem is relevant  in cortical networks~\cite{Landau}, where significant heterogeneity of in-degrees as been reported~\cite{daCosta2011,Furuta2011,Xue2014,Schoonover2014,Okun2015}, and different mechanisms have been proposed to solve it~\cite{Landau}.
Here we showed that networks of conductance-based neurons also generate irregular activity without any need for finite tuning, and furthermore can support irregular activity with substantial structural heterogeneity, up to order $1/\log (K)$.
Therefore, these networks are more robust to connection heterogeneity than the current-based model, and do not need the introduction of additional mechanism to sustain the asynchronous irregular state.

The strength of coupling in a network, both in the current-based model~\cite{Ahmadian2013,Sanzeni856831} and in the conductance-based model (e.g. Fig.~\ref{fig:compare_r_solutions}) determines the structure of its response and hence the computations it can implement. 
Recent theoretical work, analyzing experimental data in the framework of current-based models, has suggested that cortex operates in a regime of moderate coupling~\cite{Ahmadian2019,SanzeniISN}, where response nonlinearities are prominent. 
In conductance-based models, the effective membrane time constant can be informative on the strength of coupling in a network, as it decreases  with coupling strength. 
Results from  \textit{in vivo} recordings in cat parietal cortex~\cite{Destexhe1999} showed evidence that single neuron response is sped up by network interactions. 
In particular, measurements are compatible with inhibitory conductance approximately 3 times larger than leak conductance and support the idea that cortex operates in a ``high-conductance state''~\cite{Destexhe2003} (but see \cite{Li2020} and discussion below). 
This limited increase in conductance supports the idea  of moderate coupling in cortical networks, in agreement with what found in previous work~\cite{Ahmadian2019,SanzeniISN}.

When the synaptic time constant is much larger than the membrane time constant,  we showed that, regardless of synaptic strength, the size of membrane potential fluctuations decreases and firing in the network is preserved by a reduction of the distance from threshold of the mean membrane potential. 
Moreover, the robustness to heterogeneity in connection fluctuations decreases substantially (the maximum supported heterogeneity becomes of order $1/\sqrt{K}$) and the membrane potential dynamics becomes correlated over a time scale fixed by the synaptic time constant.
For really strong coupling, the regime of large synaptic time constant is reached for low input rates. 
In the case of moderate coupling, which is consistent with experimental data on cortical networks~\cite{SanzeniISN,Ahmadian2019}, the network response  at low rates is well approximated by that of networks with instantaneous synapses, and 
the regime of large synaptic time constant is reached gradually, as the input to the network increases (Fig.~\ref{fig:synaptic_time_constant}). 
This observation provides a list of prediction on how properties of cortical networks should evolve with input strength (summary in Tab.~\ref{tab:summary}), that are testable experimentally.

Experimental evidence suggests that the response to multiple inputs in cortex is non-linear (for an overview, see~\cite{Rubin2015}). 
Such nonlinearities, which are thought to be fundamental to perform complex computations, cannot be captured by the classical balanced state model, as it features a linear transfer function~\cite{vanvreeswijk96b,vanvreeswijk98}. 
Alternative mechanisms have been proposed~\cite{Ahmadian2013,Rubin2015,Baker2019}, but their biophysical foundations~\cite{Ahmadian2013,Rubin2015} or their ability to capture experimentally observed nonlinearities~\cite{Baker2019} are still not fully understood.
We have recently shown~\cite{Sanzeni856831} that, in networks of current-based spiking neurons,  nonlinearities compatible with those used in~\cite{Ahmadian2013, Rubin2015} to explain  phenomenology of inputs summation in cortex emerge at moderate coupling.
Here we have shown that, as in the case of networks of current-base neurons~\cite{Sanzeni856831}, nonlinear responses appears in networks of conductance-based neurons at moderate coupling,  both at response onset and close to single neuron saturation.
In addition, we have found that synaptic time constants provide an additional source of nonlinearity, with nonlinear responses emerging as the network transitions between the two regimes described above.
A full classification of the nonlinearities generated in these networks is outside the scope of this work, but  could be performed generalizing the approach developed in~\cite{Sanzeni856831}.

Recent whole cell recording have reported that an intrinsic voltage-gated conductance, whose strength decreases with membrane potential, contributes to the modulation of neuronal conductance of cells in primary visual cortex of awake macaques and anesthetized mice~\cite{Li2020}. 
For spontaneous activity, this intrinsic conductance is the dominant contribution to the cell conductance and drives its (unexpected) decrease with increased depolarization. 
For activity driven by sensory stimuli, on the other hand, modulations coming from synaptic interactions overcome the effect of the intrinsic conductance and 
neuronal conductance increases with increased depolarization.  
Our analysis shows that voltage-dependent currents, such as that produced by the voltage-gated channels~\cite{Li2020} or during spike generation~\cite{Fourcaud2003}, affect quantitatively, but not qualitatively, the single neuron response and the scaling relation allowing firing. 
Therefore, the results we described in this contribution seem to be a general property of networks of strongly coupled  integrate-and-fire neurons with conductance-based synapses.

Understanding the dynamical regime of operation of the cortex is an important open question in neuroscience, as it constrains which computations can be performed by a network~\cite{Ahmadian2013}.
Most of the theories of neural networks have been derived using rate models or current-based spiking neurons.
Our work provides the first theory of the dynamics of strongly coupled conductance-based neurons, it can be easily related to measurable quantities because of its biological details,  and suggests   predictions that could be  tested experimentally.

\section*{Acknowledgments} 
Supported by the NIMH Intramural Research Program and by NIH BRAIN U01 NS108683 (to M.H. and N.B.). We thank Alex Reyes for providing his data, and Vincent Hakim and Magnus Richardson for comments on the manuscript. This work used the computational resources of the NIH HPC Biowulf cluster (http://hpc.nih.gov) and the Duke Compute Cluster (https://rc.duke.edu/dcc).

\bibliographystyle{unsrt}
\bibliography{mybib}

\onecolumngrid


\appendix


\clearpage
\section{Calculations in the multiplicative noise case}\label{SI:mean_field_response}

In the main text, we analyze the distribution of membrane potential, firing rate and CV using the effective time constant approximation, which neglects the dependence of the noise amplitude on the membrane potential. 
This approximation is motivated by the fact that corrections to this approximation are of the same order of shot noise corrections to the diffusion approximation used to describe synaptic inputs~\cite{Richardson2007}.
In this section, we derive results without resorting to the effective time constant approximation (i.e.~keeping the voltage dependence of the noise term), and  show that the results derived in the main text remain valid, even though it complicates the  calculations.
The inclusion of shot noise corrections is outside the scope of this contribution.

\subsection{Equations for arbitrary drift and diffusion terms}

In this section, we compute the probability distribution of the membrane potential,  the firing rate, and the CV of ISI of a neuron whose membrane potential follows the equation 
\begin{equation}\label{eq:general_equation}
\begin{split}
\frac{d{V}}{dt}={A}({V})+{B}(V) \bf{\zeta}\, .
\end{split}
\end{equation}
Eq.~\eqref{eq:intro_V} of the main text is a special form of Eq.~\eqref{eq:general_equation} with 
\begin{equation}\label{eq:A_B_dyna}
\begin{split}
{A}({V})=\frac{\mu-V}{\tau} \, ,\quad
{B}(V)=\frac{\sigma(V)}{\sqrt{\tau}} \, .
\end{split}
\end{equation}

The Fokker-Plank equation associated to  Eq.~\eqref{eq:general_equation}, in the Stratonovich regularization scheme, is given by 
\begin{equation*}\label{eq:fokker_plank}
\begin{split}
\frac{d P}{d t}=- \frac{\partial J}{\partial V} \, , 
\end{split}
\end{equation*}
where $P$ is the probability of finding a neuron with membrane potential $V$ and $J$ is the corresponding probability current given by
\begin{equation}\label{eq:current}
\begin{split}
J=\left(A +\frac 12  B\frac{ \partial B  }{\partial V}\right) P-\frac 12  \frac{ \partial B^2 P  }{\partial V}\, .
\end{split}
\end{equation}

We are interested in the stationary behavior of the system in which $P$ does not depend on time and the current $J$ is piecewise constant. In particular,  for $V$ between the activation threshold $\theta$ and the resting potential $V_r$, $J$
 is equal to the neuron firing rate  $\nu$ and the normalization condition implies 
\begin{equation*}
\int_{V_r}^{\theta} P(V) dV +\nu \,  \tau_{rp}=1 \, ,
\end{equation*}
where $\tau_{rp}$ is the refractory period.

To derive the probability distribution of the neuron potential, we introduce in Eq.~\eqref{eq:current} the integrating factor
\begin{equation*}
W(V)=\exp\left[-2\int^V du\frac{A(u)+\frac12B(u)\frac{ \partial B(u)  }{\partial u}}{B(u)^2}\right] \, 
\end{equation*}
and obtain 
\begin{equation*}
-2 \nu W (V) \theta(V-V_r) =\frac{ \partial }{\partial V} \Bigg[ W (V)B(V)^2 P(V)\Bigg] \, .
\end{equation*}
Using the boundary condition $P(\theta)=0$, we find
\begin{equation}\label{eq:solution_P_of_V}
P(V)=\frac{2 \nu}{W (V) B(V)^2 } \int^{\theta}_{V} du  W (u)\theta(u-V_r)
\end{equation}
and 
\begin{equation*}
\frac 1{\nu}=\tau_{rp}+2\int^{\theta}_{-\infty}dx\frac{1}{W(x) B(x)^2} \int^{\theta}_{x} du  W(u)\theta(u-V_r)
\end{equation*}
Integrating by parts, we obtain
\begin{equation}\label{eq:general_solution}
\frac 1{\nu}=\tau_{rp}+2\int^{\theta}_{V_r}dvW(v)\int^{v}_{-\infty} dx \frac{1}{W(x)B(x)^2}   
\end{equation}
This solution has been obtained in general form in~\cite{Siegert1951} and for the specific form of Eq.~\eqref{eq:A_B_dyna} in~\cite{Richardson2004}.

We now compute the coefficient of variation of the interspike interval.
The moments $T_k$ of the interspike intervals of the stochastic process defined by Eq.~\eqref{eq:general_equation} are given by (see~\cite{GardinerBook09})
\begin{equation*}
\frac{B(x)^2}2 \frac{d^2 T_k(x)}{d x^2}+\left(A(x) +\frac 12  B(x)\frac{ \partial B(x)  }{\partial x}\right) \frac{d T_k(x)}{d x}=-k T_{k-1}(x)
\end{equation*} 
with boundary conditions 
\begin{equation*}
T_k(\theta)=0 \, , \quad \frac{d T_k(b)}{d x}=0 \, ,
\end{equation*}
i.e. $\theta$ is an absorbing boundary and $b$ is a reflective boundary (we will then consider the limit $b\rightarrow-\infty$). 
The general solution of an equation of the form 
\begin{equation}\label{eq:general_form}
\frac{d^2 f(x)}{d x^2}+P(x)\frac{d f(x)}{d x}=Q(x)
\end{equation} 
is 
\begin{equation*}
f(x)=\int_{\theta}^xdt\int_{-\infty}^t dz  Q(z) \exp\left( \int_t^z dwP(w)  \right) \, .
\end{equation*} 
For $T_1(x)$ we have
\begin{equation*}
P(x)=\frac{2 A(x)+B(x)\frac{ \partial B(x)  }{\partial x}}{B(x)^2}\, , \quad
Q(x)=-\frac2{B(x)^2}
\end{equation*}
For $T_2(x)$ we look for a solution of the form
\begin{equation*}
T_2(x)=T_1(x)^2+R(x)
\end{equation*}
and find that $R$ obeys to an equation of the form of Eq.~\eqref{eq:general_form} with 
\begin{equation*}
P(x)=\frac{2 A(x)+B(x)\frac{ \partial B(x)  }{\partial x}}{B(x)^2}\, , \quad
Q(x)=-2\left( \frac{d T_1(x)}{d x} \right)^2
\end{equation*}
Combining the previous results, the CV of ISI is obtained as
\begin{equation}\label{eq:solution_CV}
CV^2=\frac{R(x)}{T_1(x)^2}\, ;
\end{equation}
the explicit expression of the $CV$ is given in the following section.

Eqs.~\eqref{eq:P_of_V}, \eqref{eq:nu} and \eqref{eq:CV} of the main text  have been obtained from Eqs.~\eqref{eq:solution_P_of_V}, \eqref{eq:general_solution} and \eqref{eq:solution_CV}  using Eq.~\eqref{eq:A_B_dyna}. 

\subsection{Equations for conductance-based LIF neurons}
\label{SI:model_def_no_approx}

\begin{figure*}[!h]
\includegraphics[width=15cm]{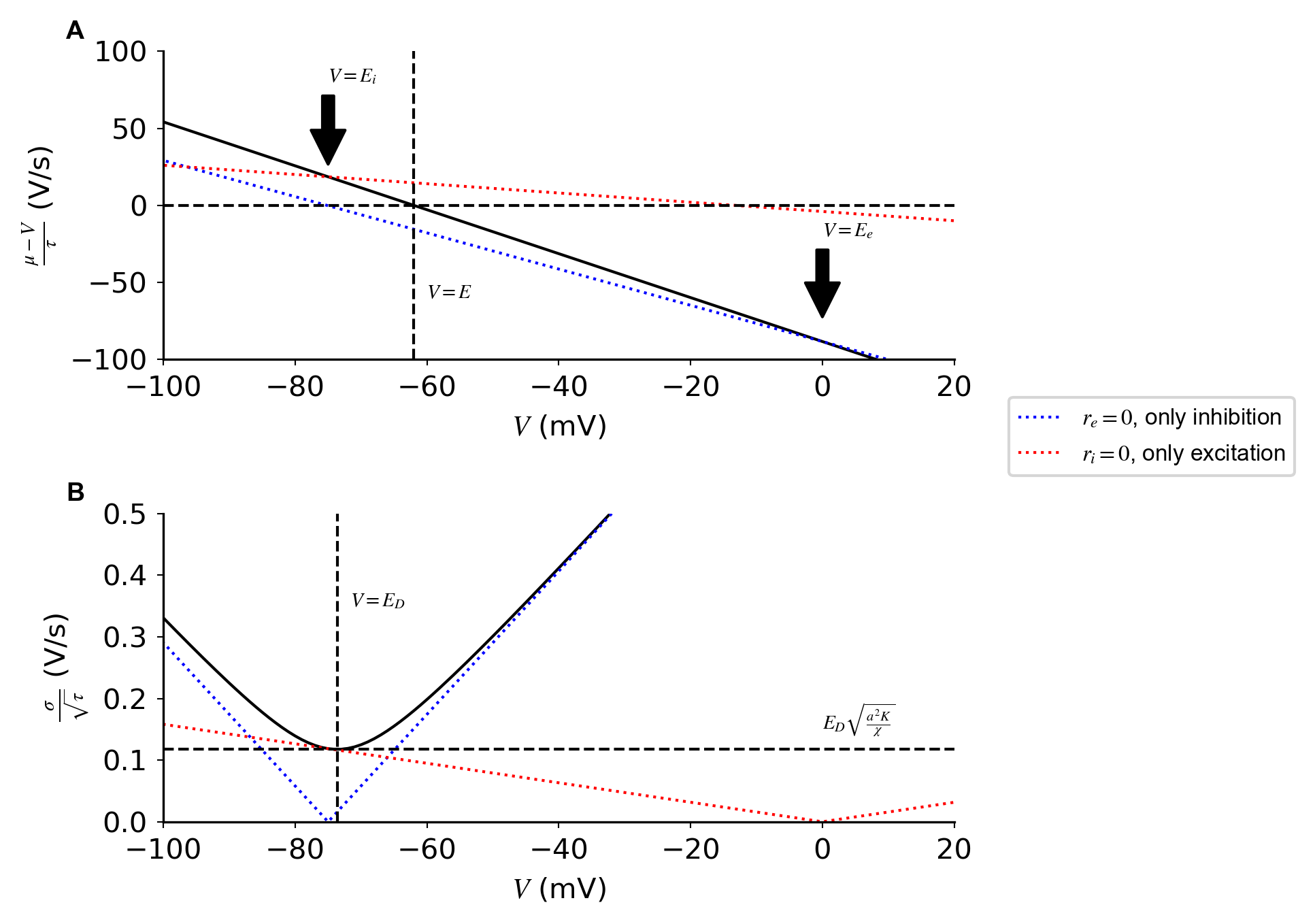}
\caption{{\bf Drift and diffusion terms of Eq.~\eqref{eq:intro_V} as a function of voltage.}
({\bf A}) Input drift as a function of membrane potential $V$ produced with both inhibitory and excitatory inputs (black line), excitatory inputs only (red dotted line), or inhibitory inputs only (blue dotted line).
The drift term decreases monotonically with  $V$ and it is zero at  $V=\mu$, which is a stable fixed point of the deterministic dynamics.
({\bf B}) The noise variance is quadratic in $V$.  Its minimum at $V=\mathcal{E_S}$  is equal to $\mathcal{E_D}\sqrt{a^2K/\chi}$. Note that the minimum amplitudes of  drift and  variance are obtained at different values of $V$.
}
\label{SIfig:Interpretations}
\end{figure*}

\medskip
Starting from Eqs.~(\ref{eq:intro_V},\ref{eq:param_MM}) of the main text, we write the different terms as
\begin{align}\label{SIeq:terms}
\begin{split}
\tau^{-1}={\tau_L}^{-1}+ a K \, {\omega}^{-1}\, , \quad
\mu=\tau\{E_L/\tau_L+a K [  r_E E_E +r_I g \gamma E_I]\}\, , \\
\sigma^2=  a^2K \frac{\tau}{\chi} \left[ \left( V-\mathcal{E_S}\right)^2 +\mathcal{E_D}^2\right] \, ,
\end{split}\end{align}
where, to  shorten the expressions, we have introduced two auxiliary variables with time dimension
\begin{align}
\begin{split}
\omega^{-1}={r_E  +r_I g \gamma} \, ,\quad
\chi^{-1}={r_E  +r_I g^2 \gamma} \, ,
\end{split}\end{align}
as well as two variables with voltage dimensions,
\begin{align}
\begin{split}
\mathcal{E_S}=\chi \left(r_E E_E +r_I g^2 \gamma E_I\right) \, ,\quad  
\mathcal{E_D}=\chi \sqrt{r_E r_I  g^2\gamma }\left( E_E-E_I\right)\, .
\end{split}\end{align} 
The terms $-\left( V-\mu\right)/\tau$ and $\sigma(V) \zeta /\sqrt{\tau}$ of Eq.~\eqref{eq:intro_V} represent the input drift and noise to the membrane dynamics respectively. The voltage dependence of these terms is sketched in Fig.~\ref{SIfig:Interpretations}. 

In the large $K$ limit, the different terms in Eq.~\eqref{eq:intro_V} scale as
\begin{equation}\label{eq:large_K_terms}
\tau\sim \frac{ \omega}{aK}\, , \quad \mu\sim \omega \left(r_E E_E +r_Ig \gamma E_I \right)\, , \quad \sigma\sqrt{\tau}\sim \frac{\omega}{\sqrt{\chi K}} \sqrt{ \left( V-\mathcal{E_S}\right)^2 +\mathcal{E_D}^2} \, ;
\end{equation} 
while the values of  $\omega$, $\mu$, $\mathcal{E_S}$, and $\mathcal{E_D}$ are independent of $K$. 
It follows that the noise term $\sigma\sqrt{\tau}$ and the time constant $\tau$ in Eq.~\eqref{eq:intro_V} become small in the strong coupling limit. This result is analogous to what we obtained in the main text with the effective time constant approximation, since this approximation does not change how these terms scale with $a$ and $K$.

We now insert the drift and diffusion terms of the conductance-based LIF neuron in
Eqs.~\eqref{eq:solution_P_of_V}, \eqref{eq:general_solution}, and \eqref{eq:solution_CV}, and obtain
\begin{equation}\label{SIeq:P_of_V}
P(V)=\frac{2  \nu \chi \mathcal{E_D}  e^{\frac{ -\mathcal{F}\left(V\right)}{a}} }{a^2K  \left[\left( V-\mathcal{E_S}\right)^2+\mathcal{E_D}^2\right]}
\int_{u(V)}^{v_{max}}dx \theta(x-u(V_r)) e^{\frac{ \mathcal{F}\left(x\right)}{a}} \, ,  
\end{equation}
\begin{equation}\label{SIeq:nu}
\frac1{\nu}=\tau_{rp}+ \frac{2 \chi}{a^2K}  \int_{v_{min}}^{v_{max}}dv \int_{-\infty}^{v}dx\frac{1}{x^2+1} \, \exp\left[\frac{\mathcal{F}\left(v\right)-\mathcal{F}\left(x\right)}{a} \right]\, ,
\end{equation}
and 
\begin{equation}\label{SIeq:CV}
CV^2=\frac{8\chi^2 \nu^2}{a^4K^2} \int_{v_{min}}^{v_{max}} dv \int_{-\infty}^v dz \exp\left[\frac{\mathcal{F}(v)-\mathcal{F}(z)}a \right] \Bigg\{\int_{-\infty}^zdw \frac1{w^2+1} \exp\left[\frac{\mathcal{F}(z)- \mathcal{F}(w)}a \right] \Bigg\}^2
\end{equation}
where
\begin{equation}\label{SIeq:terms_model_A_MFT}
\begin{split}
\mathcal{F}(x)=\frac{2 \chi }{a K \tau } \left[\frac12 \left( 1-\frac{a^2K\tau}{2\chi}\right)\log(x^2+1)-\alpha\arctan(x)\right] \, , \quad  u({V})=\frac{V-\mathcal{E_S}}{\mathcal{E_D}} \, , \\
v_{min}=u({V_r})\, , \quad
v_{max}=u(\theta) \, , \quad
\alpha=u(\mu) \, .
\end{split}
\end{equation}
Eqs.~\eqref{SIeq:P_of_V} and \eqref{SIeq:nu}  are analogous to those derived in~\cite{Richardson2004}. 
To simplify the following analysis, we will neglect the contribution of the term ${a^2K\tau}/{2\chi}$, which derives from the regularization scheme.
This assumption is justified by the fact that, for large $K$, $\tau\sim1/aK$ and the factor ${a^2K\tau}/{2\chi}$ is of order $a\ll1$.

\section{Calculations in the strong coupling regime - Single neurons}\label{SI:single_neuron_response_at_strong_coupling}
\begin{figure*}[htb!]
\centering
\includegraphics[width=15cm]{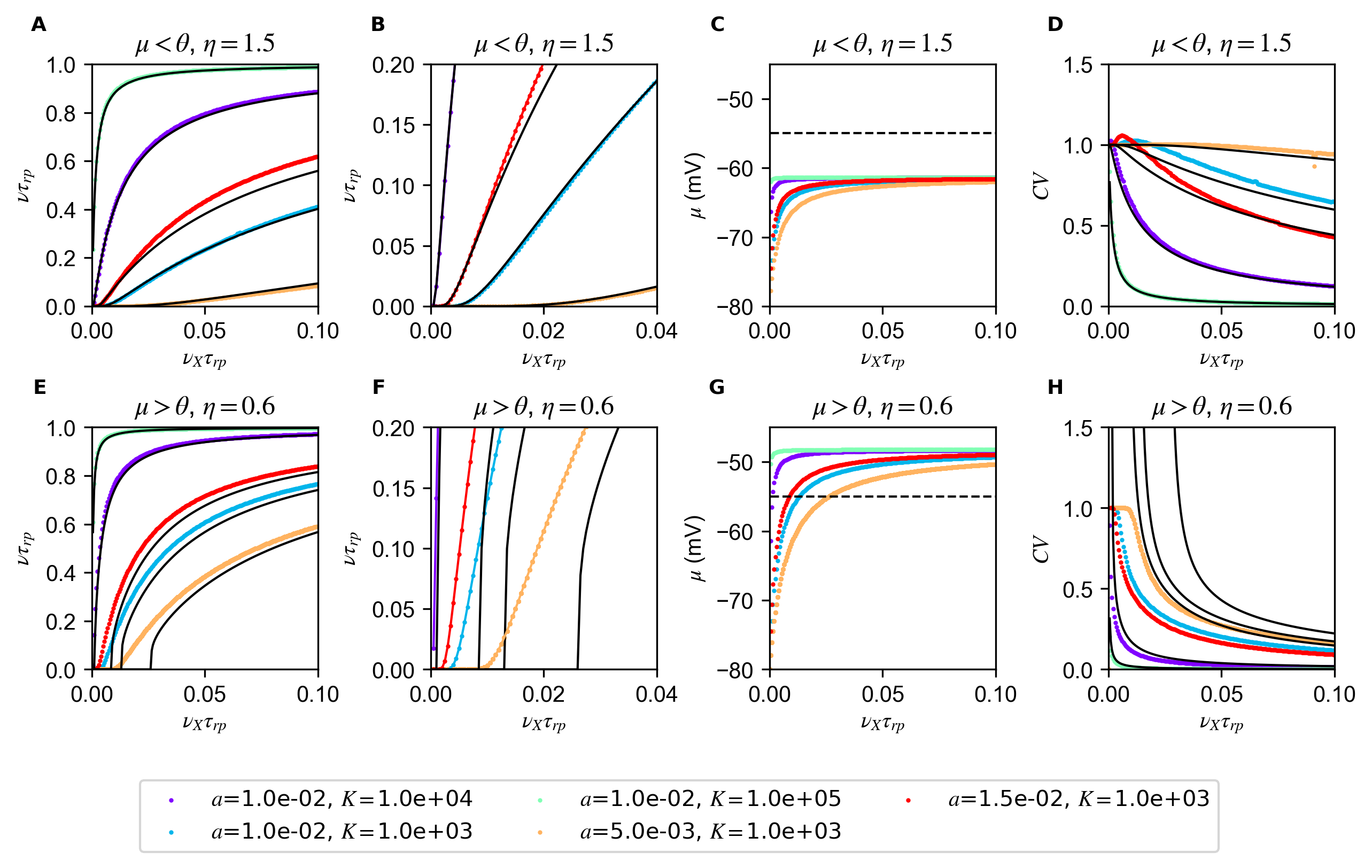}
\caption{\textbf{Response of single conductance-based neuron to noisy inputs.}
Estimates of firing rate  (\textbf{A}, \textbf{B}, \textbf{E}, \textbf{F}), $\mu$ (\textbf{C}, \textbf{G}) and $CV$ (\textbf{D}, \textbf{H}) obtained with numerical integration of Eqs~\eqref{SIeq:nu}, \eqref{eq:param} and \eqref{SIeq:CV} for different values of $a$ and $K$ (colored dots). 
 For the two regimes  $\mu<\theta$  (first row) and $\mu> \theta$ (second row), the transfer function saturates as $K$ increases.
Note the same change in $a$ has a more drastic effect if $\mu<\theta$, this is due to the exponential dependence that appears in Eq.~\eqref{SIeq:E_sub}. 
The approximated expressions (continuous lines) capture the properties of transfer function  (\textbf{A} Eq.~\eqref{SIeq:E_sub} and \textbf{E}, Eq.~\eqref{SIeq:E_sup}) and $CV$ (\textbf{C}, Eq~\eqref{SIeq:CV_sub} and \textbf{G}, Eq~\eqref{SIeq:CV_sup}).
For small inputs (\textbf{F}), Eq.~\eqref{SIeq:E_sup}  fails to describe the transfer function for some values of $K$ because the corresponding $\mu$ is below threshold.
Simulations parameter are: $g=12$; $\gamma=1/4$;  $\eta=1.5$ (top)  or $0.6$ (bottom).
}
\label{SIfig:transfer_function_single_neuron_test}
\end{figure*}

In the main text, we derived a simplified expression for the single neuron response neglecting the dependency of noise on membrane potential.
In this section we generalize this result to the case in which the full noise expression is considered.
We compute simplified expressions of the single neuron transfer function and CV, both in the sub-threshold regime $\mu<\theta$, and the supra-threshold regime $\mu>\theta$.
These expressions are validated numerically in Fig.~\ref{SIfig:transfer_function_single_neuron_test}  and  used in the last part of this section to define a scaling relation between $a$ and $K$ which preserves single neuron  firing in the strong coupling limit.

\subsection{Single neuron transfer function at strong coupling} 
\begin{figure*}[!h]
\includegraphics[width=10cm]{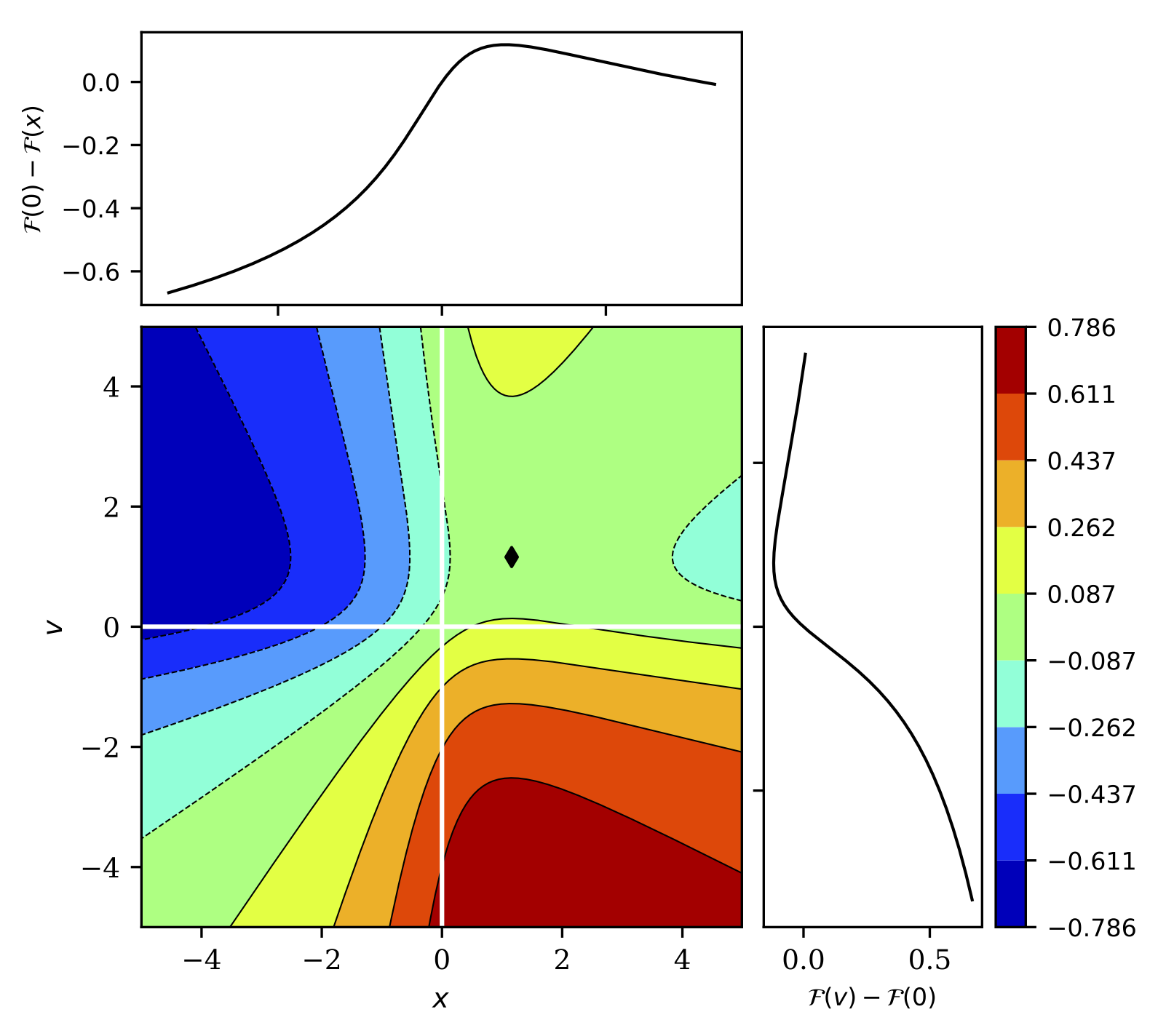}
\caption{{\bf Graphical representation of  the exponent  in Eq.~\eqref{SIeq:nu}}
The function $\mathcal{F}\left(v\right)-\mathcal{F}\left(x\right)$ is stationary at $x=v=\alpha$, this point is a maximum for $x$ and a minimum for $v$.
Parameters are as in Fig.~\ref{SIfig:Interpretations}.
In this figure, $\alpha=1.2$ {(black diamond)}.
}
\label{SIfig:saddle_point}
\end{figure*}

The starting point of  our analysis is the observation that the integrand in Eq.~\eqref{SIeq:nu} depends exponentially on $1/a\gg1$.
This suggests to perform the integration with a perturbative expansion of the exponent.
We will show below that, since the exponent has a stationary point at $x=v=\alpha$ (see Fig.~\ref{SIfig:saddle_point}),  the integration gives two qualitatively different results if $\alpha$ is larger or smaller than the upper bound of the integral $v_{max}$.
Moreover, since the condition $\alpha \lessgtr v_{max}$ corresponds to  $\theta \lessgtr \mu$, the two behaviors  correspond to supra/sub-threshold regimes, respectively.

\bigskip

\textit{\textbf{Supra-threshold regime $v_{max}<\alpha$ ($\theta<\mu$)}}\\
The  exponent in Eq.~\eqref{SIeq:nu} is negative for every value of $x$, except for $x=v$ in which it is zero.
The integral in $x$ can be written has
\begin{equation}
I=\int_{-\infty}^{v}dx
\, g(x) \, 
 e^{ \frac{f_v(x)}a }=\int_{-\infty}^{v}dx
\, g(x) \, 
 e^{ \frac{1}a  \left(f'_v(v) (x-v)+f''_v(v)/2 (x-v)^2+\dots \right) }
\end{equation}
With a change of variable $z= (x-v)/a$ we obtain
\begin{equation}\label{SIeq:integral}
I=a \, \int_{-\infty}^{0}dz
\, g(v+a z) \, 
 e^{ f'_v(v) z+a f''_v(v)\frac{z^2}{2}+\dots  }
\end{equation}
Neglecting all the terms of order $a$ we get
\begin{equation}
I=a \, \frac{g(v)}{f'_v(v)}  \, .
\end{equation}
Performing the integration in $v$  we obtain  
\begin{equation}\label{SIeq:E_sup}
\frac1{\nu}=\tau_{rp}+ \tau \log\left( \frac{\mu-V_r}{\mu-\theta}\right)  \, .
\end{equation}
Eq.~\eqref{SIeq:E_sup} is the transfer function of a deterministic conductance-based neuron with the addition of the refractory period. This is not surprising since the noise term becomes negligible compared to mean inputs in the small $a$ limit. In Fig.~\ref{SIfig:transfer_function_single_neuron_test}B we show that Eq.~\eqref{SIeq:E_sup} gives a good description of the transfer function predicted by the mean field theory in the supra-threshold regime.

\textit{\textbf{Sub-threshold regime $v_{max}>\alpha$ ($\theta>\mu$)}}\\
First we consider $\alpha<v_{min}$ ($\mu<V_r$).
For every value of $v$, the integral in $x$ in Eq~\eqref{SIeq:nu} has a maximum in the integration interval, hence it can be performed through saddle-point method and gives
\begin{equation}
\frac 1{\nu}-\tau_{rp}=\sqrt{\frac{4 \pi \chi \tau}{a^2 K( \alpha^2+1)}} \int_{v_{min}}^{v_{max}}dv 
 \, 
\exp \left[\frac{\mathcal{F}(v)-\mathcal{F}(\alpha)}a\right] \, .
\end{equation}
In the last equation, the exponent in the integrand has a minimum for $v=\alpha$ and is maximum at $v=v_{max}$; 
we expand the exponent around $v=v_{max}$ and, keeping term up to the first order, obtain
 \begin{equation}\label{SIeq:E_sub}
\frac 1{\nu}-\tau_{rp}=\tau \sqrt{\frac{\pi a^2 K \tau}{\chi( \alpha^2+1)}}  \frac{v_{max}^2+1}{\mid v_{max}-\alpha\mid } 
 \, \exp \left[\frac{\mathcal{F}(v_{max})-\mathcal{F}(\alpha)}a\right] \,   .
 \end{equation}
In the regime $v_{min}<\alpha<v_{max}$, the integral in $v$ of Eq.~\eqref{SIeq:nu} can be divided into three parts
\begin{equation}
\int_{v_{min}}^{v_{max}}dv=\int_{v_{min}}^{\alpha-\epsilon}dv+\int_{\alpha-\epsilon}^{\alpha+\epsilon}dv+\int_{\alpha+\epsilon}^{v_{max}}dv \, ;
\end{equation}
the third integral is analogous to case $\alpha<v_{min}$, hence it has an exponential dependency on the parameters and dominates the other terms.
In Fig.~\ref{SIfig:transfer_function_single_neuron_test}A we show that Eq.~\eqref{SIeq:E_sub} gives a good description of the transfer function predicted by the mean field theory for $\mu<\theta$.

\subsection{Single neuron  $CV$ of ISI  at strong coupling}
In this section we provide details of the derivation the  approximated expressions of the response $CV$.
Starting from the  mean field result of Eq.~\eqref{SIeq:CV}, we compute integrals using the approach discussed above. 

\textit{\textbf{Suprathreshold regime $v_{max}<\alpha$ ($\theta<\mu$)}}\\
The inner integral in Eq.~\eqref{SIeq:CV} yields in the small $a$ limit
\begin{equation}
 \int_{-\infty}^zdw \frac1{w^2+1} \exp\left[ \frac{\mathcal{F}(z)- \mathcal{F}(w)}a\right]=\frac{a}{z^2+1}\frac1{\frac{d\mathcal{F}(z)}{dz}}
\end{equation}
from which we obtain
\begin{equation}\label{SIeq:CV_sup}
CV^2=a \frac{\nu^2(aK\tau)^3}{a^2K^2\chi} \left[\log\left(\frac{v_{min}-\alpha}{v_{max}-\alpha}\right)+\frac{-3\alpha^2+4\alpha v_{max}+1}{2(\alpha-v_{max})^2}-\frac{-3\alpha^2+4\alpha v_{min}+1}{2(\alpha-v_{min})^2}\right]
\end{equation}
hence the rescaling needed to preserve the deterministic component $a\sim1/K$ produces $CV^2\sim a\ll 1$.
We validated this result numerically in Figs.~\ref{SIfig:transfer_function_single_neuron_test}H  and ~\ref{SIfig:scal_single_neuron}F. 

\textit{\textbf{Subthresold regime $v_{max}>\alpha$ ($\theta>\mu$)}}\\
The integral defining the $CV$, Eq.~\eqref{SIeq:CV}, can be expressed as
\begin{equation}
\int_{-\infty}^v dz \exp\left[\frac{\mathcal{F}(v)-\mathcal{F}(z)}a \right]g(z) =\int_{-\infty}^{v^*} dz \exp\left[\frac{\mathcal{F}(v)-\mathcal{F}(z)}a\right] g(z)+\int_{v^*}^v dz \exp\left[\frac{\mathcal{F}(v)-\mathcal{F}(z)}a\right] g(z)
\end{equation}
with 
\begin{equation}
g(z)=  \Bigg\{ \int_{-\infty}^zdw \frac1{w^2+1} \exp \left[\frac{\mathcal{F}(z)- \mathcal{F}(w)}a \right]\Bigg\}^2 \, , \quad v^*=\alpha-\epsilon \, .
\end{equation}
The first integral gives
\begin{equation}
\int_{-\infty}^{v^*} dz \exp\left[\frac{\mathcal{F}(v)-\mathcal{F}(z)}a\right] g(z)=\frac{a^3}{(v^*+1)^2\left[\frac{d\mathcal{F}(v^*)}{dz} \right]^3}
\end{equation}
In the second integral 
\begin{equation}
g(z)=  \frac{a \pi}{(\alpha^2+1)^2\frac{d^2\mathcal{F}(\alpha)}{dz^2}} \exp\left[\frac{2\mathcal{F}(z)-2\mathcal{F}(\alpha)}a\right]\, .
\end{equation}
from which we get 
\begin{equation}
\int_{v^*}^v dz \exp\left[\frac{\mathcal{F}(v)+\mathcal{F}(z)-2\mathcal{F}(\alpha)}a\right]  \frac{a \pi}{(\alpha^2+1)^2\frac{d^2\mathcal{F}(\alpha)}{dz^2}}  \, .
\end{equation}
Integrating in $z$ we obtain 
\begin{equation}
\int_{v^*}^v dz \exp\left[\frac{\mathcal{F}(z)}a\right]=\frac{a}{\frac{d\mathcal{F}(v)}{dz}}  \exp\left[\frac{\mathcal{F}(v)}a \right]\, .
\end{equation}
Integrating in $v$ we obtain 
\begin{equation}
CV^2=\frac{8\chi^2 \nu^2 \pi}{(\alpha^2+1)^2\frac{d^2\mathcal{F}(\alpha)}{dz^2}\left( \frac{d\mathcal{F}(v_{max})}{dz}\right)^2}\left(\frac{\exp\left[\frac{\mathcal{F}(v_{max})-\mathcal{F}(\alpha)}a\right]}{\sqrt{a}K} \right)^2 \, .
\end{equation}
Using Eq.~\eqref{SIeq:E_sub}  we obtain 
\begin{equation}\label{SIeq:CV_sub}
CV=1-\nu \tau_{rp}\, ,
\end{equation}
which corresponds to the CV of the ISIs of a Poisson process with dead time, with rate $\nu$ and refractory period $\tau_{rp}$.
We validated this result numerically in Figs.~\ref{SIfig:transfer_function_single_neuron_test}D and ~\ref{SIfig:scal_single_neuron}C.

\subsection{Scaling relations preserving firing in the strong coupling limit}
In this section we use the simplified expressions derived above to define scaling relations of  $a$ with $K$ which preserves  neural response in the strong coupling limit. Importantly, the scaling defined here depends on the operating regime of the neuron, i.e. on the asymptotic value of $\mu$.
\begin{figure*}[!htb]
\centering
\includegraphics[width=15cm]{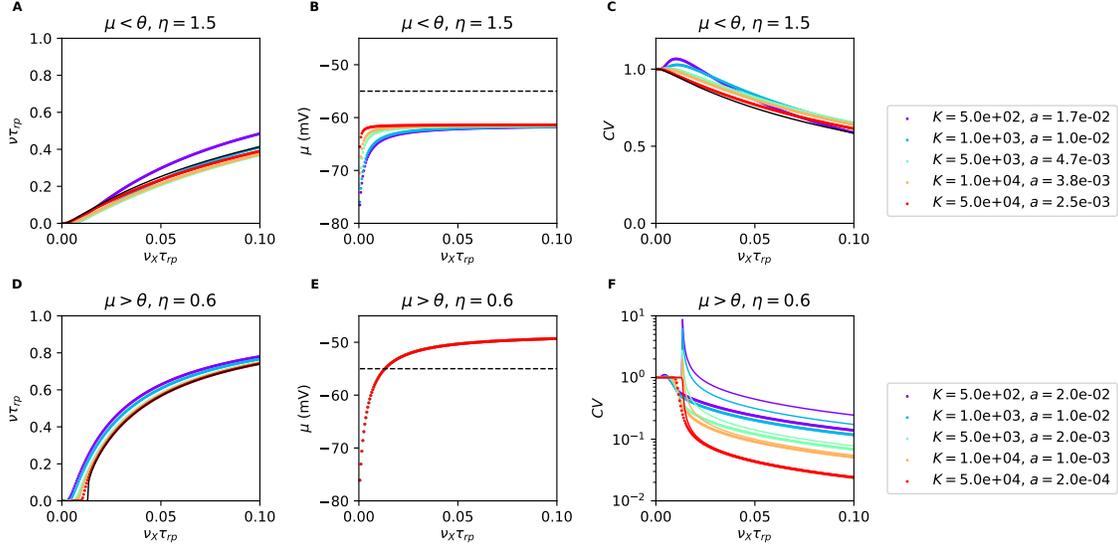}
\caption{\textbf{Scaling relationships preserving firing in the large $K$ limit.} 
Colored dots represent  mean field transfer function (\textbf{A}, \textbf{B}), CV (\textbf{C}, \textbf{D}) and membrane potential (\textbf{E}, \textbf{F}) obtained from Eqs.~\eqref{SIeq:nu}, \eqref{SIeq:CV} and \eqref{SIeq:terms}, respectively. 
Different colors correspond to different values of $a$ and $K$ which are scaled according to Eqs.~\eqref{SIeq:scaling_sub} (first row) and \eqref{SIeq:scaling_sup} (second row).
Mean field predictions are well described by the relevant approximated expressions (continuous lines).
For $\mu<\theta$  transfer function and  $CV$ are described by Eqs.~\eqref{SIeq:scaling_sub} (\textbf{A}) and~\eqref{SIeq:CV_sub} (\textbf{C}); both quantities are invariant as $K$ increases.
For $\mu>\theta$,  transfer function and  $CV$ are described by Eqs.~\eqref{SIeq:scaling_sup} (\textbf{A}) and~\eqref{SIeq:CV_sup} (\textbf{C}); note that, as explained in the text, the firing is preserved while the $CV$ becomes smaller as $K$ increases (different line colors correspond to different values of $K$).
Parameters: $g=12$;  $\gamma=1/4$.}
\label{SIfig:scal_single_neuron}
\end{figure*}

In the limit of large $K$,  terms in Eq.~\eqref{SIeq:terms} can be written as
\begin{align}\begin{split}
\tau^{-1}=a K \nu_X \, \left(1  +\eta g \gamma \right)\, , \quad
\omega^{-1}=\nu_X   \left(1  +\eta g \gamma \right) \, ,\quad
\chi^{-1}= \nu_X  \left(1  +\eta g^2 \gamma \right) \, ,
\end{split}\end{align}
while $\mu$, $\mathcal{E_D}$, $\mathcal{E_S}$, $v_{max}$, $\alpha$ and the function $\mathcal{F}(x)$ are independent of  $K$, $a$ and $\nu_E$.
We have shown in  the previous section that the single neuron  transfer function  is given by
\begin{equation}\label{SIeq:neuron_solution}
\frac1{\nu}=\tau_{rp}+\frac{\mathcal{Q}}{\nu_E}
\end{equation}
with 
\begin{equation}\label{SIeq:Q_definition}
\mathcal{Q}=
\begin{cases}
\left(\frac{ 1}{\sqrt{a}K}\exp \frac{\mathcal{F}(v_{max})-\mathcal{F}(\alpha)}a  
\right) \sqrt{\frac{\pi \left(1  +\eta g^2 \gamma \right)}{\left(1  +\eta g \gamma \right)^3( \alpha^2+1)}}  \frac{v_{max}^2+1}{\mid v_{max}-\alpha\mid } 
 \, \, \quad &\text{for $\mu<\theta$} \\
\frac{1}{  aK \left(1  +\eta g \gamma \right)}\log\left( \frac{\mu-\theta}{\mu-V_r}\right) \, \quad &\text{for $\mu>\theta$}
\end{cases}  
 \end{equation}
For  $\mu>\theta$, the parameters $a$ and $K$ in Eq.~\eqref{SIeq:Q_definition} appear only in the combination  $aK$.
It follows that a rescaling
\begin{equation}\label{SIeq:scaling_sup}
a\sim \frac1K
\end{equation}
leaves invariant the neural response for large $K$.
For  $\mu<\theta$,  Eq.~\eqref{SIeq:Q_definition}, and hence the transfer function, is invariant under the rescaling
 \begin{equation}\label{SIeq:scaling_sub}
K\sim \frac{ 1}{\sqrt{a}}\exp \left[\frac{\mathcal{F}(v_{max})-\mathcal{F}(\alpha)}a  \right]
\end{equation}
In Fig.~\ref{SIfig:scal_single_neuron}A,D we show neural responses computed for different values of $K$ with $a$ rescaled according to Eqs~\eqref{SIeq:scaling_sup} or~\eqref{SIeq:scaling_sub}; as predicted the network transfer function remains invariant as  $K$ increases.
Note that the response remains nonlinear in the limit of large $K$; we will show in the next section that in the network case, because of the self consistency relation,  nonlinearities are suppressed by the scaling relation. 

Finally, from Fig.~\ref{SIfig:scal_single_neuron}C,F, we see that the rescaling  preserves the $CV$ for $\mu<\theta$ and suppresses it for $\mu>\theta$. In the case $\mu<\theta$, the $CV$ is given by  Eq.~\eqref{SIeq:CV_sub}. This expression shows that the scaling relation of Eq.~\eqref{SIeq:scaling_sub} also leaves invariant the $CV$.
Interestingly, in some parameter regime, the $CV$ in Figs.~\ref{SIfig:transfer_function_single_neuron_test}D and ~\ref{SIfig:scal_single_neuron}C  shows a  non-monotonic behavior with $\nu_X$ which is not captured by Eq.~\eqref{SIeq:CV_sub}. 
In particular, a $CV$ above one 1 is observed when $\mu$ is below the reset $V_r$. 
As pointed out in~\cite{Barbieri2007}, this  supra-Poissonian firing is explained by the fact that, when $\mu<V_r$, spiking probability is higher just after firing that it is afterwards. In agreement with this interpretation, we find that the non-monotonic behavior of the  $CV$ is disappears in the large $K$ limit, where the region of inputs for which $\mu<V_r$ becomes negligible.
Thus, our analysis shows that the irregularity of firing is preserved in the strong coupling limit  of  a single neuron with $\mu<\theta$.

In the case $\mu>\theta$, the $CV$ is given by Eq.~\eqref{SIeq:CV_sup}.
This expression shows that the scaling relation of Eq.~\eqref{SIeq:scaling_sup} produces a $CV$ which decreases as $1/K$ in the strong coupling limit.
It follows that, in a single neuron with $\mu>\theta$, the strong coupling limit produces finite firing that is regular.

Starting from the next section we will focus our attention to network of conductance-based neurons. Since we are interested in describing the irregular firing observed in the cortex, we will focus our study on networks with $\mu<\theta$.

\clearpage
\section{Firing rate and scaling relation in leaky integrate-and-fire neuron models with voltage-dependent currents}\label{SI:general_single_neuron}
In the main text, we have shown that, when coupling is strong and $a\ll1$, the response of a single LIF neuron with conductance-based synapses is well approximated by Eq.~\eqref{eq:single_neuron}, i.e.~Kramers escape rate. Using this expression, we have show that the scaling relation of Eq.~\eqref{eq:scaling_sub} allows finite firing in single neuron and in networks of neurons.
Here, we show that the first order approximation of this scaling, i.e.  $a\sim1/\log(K)$,  appears also in neuron models with additional biophysical details, such as  spike generating currents~\cite{Fourcaud2003} and voltage-gated subthreshold currents~\cite{Li2020}, as long as coupling is strong, $a$ is small, and synapses are conductance-based.

We consider integrate-and-fire models featuring voltage-dependent currents, indicated here as $\phi(V)$, and conductance-based synapses. 
In these models, the membrane potential dynamics can be written as
\begin{equation}\label{Seq:membrane_potential_full}
\mathcal{C}_j \frac{dV_j}{dt}=- \sum_{A=L,E,I}g^j_A \left( V_j -E_{A}\right) \, +\psi(V).
\end{equation}
In the leaky integrate-and-fire model (LIF), $\psi(V)=0$ and Eq.~\eqref{Seq:membrane_potential_full} reduces to  Eq.~\eqref{eq:membrane} analyzed in the main text. 
In the exponential integrate-and-fire model (EIF)~\cite{Fourcaud2003}, the function $\psi(V)=\Delta T g_L \exp[(V-\theta)/\Delta T]$ describes the spike generation current; in this model, once the membrane potential crosses the threshold $\theta$ it diverges to infinity in finite time. 
The current generated by inward rectifier voltage-gated channels, such as the one recently reported in~\cite{Li2020}, is captured by an expression of the form $\psi(V)=-g_{in}(V)(V-E_{in})$, where  $g_{in}(V)$ and  $E_{in}$ represent the conductance and the reversal potential of the channels, respectively; in the case of~\cite{Li2020},  $1/g_{in}(V)$ was shown to well approximated by a linear increasing function of $V$.

The dynamics Eq.~\eqref{Seq:membrane_potential_full}, following an approach analogous to the one we used for the derivation of Eq.~\eqref{eq:intro_V}, can be approximated by 
\begin{equation}\label{Seq:membran_potential_approx}
\begin{split}
\tau \frac{d V}{d t}=-\frac{\partial \mathcal{H}(V) }{\partial V}+ \sigma \sqrt{\tau}\zeta\, , \quad
\mathcal{H}(V)=\frac12\left(V-\mu \right)^2-\frac{\tau}{\tau_L g_L}\int^V\psi(x)dx
\end{split}
\end{equation}
where $\zeta$ is a white noise term, with zero mean and unit variance density, while $\tau$, $\mu$ and $\sigma(V)$ are as in Eq.~\eqref{eq:param_MM}. 
In what follows, as in the main text, we use the effective time constant approximation~\cite{Richardson2005}, i.e. we neglect the multiplicative component of the noise term in Eq.~\eqref{Seq:membran_potential_approx}, and  make the substitution $\sigma(V)\rightarrow \sigma(\mu^*) $, where $\mu^*$ is the mean value of the membrane potential dynamics.

The firing rate of a neuron following Eq.~\eqref{Seq:membran_potential_approx} can be computed exactly using Eq.~\eqref{eq:general_solution} and is given by
\begin{equation}
    \nu=\left[\tau_{rp}+\frac{2\tau}{\sigma^2}\int_{-\infty}^{\infty}dx\int^{\infty}_{\max(V_r,x)}\exp\left(\frac{\mathcal{H}(z)-\mathcal{H}(x)}{\sigma^2}\right)dz\right]^{-1} \, . 
\end{equation}
In what follows, we provide a more intuitive derivation of the single neuron response, which is valid in the biologically relevant case of $a\ll1$. 
The function $\mathcal{H}$ in Eq.~\eqref{Seq:membran_potential_approx} can be though of as an energy function which drives the dynamics of the membrane potential.
In the case of LIF neurons, $\mathcal{H}$ is a quadratic function with a minimum at $V=\mu$. In neuron models with a spike generation current, such as the EIF model~\cite{Fourcaud2003}, the shape of the function $\mathcal{H}$ far from threshold is qualitatively similar to that of the LIF model (with a minimum at $V=\mu^*$), but becomes markedly  different close to threshold, where the potential energy has a maximum at $V=\theta^*$ and goes to  $-\infty$ for $V>\theta^*$.
Here, we focus on the case in which additional subthreshold voltage-gated currents do not lead to additional minima of the energy function, a scenario that can happen with potassium inward-rectifier currents (e.g. see~\cite{Ermentrout2010} chapter 4.4.3). 
In models in which $\mathcal{H}$ has a single minimum in the subthreshold range at $\mu^*$, and a maximum at $\theta^*$, the firing rate of a neuron when input noise is small (i.e. when $a\ll1$) can again be computed using Kramers escape rate, which gives the average time it take for the membrane potential to go from $\mu^*$ to $\theta^*$, (see~\cite{GardinerBook09} section 5.5.3)
\begin{equation}\label{Seq:Kramer}
\begin{split}
\frac1{\nu}-\tau_{rp}=\frac{2\pi \bar{\tau}\bar{\Upsilon}}{aK\nu_X}\,
\,\exp\left(\frac{\bar{\Delta}}a\right) 
\end{split}
\end{equation}
where 
$$
\bar{\Upsilon}=\left(\frac{d^2\mathcal{H}}{dV^2}\Big\vert_{\theta^*}\frac{d^2\mathcal{H}}{dV^2}\Big\vert_{\mu^*}
\right)^{-\frac12}\,, \quad
\bar{\Delta}=\frac{\mathcal{H}(\theta^*)-\mathcal{H}(\mu^*)}{\bar{\sigma}}\,, \quad \bar{\tau}={aK\nu_X}\tau \,, \quad \bar{\sigma}=\frac{\sigma}{\sqrt{a}}\,,
$$
while $\bar{.}$ indicates quantities  that  remain of order 1 in the small $a$ limit, provided the external inputs $\nu_X$ are at least of order $1/(aK\tau_L)$.
Eq.~\eqref{Seq:Kramer} is the generalization of Eq.~\eqref{eq:single_neuron} to the case of integrate-and-fire neuron models with voltage-dependent currents; it shows that, at the dominant order, finite firing emerges if  $a\sim1/\log(K)$. Moreover, Eq.~\eqref{Seq:Kramer} shows that corrections to the logarithmic scaling depend on the specific type of  voltage-dependent currents used in the model.

\section{Calculations in the strong coupling regime - Networks}\label{SI:response_strong_coupling}
In this section, we show how the results on the strong coupling limit of single neuron response can be generalized to the network case.
First, we analyze the problem in the case in which excitatory and inhibitory neurons have the same biophysical properties (model A). In this model we start by discussing the results using the effective time constant approximation, and then discuss the full results. Then we study the case in which excitatory and inhibitory neurons have different biophysical properties (model B). 

\subsection{Model A, effective time constant approximation}
As discussed in the main text, the network response in model A with the effective time constant approximation is obtained solving the self-consistency condition given by  \eqref{eq:net_rates} and Eq.~\eqref{eq:nu}. 
At strong coupling, this condition can be simplified to the form of Eq.~\eqref{eq:single_neuron}. 
In the strong coupling limit, when $\nu_X\gg 1/{aK \tau_L }$ and $\nu \gg 1/\tau_{rp}$, the right hand side of Eq.~\eqref{eq:nu} depends on $\nu$ and $\nu_X$ only through their ratio.
Therefore, we look for solutions of the simplified self-consistency condition with a Taylor expansion
\begin{equation}\label{eq:taylor_model_A}
\frac{\nu}{\nu_X}=\sum_{k=1}^{k=\infty} \rho_{k}  x ^{k-1} \, , \quad{x=\tau_{rp}\nu_X}
\end{equation} 
Keeping only terms up to first order in $x$,  the self-consistency condition becomes
\begin{equation*}
\frac 1{\rho_1}-\left( 1+\frac{\rho_2}{\rho_1^2}\right)x =\mathcal{Q}(\rho_1)+\rho_2\frac{d\mathcal{Q}(y)}{dy}\Big\vert_{y=\rho_1}x
\end{equation*}
from which we find 
\begin{equation}\label{eq:eta_Q_eta}
\rho_1=\frac 1{\mathcal{Q}(\rho_1)} \, .
\end{equation}
The solution of Eq.~\eqref{eq:eta_Q_eta} provides the linear component of the network response; this is preserved in the strong coupling limit with an expression analogous to Eq.~\eqref{eq:scaling_sub} but with 
\begin{equation*}
    \frac{r_E}{\nu_X}=1 +\rho_1\, , \quad \frac{r_I}{\nu_X}=\rho_1 \, .
\end{equation*}
This uniquely defines a scaling between $a$ and $K$ (see Fig.~\ref{fig:compare_r_solutions}A for an example of the scaling function).
We test the validity of our result  in Fig.~\ref{fig:compare_r_solutions}B.
The numerical analysis shows that, as $K$ increases, the scaling relation prevents saturation and suppression of the network response.
However, unlike what happens in the single neuron case, the shape of the transfer function is not preserved and becomes increasingly linear as $K$ becomes larger. 
This is analogous to what happens in the balanced state model~\cite{vanvreeswijk96b,vanvreeswijk98,brunel00,Sanzeni856831}, where the network transfer function becomes linear in the strong coupling limit. 
For the case under investigation here, we can understand this suppression of nonlinearities by looking at the second order terms in the expansion of Eq.~\eqref{eq:taylor_model_A}.  Keeping the dominant contribution in $a$, we find
\begin{equation}\label{eq:r_Q_2}
\rho_2 \sim a \frac{ \rho_1 \bar{\sigma}^2}{2 \bar{v}_{max} \left( \bar{\sigma} \frac{d\mu}{dy} +(\theta-\mu)\frac{d\bar{\sigma}}{dy}\right)} \, .
\end{equation}
Hence $\rho_2$ goes to zero as $a$ decreases, producing a linear transfer function.
This follows directly from the self-consistency relation and is not present in the single neuron case, where in fact a nonlinear  transfer function is observed in the large $K$ limit. 
Fig.~\ref{fig:compare_r_solutions}B shows that linearity is reached really slowly with $K$; this follows directly from Eq.~\eqref{eq:r_Q_2} where the suppression of nonlinear terms is controlled by $a$,  which slowly goes to zero with $K$ (approximately logarithmically).

\subsection{Model A, multiplicative noise}
In this section, we generalize the approach used above, relaxing the effective time constant approximation.
\begin{figure*}[!htb]
\centering
\includegraphics[width=15cm]{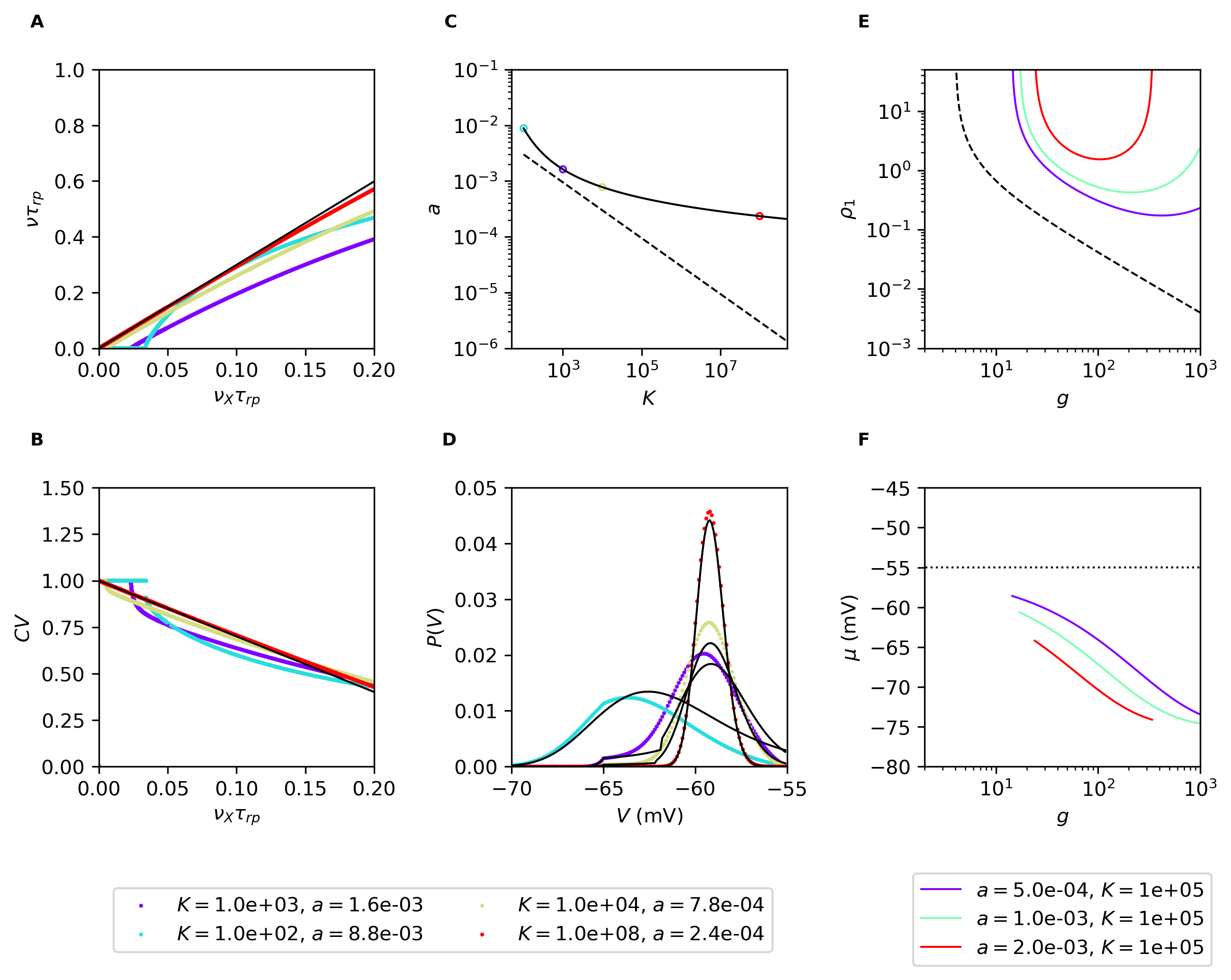}
\caption{\textbf{Strong coupling limit of networks of conductance-based neurons in model A.} 
Numerically computed network transfer function (\textbf{A}), $CV$ (\textbf{B}) and probability distribution of the membrane potential (\textbf{D}) obtained from Eqs.~\eqref{SIeq:network_solution}, \eqref{SIeq:P_of_V} and \eqref{SIeq:CV_sub}. 
Different colors correspond to different values of $a$ and $K$ which have been changed according to the scaling relation~\eqref{SIeq:scaling_network} (\textbf{C}). 
As $K$ increases the network transfer function and $CV$ converges to the expression derived in the main text (black lines). 
Note that, unlike the case of single neuron, the network transfer functions becomes linear. 
The probability distribution of the membrane potential becomes Gaussian and slowly converges to a delta function. 
Panels (\textbf{E}) and (\textbf{F}) show the network gain and membrane potential for different values of $a$ at fixed $K$. Note that, unlike what happens in current-based networks (black dashed lines), the gain is not monotonic with $g$.
Simulation parameters are as in Fig.~\ref{SIfig:transfer_function_single_neuron_test}; in panels \textbf{A}-\textbf{D}  $g=20$.
}
\label{SIfig:compare_r_solutions}
\end{figure*}

As discussed in Appendix~\ref{SI:single_neuron_response_at_strong_coupling}, Eq.~\eqref{SIeq:nu} in the strong coupling limit  becomes
\begin{equation}\label{SIeq:network_solution}
\frac1{\nu}=\tau_{rp}+\frac{\mathcal{Q}}{\nu_X}
\end{equation}
with
\begin{equation}
\mathcal{Q}=
\Bigg\{\frac{ 1}{\sqrt{a}K}\exp \left[\frac{\mathcal{F}(v_{max})-\mathcal{F}(\alpha)}a\right]  
\Bigg\} \sqrt{\frac{\pi \left[1+\frac{\nu}{\nu_X} (1+g^2 \gamma )\right]}{\left(1+\frac{\nu}{\nu_X} (1+g \gamma )\right)^3( \alpha^2+1)}}  \frac{v_{max}^2+1}{\mid v_{max}-\alpha\mid } 
 \, ,
 \end{equation}
and
\begin{align}
\begin{split}
\tau^{-1}=a K \omega^{-1}\, , \quad
\omega^{-1}=\nu_X \, \left[1+\frac{\nu}{\nu_X} (1+g \gamma )\right] \, ,\quad
\chi^{-1}= \nu_X \, \left[1+\frac{\nu}{\nu_X} (1+g^2 \gamma )\right] \, , \\
\mu=\frac{    E_E+\frac{\nu}{\nu_X} (E_E+g \gamma E_I)}{1+\frac{\nu}{\nu_X} (1+g \gamma )}\, , \quad
\mathcal{E_S}=\frac{ E_E+\frac{\nu}{\nu_X} (E_E+g^2 \gamma E_I)}{1+\frac{\nu}{\nu_X} (1+g^2 \gamma )}\, ,\\  
\mathcal{E_D}=\frac{  \left( E_E-E_I\right) \sqrt{(1+\frac{\nu}{\nu_X})\frac{\nu}{\nu_X}g^2 \gamma }}{1+\frac{\nu}{\nu_X} (1+g^2 \gamma )}\, .
\end{split}\end{align}
Here we assumed $aK\gg1/{\tau_L \nu_X}$ so that the function $\mathcal{Q}$ depends on $\nu$ and $\nu_X$ only through the combination $\nu/\nu_X$.  We will show below that a scaling relation analogous to that of single neurons holds, hence for $K$ large enough $aK\gg1/{\tau_L \nu_X}$ is automatically implemented.
To solve the self consistency condition, we express the firing rate $\nu$ with a Taylor expansion 
\begin{equation}\label{SIeq:taylor_model_A}
\tau_{rp}\nu=\sum_{k=1}^{k=\infty} \rho_{k}  x ^k \, , \quad x=\tau_{rp}\nu_{X} \, .
\end{equation} 
Note that in Eq.~\eqref{SIeq:taylor_model_A} we  assumed $ \rho_{0}=0$, we will come back to this point at the end of the section.  
Under this assumption $y:={\nu}/{\nu_X}=\sum_{k=1}^{k=\infty} \rho_{k}  x ^{k-1}$  and the  function $\mathcal{Q}$ depends only on  powers  of the dimensionless variable $x$.
Keeping only terms up to  first order in $x$,  Eq.~\eqref{SIeq:network_solution} becomes
\begin{equation}
\frac 1{\rho_1}-\left( 1+\frac{\rho_2}{\rho_1^2}\right)x =\mathcal{Q}(\rho_1)+\rho_2\frac{d\mathcal{Q}(y)}{dy}\Big\vert_{y=\rho_1}x
\end{equation}
from which we find 
\begin{equation}\label{SIeq:eta_Q_eta}
\rho_1=\frac 1{\mathcal{Q}(\rho_1)} \, .
\end{equation}
The solution of Eq.~\eqref{SIeq:eta_Q_eta} provides the  linear component of the network response, i.e. its gain; we will discuss  this function in more detail at the end of this section.

From Eq.~\eqref{SIeq:eta_Q_eta} we find that the network gain $\rho_1$  is preserved in the strong coupling limit if the factor 
\begin{equation}\label{SIeq:scaling_network}
 \frac{ 1}{\sqrt{a}K}\exp\left[ \frac{\mathcal{F}(v_{max})-\mathcal{F}(\alpha)}a\right] \, ,
\end{equation}
is constant.
Eq.~\eqref{SIeq:scaling_network} uniquely defines a scaling between $a$ and $K$ (see Fig.~\ref{SIfig:compare_r_solutions}C for an example of the scaling function).
We test the validity of the scaling  in Fig.~\ref{SIfig:compare_r_solutions} as follows: given a set of parameters $a$, $K$  and $\rho_1$, we compute numerically the transfer function from Eq.~\eqref{SIeq:nu}, then we increased $K$, determined the corresponding change in $a$ using Eq.~\eqref{SIeq:scaling_network} and compute again the transfer function; results of this procedure are shown in Fig.~\ref{SIfig:compare_r_solutions}A.
The numerical analysis shows that, as $K$ increases our scaling relation prevent saturation and the network response remains finite.

As in the case with diffusion approximation, the shape of the transfer function is not preserved by the scaling and an increasing linear response is observed. We can understand this suppression of nonlinearities by  looking at the second order terms in the expansion of Eq.~\eqref{SIeq:network_solution}; we find 
\begin{equation}\label{SIeq:r_Q_2}
\rho_2=\frac{-\rho_1^2}{\rho_1\frac{d\log \left( \mathcal{Q}(y)\right)}{dy}+1}  \, ,
\end{equation}
and, keeping the dominant contribution in $1/a$ at the denominator, 
\begin{equation}\label{SIeq:r_Q_2}
\rho_2  \sim \frac{ - a \, \rho_1}{\frac{d\mathcal{F}(v_{max}(y),y)}{dy}\Big\vert_{\rho_1}+\frac{d\mathcal{F}(\alpha(y),y))}{dy}\Big\vert_{\rho_1}}\, .
\end{equation}
Hence $\rho_2$ goes to zero as $a$ decreases, producing a linear transfer function.
The nonlinearities at low rate  in Fig.~\ref{SIfig:compare_r_solutions}A (e.g. see red and yellow lines) show that  our assumption $\rho_0=0$ is not valid in general. However  it turns out that the above defined scaling relation suppresses also these nonlinearities  in the limit of strong coupling (e.g. blue and cyan lines).

\medskip

We now characterize the dependency of the transfer function gain, i.e. its slope, on network parameters.
For fixed network parameters, the network gain $\rho_1$ is defined as the solution of Eq.~\eqref{SIeq:eta_Q_eta}; solutions as a function of $a$ and $g$ are shown in Fig.~\ref{SIfig:compare_r_solutions}E. 
At fixed values of $a$, the gain initially decreases as $g$ increases and, for $g$ large enough, the opposite trend appears.
This behavior is due to two different effects which are produced by the increase of $g$: on one hand, it increases the strength of recurrent inhibition; on the other hand, it decreases the equilibrium membrane potential $\mu$ and bring it closer to the inhibitory reversal potential $E_i$, which in turn  weakens inhibition (see Fig.~\ref{SIfig:compare_r_solutions}F). 
Fig.~\ref{SIfig:compare_r_solutions}E shows that the gain is finite only for a finite range of the parameter $g$; divergences appear because recurrent inhibition is not sufficiently strong to balance excitation.
At small $g$, the unbalance is produced by  week efficacy of inhibitory synapses; at large $g$, inhibition is suppressed by the approach of the membrane potential to the reversal  point of inhibitory synapses.
Increasing the value of $a$  produces an upward shift in the curve and, at the same time, decreases the range of values in which the gain is finite. 
The  observed decrease in gain generated at low values of $g$ is observed also networks of current-based neurons~\cite{brunel00} where the gain is found to be $1/(g\gamma-1)$.
Finally, we note that the difference between conductance and current-based model decreases with $a$.

\medskip

To conclude this analysis, we give an approximated expression of the probability distribution of the membrane potential of Eq.~\eqref{SIeq:P_of_V} which, in the strong coupling limit, becomes
\begin{equation}
P(V)=\frac{\nu \omega}{\vert  v_{max}-\mu\vert} \left[\frac{u\left( V_{max}\right)^2+1}{u\left( V\right)^2+1}\right]\frac{e^{\frac{ \mathcal{F}\left(v_{max}\right)- \mathcal{F}\left(V\right)}{a}}}{aK}
\end{equation}
where $V_{max}$ is the value of the membrane potential $V$ which maximizes the integrand of Eq.~\eqref{SIeq:P_of_V} while the function $u()$ has been defined in Eq.~\eqref{SIeq:terms_model_A_MFT}. 
Examples of the probability distribution and the corresponding approximated expressions are given in  Fig.~\ref{SIfig:compare_r_solutions}D.

\subsection{Model B, multiplicative noise}

In this section we generalize the results obtained so far to the case of networks with  excitatory and inhibitory neurons with different biophysical properties.
\begin{figure*}[h!]
\centering
\includegraphics[width=15cm]{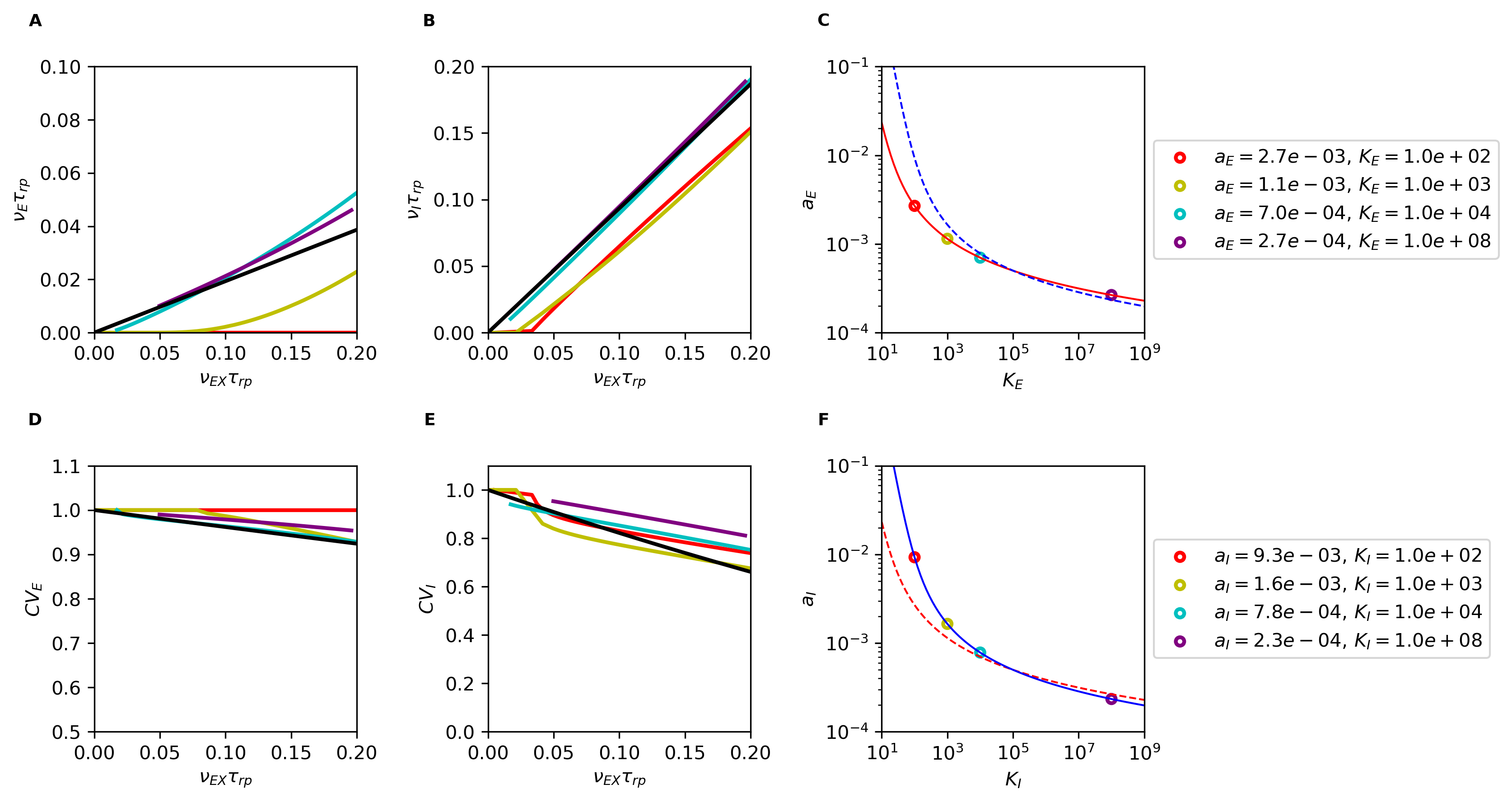}
\caption{\small{\textbf{Limit of large $K$ for networks, model B}.
Firing rate and $CV$ of excitatory and inhibitory neurons in  a  network  predicted by the mean field model for different values of inputs and $K$;
the expected asymptotic behavior is shown in black.
On the left ({\bf C}, {\bf F}),we show the corresponding scaling relations with dots associated to the connectivity parameters.
Simulations parameter: the two populations have $g_e=20.0$ and $g_i=19.0$; for both populations the $a=0.0005$ for 
$K=10^5$; other parameters as in Fig.~\ref{SIfig:transfer_function_single_neuron_test}. 
} 
}
\label{SIfig:a_of_K_diff}
\end{figure*}

\subsubsection{Model definition}
Here  we take into account the diversity of the two types of neurons with   
\begin{equation}
\tau_j =\tau_{E} \, , \quad   a_{jm} =a_{EX} \, , a_{EE} \, , a_{EI} ;
\end{equation} 
for excitatory neurons and
\begin{equation}
\tau_j =\tau_{I} \, , \quad   a_{jm} =a_{IX} \, , a_{IE} \, , a_{EE} ;
\end{equation} 
for inhibitory neurons.
We  use  the parametrization
\begin{align}\begin{split}
a_{EX}=a_E  \, , \quad  a_{EE}=a_E  \, , \quad  a_{EI} =g_E a_E\, , \\ 
a_{IX}=a_I  \, , \quad  a_{IE}=a_I  \, , \quad  a_{II} =g_I a_I\, ,
\end{split}\end{align}
and
\begin{align}\begin{split}
K_{EX}=K_E  \, , \quad  K_{EE}=K_E  \, , \quad  K_{EI} =\gamma_E K_E\, , \\ 
K_{IX}=K_I  \, , \quad  K_{IE}=K_I  \, , \quad  K_{II} =\gamma_I K_I\, .
\end{split}\end{align}
Eq.~\eqref{eq:membrane} becomes
\begin{equation}\label{SIeq:intro_V_2d}
\begin{cases}
\tau_E \frac{dV_E}{dt}=-\left( V_E-\mu_E\right) -\sigma_E(V_E) \sqrt{\tau_E}\zeta_E   ,\\
\tau_I \frac{dV_I}{dt}=-\left( V_I-\mu_I\right) -\sigma_I(V_I) \sqrt{\tau_I}\zeta_I \, .
\end{cases}
\end{equation}
The expressions for excitatory neurons are
\begin{align}\begin{split}
\tau_E^{-1}=\tau_{L,E}^{-1}+ a_E K_E \omega_E^{-1}\, , \quad \omega_E^{-1}= \nu_{EX}+\nu_E+g_E \gamma_E \nu_I  \, ,\\ 
\mu_E=\tau_E\{E_L+a_E K_E \tau_{L,E} [\nu_{EX} E_E +\nu_E E_E +\nu_I g_E \gamma_E E_I ]\}\, , \\
\sigma_E^2=  a_E^2K_E\frac{\tau_E}{\chi_E}\left[ \left( V-\mathcal{E_{S\textit{,E}}}\right)^2 +\mathcal{E_{D\textit{,E}}}^2\right] \, , \quad \chi_E^{-1}= \nu_{EX}+\nu_E+g_E^2 \gamma_E \nu_I\\
\mathcal{E_{S\textit{,E}}}=\chi_E \left[ \nu_{EX} E_E +\nu_E E_E +\nu_I g_E^2 \gamma_E E_I)\right] \, ,\\  
\mathcal{E_{D\textit{,E}}}= \chi_E\sqrt{\left( \nu_{EX}+\nu_E \right)  g_E^2 \gamma_E \nu_I}\left( E_E-E_I\right) \, ;
\end{split}\end{align}
analogous expressions are valid for inhibitory neurons.

The firing rate is given by solving a system of two equations
\begin{equation}\label{SIeq:nu_two}
\begin{cases}
\frac1{\nu_E}-\tau_{rp}= \frac{2 \chi_E}{a_E^2K_E}  \int_{v_{min,E}}^{v_{max,E}}dv \int_{-\infty}^{v}dx\frac{1}{x^2+1} \, \exp\left[\frac{\mathcal{F}_E\left(v\right)-\mathcal{F}_E\left(x\right)}{a_E} \right] \, , \\
\frac1{\nu_I}-\tau_{rp}= \frac{2 \chi_I}{a_I^2K_I}  \int_{v_{min,I}}^{v_{max,I}}dv \int_{-\infty}^{v}dx\frac{1}{x^2+1} \, \exp\left[\frac{\mathcal{F}_I\left(v\right)-\mathcal{F}_I\left(x\right)}{a_I} \right]
\end{cases}
\end{equation}
with
\begin{equation}
\begin{split}
\mathcal{F}_E(x)=\frac{2 \chi_E }{a_E K_E \tau_E } \left[\frac12 \log(x^2+1)-\alpha_E\arctan(x)\right] \, , \\
v_{min,E}=\frac{V_r-\mathcal{E_{S\textit{,E}}}}{\mathcal{E_{D\textit{,E}}}}\, , \quad
v_{max,E}=\frac{\theta-\mathcal{E_{S\textit{,E}}}}{\mathcal{E_{D\textit{,E}}}}  \, , \quad
\alpha_E=\frac{\mu_E-\mathcal{E_{S\textit{,E}}}}{\mathcal{E_{D\textit{,E}}}} \, .
\end{split}
\end{equation}
and analogous expressions for the inhibitory population.
The probability distribution of the membrane potential and the $CV$ are straightforward generalizations of Eqs.~\eqref{SIeq:P_of_V} and~\eqref{SIeq:CV}.

\subsubsection{Scaling analysis}
We parametrize inputs  to the two populations as $\nu_{EX}$ and $\nu_{IX}=\eta \nu_{EX} $.
Using an analysis analogous to the one depicted above, we obtain a simplified expression for the self-consistency Eq.~\eqref{SIeq:nu_two} that is
 \begin{equation}
\begin{cases}
\frac1{\nu_E}-\tau_{rp}= \frac{\mathcal{Q}_E(\nu_E/\nu_{EX},\nu_I/\nu_{EX})}{\nu_{EX}} \, , \\
\frac1{\nu_I}-\tau_{rp}= \frac{\mathcal{Q}_i(\nu_E/\nu_{EX},\nu_I/\nu_{EX})}{\nu_{EX}} \, , \\
\end{cases} 
\end{equation}
where
\begin{equation}
\mathcal{Q}_E=
\left[\frac{ 1}{\sqrt{a_E}K_E}\exp \frac{\mathcal{F}_E(v_{max,E})-\mathcal{F}_E(\alpha_E)}{a_E}  
\right] 
\sqrt{\frac{\pi \left[1+ \frac{\nu_E}{\nu_{EX}}+g_E^2 \gamma_E \frac{\nu_I}{\nu_{EX}}\right]}{\left[1+\frac{\nu_E}{\nu_{EX}} +g_E \gamma_E \frac{\nu_I}{\nu_{EX}} \right]^3( \alpha_E^2+1)}}  \frac{v_{max,E}^2+1}{\mid v_{max,E}-\alpha_E\mid } 
 \, ,
 \end{equation}
 and
\begin{equation}
\mathcal{Q}_I=
\left[\frac{ 1}{\sqrt{a_I}K_I}\exp \frac{\mathcal{F}_I(v_{max,I})-\mathcal{F}_I(\alpha_I)}{a_I}  \right] 
\sqrt{\frac{\pi \left[\eta+ \frac{\nu_E}{\nu_{EX}}+g_I^2 \gamma_I \frac{\nu_I}{\nu_{EX}}\right]}{\left[\eta+\frac{\nu_E}{\nu_{EX}} +g_I \gamma_I\frac{\nu_I}{\nu_{EX}} \right]^3( \alpha_I^2+1)}}  \frac{v_{max,I}^2+1}{\mid v_{max,I}-\alpha_I\mid } 
 \, .
 \end{equation}
We investigate the solution in the strong coupling limit using an expansion 
\begin{equation}
\tau_{rp}\nu_E=\sum_{k=1}^{k=\infty} \rho_{k}^E  x^k \, , \quad \tau_{rp}\nu_I=\sum_{k=1}^{k=\infty} \rho_{k}^I  x^k \, , \quad x=\tau_{rp}\nu_{EX} \, ,
\end{equation}
and obtain 
\begin{equation}\label{SIeq:gain_2D}
\begin{cases}
\rho_{1}^E= \frac{1}{\mathcal{Q}_E(\rho_{1}^E,\rho_{1}^I)} \\
\rho_{1}^I= \frac{1}{\mathcal{Q}_I(\rho_{1}^E,\rho_{1}^I)} 
\end{cases}\, .
 \end{equation}
Eq.~\eqref{SIeq:gain_2D} defines the gain of the excitatory and inhibitory populations. As for model A, requiring that network gain is preserved in the large $K$ limit is equivalent to assuming the products
\begin{equation}
\frac{ 1}{\sqrt{a_j}K_j}\exp \frac{\mathcal{F}_j(v_{max,j})-\mathcal{F}_j(\alpha_j)}{a_j} \, 
\end{equation}
constant; these constraints defines how synaptic strength should scale with $K$ to preserve the response gain.
We note that, since $\mathcal{F}_j(v_{max,j})-\mathcal{F}_j(\alpha_j)$ is different for the two populations, in the general case there are two different scalings for the two populations; in Fig~\ref{SIfig:a_of_K_diff} we verify this prediction.


\section{Simulations vs theory}\label{SI:MFT_vs_SIM}
All the results showed in the main text are based on the mean field analysis of the network dynamics. in this section we investigate how the predictions of the mean field theory compare to numerical simulations of networks of conductance-based neurons.
\begin{figure*}[!h]
\includegraphics[width=15cm]{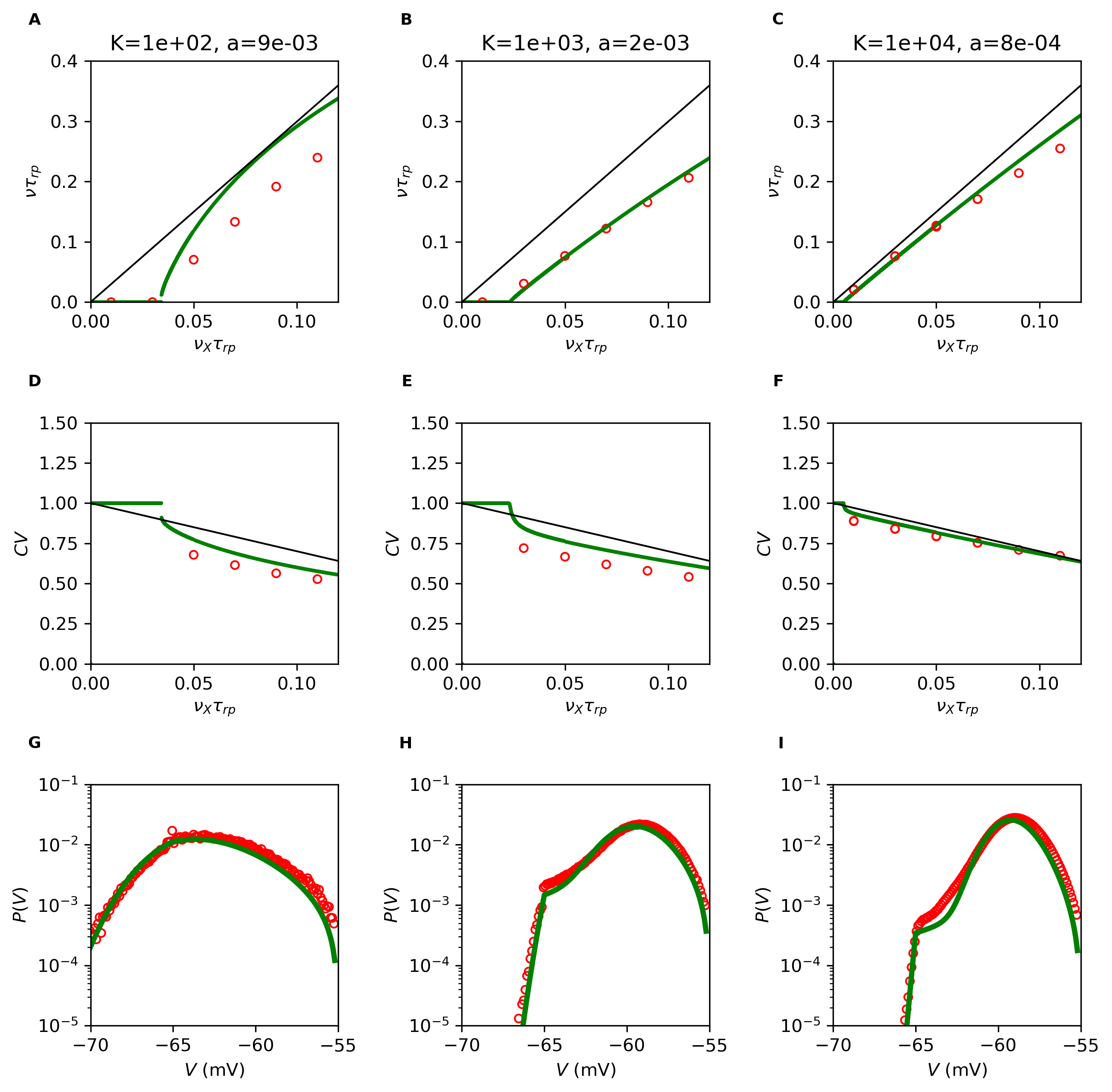}
\caption{{\bf Comparison of mean field theory and numerical simulations.} 
Network transfer function (first row), $CV$ of ISI distribution (second row) and probability distribution of the membrane potential at $\nu_E=0.05 \tau_{rp}$ (third row). 
In every panel we show mean field prediction (green), results from numerical simulations (red) and value expected in the strong coupling limit (black).
Different columns correspond to different values of $K$ and $a$ which were scaled according to Eq.~\eqref{SIeq:scaling_network}. The agreement between network simulations (red) and mean field predictions (green) improves as $a$ decreases, as expected since we used the diffusion approximation to derive the results.
Simulation parameters are: $g=20$, $N_E=N_I=N_{EX}=N_{IX}=100K$.
}
\label{SIfig:simulations}
\end{figure*}

\medskip

Using the simulator Brian2~\cite{Stimberg2019}, we simulated the dynamics of networks of spiking neurons defined by  Eq.~\eqref{eq:membrane}.
We investigated networks of  $N_E$ excitatory and  $N_I$ inhibitory neurons; the two groups were driven by two populations of Poisson units of size   $N_{EX}$  and  $N_{IX}$, respectively. Simulations were performed for $N_E=N_I=N_{EX}=N_{IX}=10K$ and $100K$, with no significant differences between the two.
We used uniformly distributed delays of excitatory and inhibitory synapses. Delays were drawn randomly and independently at each existing synapse from uniform distributions in the range [0, 10]ms (E synapses) and [0, 1]ms (I synapses).
For fixed network parameters, the dynamics was simulated for 10 seconds with a time step of $10\mu$s.
We performed simulations  for different values of $K$; the values of  $a$ was rescaled according to the scaling relation of Eq.~\eqref{SIeq:scaling_network}.
From the resulting activity we measured firing rate, $CV$ and probability distribution of the membrane potential; results are shown in Fig.~\ref{SIfig:simulations}.
Mean field predictions are in qualitative agreement with numerical simulations, and the agreement improves as $a$ decreases.  
{Deviations from mean-field are expected to arise potentially from three factors: (1) Finite size of conductance jumps due to pre-synaptic action potentials; (2) Correlations in synaptic inputs to different neurons in the network due to recurrent connectivity; (3) Temporal correlations in synaptic inputs due to non-Poissonian firing behavior. In our simulations, deviations due to (1) and (2) become small when both $a$ and the connection probability are small. Deviations due to (3) become small when $\nu\ll 1/\tau_{rp}$, since as shown in Eq.~\eqref{SIeq:CV_sub} of  Appendix~\ref{SI:single_neuron_response_at_strong_coupling}, the statistics of presynaptic neurons firing tend to those of a Poisson process.}
As predicted by the mean field analysis, with increasing $K$ (and decreasing $a$) the network response becomes linear and approaches the asymptotic scaling; the firing remains irregular, as shown by the $CV$, and the membrane potential becomes Gaussian distributed.


\section{Effects of heterogeneity in the connectivity between neurons}\label{SI:conn_variability}
In this section, we describe how  fluctuations in single cell properties modify the expressions described above; in particular we investigate the effect of heterogeneities in number of connections per neuron in the simplified framework of model A. The formalism described here is a generalization to networks of conductance-based neurons of the analysis done in refs~\cite{Amit1997a,Roxin16217} for networks of current-based neurons.

We assume that the $i$-th neuron in the network receives projections from $K_X^i$, $K_E^i$ and  $K_I^i$ external, excitatory and inhibitory neurons, respectively.
These numbers are drown randomly from Gaussian distributions with mean $K$ ($\gamma K$) and variance $\Delta K^2$ ($\gamma^2 \Delta K^2$) for excitatory (inhibitory) synapses. Note that $\Delta K^2$ is assumed to be sufficiently small so that the probability to generate a negative number can be neglected.
Fluctuations in the number of connections are expected to produce a distribution of rates in the population, characterized by  mean and variance ${\nu}$ and  $\Delta{\nu}^2$. 
As a result,  the rates of incoming excitatory and inhibitory spikes differ from cell  to cell and become
\begin{equation}\label{eq:rates_z_i}
\begin{split}
 K^i_E r^i_E= K \left(r_E + \Delta_E z^i_E \right) \, , \quad
 K^i_I r^i_I= \gamma K \left(r_I + \Delta_I z^i_I \right) \, , \quad
 r_E= {\nu} +\nu_X \, , \quad r_I= {\nu}\, , \\
 \Delta_E^2= CV_K^2 \left( {\nu}^2 +\nu_X^2 \right) + \frac{\Delta \nu^2}{K}\approx  CV_K^2 \left( {\nu}^2 +\nu_X^2 \right)\, , \quad
\Delta_I^2= CV_K^2 {\nu}^2  + \frac{\Delta \nu^2}{\gamma K} \approx  CV_K^2 {\nu}^2 \, ;
 \end{split}
\end{equation}
where $r_{E,I}$ are the average presynaptic rates  and $z^i_{E,I}$ are realizations of a quenched normal noise with zero mean and unit variance,  fixed in a given realization of the network connectivity.
Starting from Eq.~\eqref{eq:rates_z_i}, the rate $\nu^i$ of the cell is derived as in the case without heterogeneities, the main difference is that it is now a function of the particular realizations of $z^i_{E}$ and $z^i_{I}$.
The quantities ${\nu}$ and  $\Delta{\nu}^2$ are obtained from population averages through the self consistency relations 
\begin{equation}\label{eq:rate_z}
\begin{cases} 
{\nu}=\langle  \nu(z_E,z_I) \rangle \, , \\
\Delta{\nu}^2=\langle  \nu(z_E,z_I)^2 \rangle -{\nu}^2\, , 
\end{cases}
\end{equation}
where $\langle  . \rangle$ represents the Gaussian average over the variables $z_E$ and $z_I$.
Once ${\nu}$ and  $\Delta{\nu}^2$ are known, the probability distribution of firing rate in the population is given by
\begin{equation}\label{eq:P_nu_integral_delta}
P(\nu)=\frac1{{2 \pi}} \int _{-\infty}^{\infty} dz_E dz_I e^{-z_E^2/2}e^{-z_I^2/2} \delta\left[\nu-\nu(z_E,z_I)\right] \, .
\end{equation}

As showed in the main text (Fig.~\ref{fig:simulations_variability}A), Eq.~\eqref{eq:P_nu_integral_delta}  captures quantitatively the heterogeneity in rates observed in numerical simulations.

In the large $K$ (small $a$) limit, the mathematical expressions derived above simplify significantly. 
First, as long as the parameter $\mu^i$  of the i-th neuron is  below threshold, its rate  is given by an expression analogous to Eq.~\eqref{eq:single_neuron} which, for small $\Delta_{E,I}$,  can be written
\begin{equation}
    \mathcal{Q}_i =\mathcal{Q} \, \exp\left(\Gamma z_i\right) \, , \quad \Gamma^2= \left(\frac{\partial {v}_{max}^2}{\partial r_E} {\Delta_E} \right)^2+\left(\frac{\partial {v}_{max}^2}{\partial r_I} {\Delta_I} \right)^2 \, ,
\end{equation}
where $z^i$ is generated from a Gaussian random variable with zero mean and unit variance. 
Moreover, if responses are far from saturation, the single rate can be written as 
\begin{equation}\label{eq:nu_exp_z}
\nu_i=\frac{\nu_X}{\mathcal{Q}_i}=\nu_0 \exp\left(-\Gamma z_i\right)\, , \quad  \Gamma^2=\Omega^2 \frac{CV_K^2}{a^2}\, , \quad 
\Omega^2=\left[  \left(a \frac{\partial {v}_{max}^2}{\partial (r_E/\nu_X)} \right)^2({\rho^2+1})^2 +\left(a \frac{\partial {v}_{max}^2}{\partial (r_I/\nu_X)} \right)^2  {\rho^2} \right]
\end{equation}
where $\nu_0$ is the rate in the absence of quenched noise (i.e. Eq.~\eqref{eq:network_solution} of the main text).
It is easy to show that, in Eq.~\eqref{eq:nu_exp_z}, $\Omega^2$ is independent of $a$, $K$ and $\nu_X$ in the large $K$ (small $a$) limit.
Finally,  as noted in~\cite{Roxin16217}, if the single neuron rate can be expressed as an exponential function of a quenched variable $z$,  Eq.~\eqref{eq:P_nu_integral_delta} can be integrated exactly and the distribution of rates is lognormal and given by
\begin{equation}\label{eq:P_nu_lognormal}
P(\nu)=\frac1{\sqrt{2 \pi} \Gamma \nu} \, \exp \left( -\frac{\left( \log(\nu)- \log(\nu_0) \right)^2}{2\Gamma^2} \right) \, .
\end{equation} 
Therefore, when the  derivation of Eq.~\eqref{eq:nu_exp_z} is valid, rates in the network should follow a log normal distribution, with parameters given by 
\begin{equation}
    \begin{cases}
    {\nu}=\nu_{0}\, \exp\left( \frac{\Gamma^2}2\right) \\
    \Delta{\nu}^2={\nu}^2\, \left[ \exp\left( \frac{\Gamma^2}2 \right)-1\right] \, 
    \end{cases}\,,
\end{equation}
For $\Gamma^2\ll1$, we find $\Delta{\nu}/\nu\approx\Gamma/2$ which scales linearly with $CV_K$, consistent with numerical results shown in Fig.~\ref{fig:simulations_variability}C.

\section{Finite synaptic time constants}\label{SI:finite_tau_S}
In this section,  we discuss the effect of synaptic time constant on single neuron and network responses. 
First, we derive an approximated expression for the single neuron membrane time constant; we then compute approximated expressions which are valid for different values of the ratio $\tau_S/\tau$; at the end of the section, we discussing the response of networks of neurons with large $\tau_S/\tau$.

The single neuron membrane potential dynamics is given by
\begin{equation}\label{eq:g_dynamics_tau_S_SI}
\begin{cases}
\mathcal{C}_j \dot{V}_j(t)=- g^j_L \left( V_j -E_{L}\right)-\sum\limits_{A=E,I}
g^j_{A}(t)\left( V_j-E_{A}\right)\, ,  \\
\tau_E \dot{g}^j_E=-{g}^j_E+{g^j_L} \tau_E \sum_m a_{jm} \sum_{n} \delta(t-t_m^n-D) \, \\
\tau_I \dot{g}^j_I=-{g}^j_I+{g^j_L} \tau_I \sum_m a_{jm} \sum_{n} \delta(t-t_m^n-D) \, 
\end{cases}
\end{equation}
Using the effective time constant approximation~\cite{Richardson2005}, we have
\begin{equation}
\begin{cases}
\mathcal{C} \dot{V}=- g_{0}\left( V -\mu\right)-g_{EF}\left( \mu-E_{E}\right)-g_{IF}\left( \mu-E_{I}\right)\, ,  \\
\tau_E \dot{g}_{EF}=-{g}_{EF}+ \sigma_E \sqrt{\tau_E} \zeta_E \, ,\\
\tau_I \dot{g}_{IF}=-{g}_{IF}+ \sigma_I \sqrt{\tau_I} \zeta_I \, ,
\end{cases}
\end{equation}
where $g_{AF}$ represents the fluctuating component of the conductance $g_{A}$, i.e.
\begin{equation}
\begin{split}
g_A(t)={g}_{A0}+{g}_{AF} (t) \, ,
\end{split}
\end{equation}
and
\begin{equation}
\begin{split}
\langle \zeta_A(t)\zeta_B(t') \rangle=\delta_{A,B}\delta(t-t') \, , \quad {g}_{0}=g_L+{g}_{E0}+{g}_{I0} \, , \\
 {g}_{A0}=a_A\tau_AR_A \, , \quad {\sigma}_{A}^2=a_A^2\tau_AR_A
\end{split}
\end{equation}
We are interested in stationary response, we introduce the term
\begin{equation}
z=\left( \mu-E_{E}\right) {g}_{EF}+\left( \mu-E_{I}\right) {g}_{IF}\end{equation}
with derivative
\begin{equation}
\dot{z}=\left(\mu-E_{E}\right)\frac{-{g}_{EF}+ \sigma_E \zeta_E}{\tau_E}+
\left(\mu-E_{I}\right)\frac{-{g}_{IF}+ \sigma_I \zeta_I}{\tau_I}
\end{equation}
Since we are interested in understanding the effect of an additional time scale, we can simplify the analysis assuming a unique synaptic time scale $\tau_E=\tau_I=\tau_S$ 
and obtain
\begin{equation}
\begin{split}
\tau_S \dot{z}=-z+\sigma_z \sqrt{ \tau_S}\zeta \, \\
\sigma_z^2=\sigma_E^2\left(\mu-E_{E} \right)^2+\sigma_I^2\left(\mu-E_{I} \right)^2
\end{split}
\end{equation}
To have the correct limit for $\tau_S\rightarrow0$, we impose $a_A=a_{A0}\tau_L/\tau_S$, where  $a_{A0}$ is the value of the synaptic efficacy in the limit of instantaneous synaptic time scale. With these assumptions the system equation becomes
\begin{equation}\label{eq:syna_time_constant}
\begin{cases}
{\tau} \frac{dV}{dt}=-\left( V-\mu\right) -\sigma \sqrt{\frac{\tau}{\tau_S}}z \,   , \\
{\tau_s} \frac{dz}{dt}=-z + \sqrt{\tau_S}\zeta \,   .
\end{cases}
\end{equation}
One can check that in the limit $\tau_S\rightarrow0$, the equations become analogous to those of the main text with $\eta=z/\sqrt{\tau_S}$. 
In what follows, we provide approximated expressions for the single neuron transfer function in three regimes: small time constant~\cite{Brunel1998}, large time constant \cite{MorenoBote2004}, and for intermediate values~\cite{Badel2011}. We also note that a numerical procedure to compute the firing rate exactly for any value synaptic time constant was introduced recently, using Fredholm theory \cite{van_Vreeswijk_2019}.

\subsection{Single neuron transfer function for different values of $\tau_S/\tau$}

\begin{figure}[htb!]
\centering
\includegraphics[width=15cm]{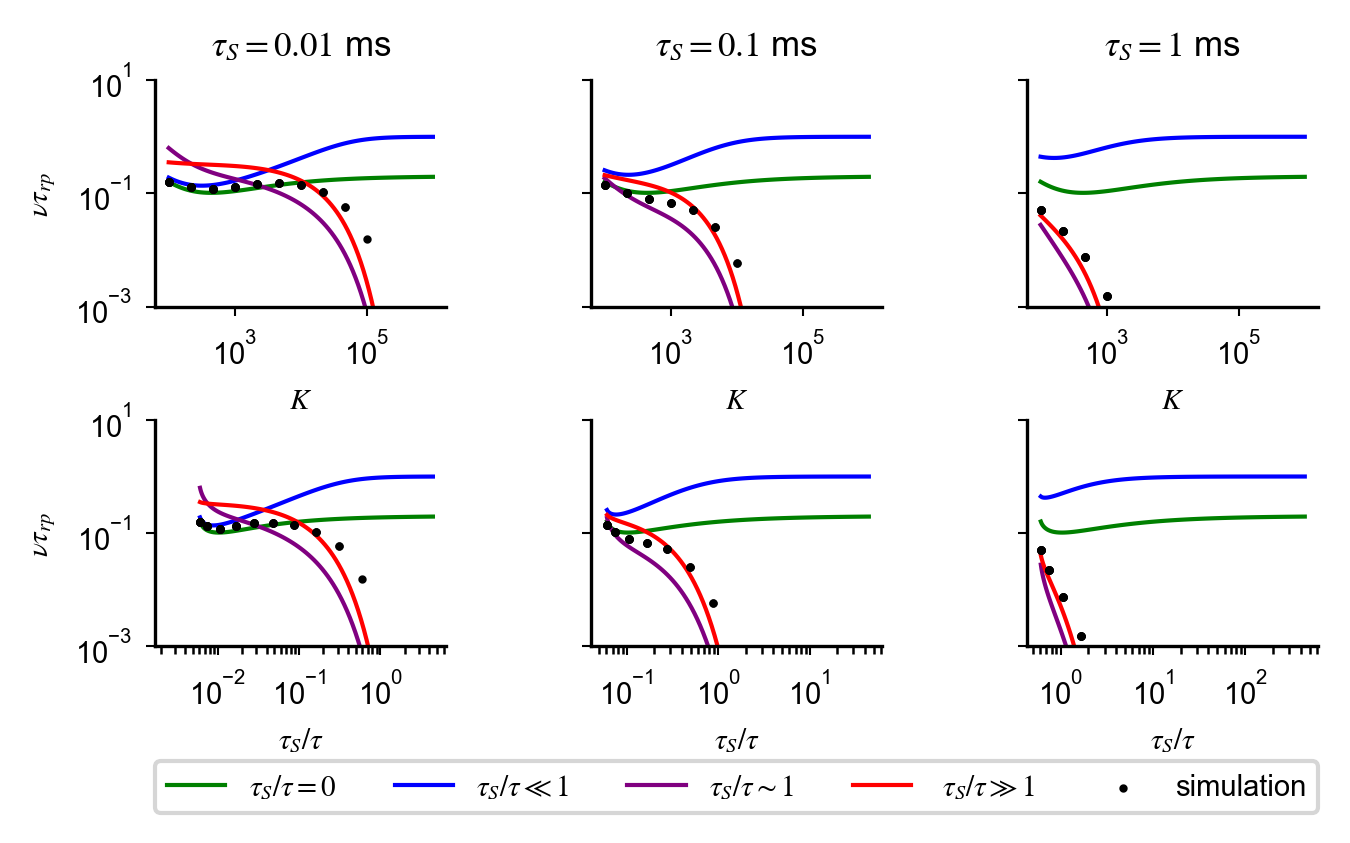}
\caption{\textbf{Synaptic time constant suppresses single neuron response in the strong coupling limit.} 
Single neuron response for different values of $K$, with $a$ rescaled according to Eq.~\eqref{eq:scaling_sub}.
Rates are plotted  as a function of $K$ (first row) and $\tau_S/\tau$ (second row); different columns correspond to different synaptic time constant $\tau_S$ (title).
As $K$ increases, because of the synaptic time constant $\tau_S$ non-negligible compared to the membrane time constant $\tau$, rates computed numerically from Eq.~\eqref{eq:g_dynamics_tau_S} (black dots) depart from the prediction of Eq.~\ref{eq:nu} (green). 
The dependency of the rate on $K$ is captured by Eq.~\eqref{eq:small_tau_S} (blue) for small values of $\tau_S/\tau$ and by Eq.~\eqref{eq:large} (red) for large values of $\tau_S/\tau$. 
This decay cannot be prevented by a new scaling relation of $a$ with $K$ and provides an upper bound to how much coupling can be increased while preserving response.
Simulations parameter: $a=0.006$ for $K=10^3$, $g=12$, $\eta=1.46$.
}
\label{fig:synaptic_time_constant_SM}
\end{figure}

For $\tau_S/\tau\ll 1$, as shown in~\cite{Brunel1998}, the firing rate can be computed with a perturbative expansion and is given by
\begin{equation}\label{eq:small_tau_S}
\frac1{\nu}=\tau \sqrt{\pi} \int_{\tilde{v}_{min}}^{\tilde{v}_{max}}dx\left(1+\text{erf}(x) \right)\, ,
\quad 
\tilde{v}(x)=\frac{x-\mu}{\sigma}-\tilde{\alpha}\sqrt{\frac{\tau_S}{\tau}}\, . 
\end{equation}
with  $\bar{\alpha}=-\zeta(1/2)\approx 1.46$.
As shown in Fig.~\ref{fig:synaptic_time_constant_SM}, Eq.~\eqref{eq:small_tau_S} generates small corrections around the  prediction obtained with instantaneous synapses,  and captures well the response for values  $\tau_S/\tau\lesssim0.1$. 

For $\tau_S/\tau\approx 1$, as shown in~\cite{Badel2011}  using Rice formula~\cite{Rice1944}, the single neuron firing rate is well approximated by the rate of upward threshold crossing of the membrane potential dynamics without reset. Starting from Eq.~\eqref{eq:syna_time_constant} and using the results of~\cite{Badel2011}, we obtain
\begin{equation}\label{eq:intermediate}
\nu=\frac{1}{2\pi \sqrt{\tau\tau_S}}\exp\left[-v_{max}
^2  \left(1+\frac{\tau_S}{\tau}\right)\right]\, .
\end{equation}

For $\tau_S/\tau\gg 1$, as shown in~\cite{MorenoBote2004}, the neuron fires only when fluctuations of $z$ are large enough for $V$ to be above threshold; the corresponding rate is given by  
\begin{equation}\label{eq:large}
\nu=\int_{v_{max}/\epsilon}^{\infty} dw \, \frac{e^{-w^2}}{\sqrt{\pi}} \frac1{\tau_{rp}+\tau\log\left( \frac{v_{min}-\epsilon w}{v_{max}-\epsilon w}\right)} \, , 
\quad 
\epsilon=\sqrt{\frac{\tau}{\tau_S}}
\end{equation}
As shown in Fig.~\ref{fig:synaptic_time_constant_SM}, Eq.~\eqref{eq:large} captures the response for values  $\tau_S/\tau\gtrsim 1$ and predicts a strong suppression of response at larger $\tau_S/\tau$.

 Higher order terms in the $\tau_S/\tau$ expansion could be computed using the approach described in \cite{van_Vreeswijk_2019}. 
However, Fig.~\ref{fig:synaptic_time_constant_SM} shows that Eqs.~(\ref{eq:small_tau_S}-\ref{eq:large}) 
are sufficient to capture quantitatively responses observed in numerical simulations for different regimes of $\tau_S/\tau$.  Eqs.~(\ref{eq:small_tau_S}-\ref{eq:large}) show that the  single neuron response is a nonlinear function of input rates, this nonlinearity  prevents  a scaling relation between $a$ and $K$ to rescue the  suppression observed in Fig.~\ref{fig:synaptic_time_constant_SM} and Fig.~\ref{fig:synaptic_time_constant}A.

\subsection{Network response for $\tau_S/\tau$ larger than one}
\begin{figure}[htb!]
\centering
\includegraphics[width=12.7cm]{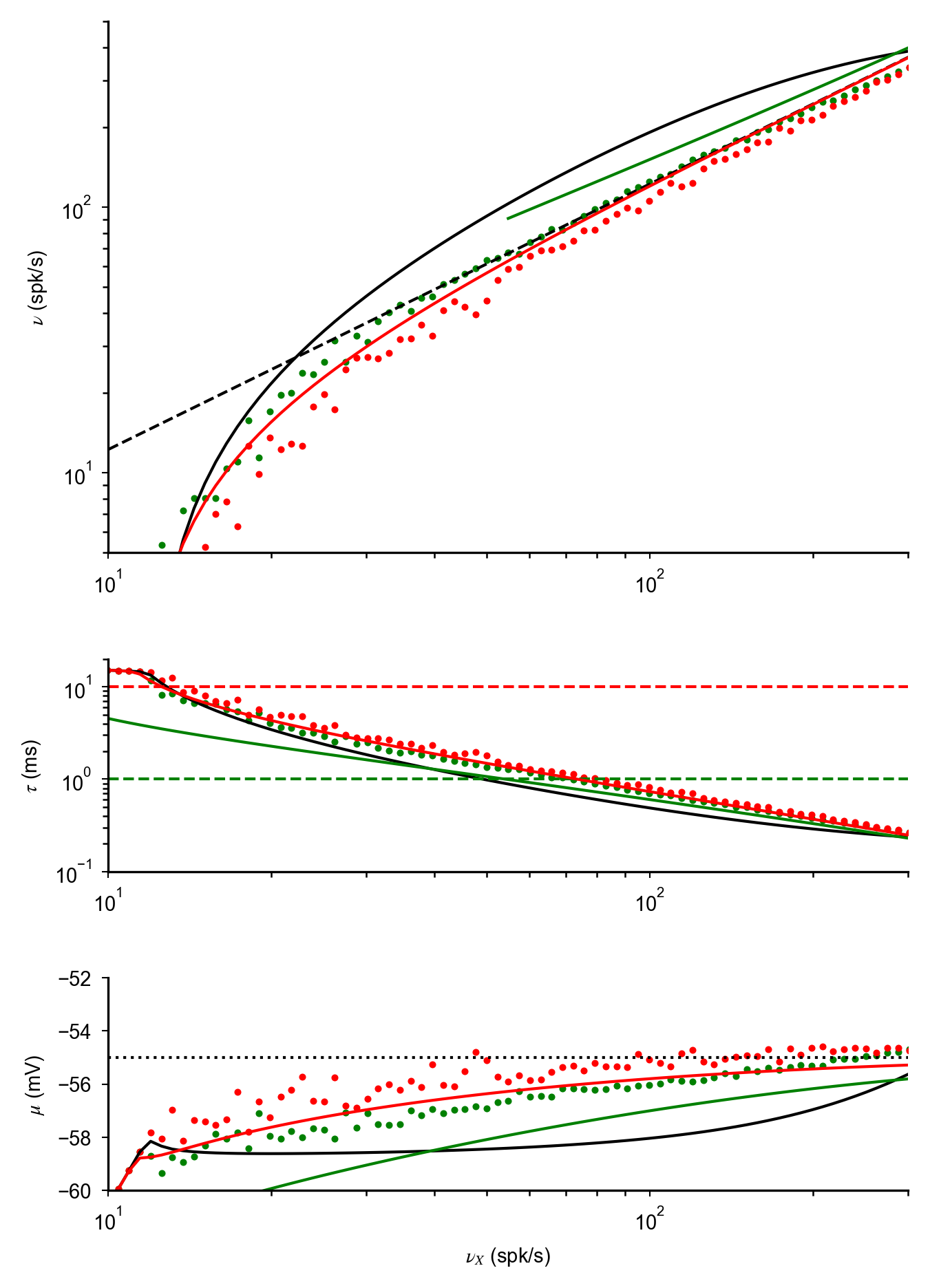}
\caption{\textbf{Approximation of network response for large $\tau_S/\tau$.} 
Plots analogous to Fig.~\ref{fig:synaptic_time_constant}B,C of the main text. 
Dots represent network response as a function of input rate $\nu_X$, computed numerically from Eqs.~\eqref{eq:membrane},
\eqref{eq:g_dynamics_tau_S} for $\tau_S=1$ms (green) and $\tau_S=10$ms (red).
Continuous lines correspond to the prediction obtained with instantaneous synapses (black) and for large synaptic time constant (Eqs.~(\ref{eq:SI_nu_tau_S},\ref{eq:param_MM}, \ref{eq:SI_E_tau_S}), colored lines). 
As explained in the text, the latter predictions are valid only for large $\tau_S/\tau$; because of this, we plotted only values obtained for $\tau_S/\tau>1$.
For $\tau_S/\tau\gg1$, the network response is well describe by Eq.~\eqref{eq:response_large_tau_S} of the main text.
}
\label{fig:synaptic_time_constant_network_SM}
\end{figure}
In this section, we study responses in networks of neurons with large $\tau_S/\tau$.
As in the case of instantaneous synapses, the network response can be obtained solving the self-consistency relation given by the single neuron transfer function using input rates
\begin{equation*}
    r_E=\nu_X+\nu\, , \quad r_I=\nu\, .
\end{equation*}
In particular, solutions of the implicit equation generated by Eq.~\eqref{eq:large} give the network response in the region of inputs for which $\tau_S/\tau\gg 1$. 
In this region of inputs,  assuming coupling to be strong, the implicit equation becomes
\begin{equation}\label{eq:SI_nu_tau_S}
    \nu=\frac{\sqrt{\tau/\tau_S}}{\tau_{rp} v_{max}\sqrt{\pi}}
    \exp\left(-v_{max}^2\frac{\tau_S}{\tau}\right) \, .
\end{equation}
Eq.~\eqref{eq:SI_nu_tau_S}, which is validated numerically in Fig.~\ref{fig:synaptic_time_constant_network_SM}, implies that  firing is preserved if $v_{max}\sqrt{{\tau_S}/{\tau}}$ is of order one,  i.e.  if
\begin{equation}\label{eq:SI_E_tau_S}
    \mu\sim \theta-\sigma\sqrt{\frac{\tau}{\tau_S}}\sim \theta-\frac1{\sqrt{K}}
    \frac{\sigma/\sqrt{a}}{\sqrt{\tau_S\left[\nu_X+ \nu \left( 1+g \gamma\right)\right]}} \, .
\end{equation}
Combining the above equation with the definition of $\mu$, we obtained Eq.~\eqref{eq:response_large_tau_S}, which captures  the behavior of network response observed in numerical simulations for $\tau_S/\tau\gg 1$ (Fig.~\ref{fig:synaptic_time_constant}B and Fig.~\ref{fig:synaptic_time_constant_network_SM}).

Eq.~\eqref{eq:SI_nu_tau_S} can be used to understand the effect of connection-heterogeneity in networks with large $\tau_S/\tau$. In particular, generalizing the analysis of Appendix~\ref{SI:conn_variability}, we found that rates in the network, in the limit of small $CV_K$  and large $K$,  are given by  
\begin{equation}
    \nu_i=\nu_0 \, \exp \left[\Omega_S \frac{CV_K}{\sqrt{K}} z_i\right]
\end{equation}
where $\nu_0$ is the population average in the absence of heterogeneity (i.e. the solution of Eq.~\eqref{eq:SI_nu_tau_S}), and $z_i$ is a Gaussian random variable of zero mean and unit variance. 
The prefactor $\Omega_S$, which is independent of $a$ and $K$, and is given by
\begin{equation}\label{eq:nu_variab_tau_S}
    \Omega_S^2=
    \left[
    \left(\frac{\partial f(r_E,r_I)}{\partial r_E} \right)^2\left( \nu^2+\nu_X^2 \right)+\left(\frac{\partial f(r_E,r_I)}{\partial r_I}\right)^2 \nu^2\right] \, , \quad
    f(r_E,r_I)=\frac{v_{max}^2 \, \tau_S}{K\, \tau} \, .
\end{equation}
Eq.~\eqref{eq:nu_variab_tau_S} is a generalization of Eq.~\eqref{eq:nu_variab} to the case of large $\tau_S/\tau$. 
It shows that, in  this limit, the state of the network is preserved with connection fluctuations up to $CV_K\sim1/\sqrt{K}$.


%
%
%

\end{document}